\documentclass[12pt]{iopart}
\usepackage[dvips]{graphics}
\usepackage{graphicx}
\usepackage{amsfonts}
\usepackage{amssymb}
\usepackage{subfigure}
\usepackage{cite}

\begin{document}

\topical{Dark solitons in atomic Bose-Einstein condensates: from theory to experiments}

\author{D J Frantzeskakis}

\address{Department of Physics, University of Athens,
Panepistimiopolis, Zografos, Athens 15784, Greece}

\begin{abstract}

This review paper presents an overview of the theoretical and experimental progress
on the study of matter-wave dark solitons in atomic Bose-Einstein condensates. Upon
introducing the general framework, we discuss the statics and dynamics of single and
multiple matter-wave dark solitons in the quasi one-dimensional setting,
in higher-dimensional settings, as well as in the dimensionality crossover regime.
Special attention is paid to the connection between theoretical results, obtained by
various analytical approaches, and relevant experimental observations.

\end{abstract}

\submitto{\JPA}

{\tableofcontents}

\maketitle

\pagestyle{plain}

\section{Introduction}

A dark soliton is an envelope soliton having the form of a density dip with a phase jump
across its density minimum. This localized nonlinear wave exists on top of a stable
continuous wave (or extended finite-width) background. Dark solitons are the most
fundamental nonlinear excitations of a universal model, the nonlinear Schr\"{o}dinger
(NLS) equation with a defocusing nonlinearity, and, as such, they have been studied in
diverse branches of physics. Importantly, apart from a vast amount of literature devoted
to relevant theoretical works, there exist many experimental results on dark solitons,
including the observation of optical dark solitons, either as temporal pulses in optical
fibers \cite{fiber1,fiber2}, or as spatial structures in bulk media and waveguides
\cite{spatial1,spatial2}, the excitation of a non-propagating kink in a
parametrically-driven shallow liquid \cite{den1}, dark soliton standing waves in a
discrete mechanical system \cite{den2}, high-frequency dark solitons in thin magnetic
films \cite{magn}, dissipative dark solitons in a complex plasma \cite{compla}, and so on.

Theoretical studies on dark solitons started as early as in 1971 \cite{tsuzuki} in the
context of Bose-Einstein condensates (BECs). In particular, in Ref.~\cite{tsuzuki},
exact soliton solutions of the Gross-Pitaevskii (GP) equation (which is a variant of
the NLS model) \cite{GP} were found and connected, in the small-amplitude limit,
with the solitons of the Korteweg-de Vries (KdV) equation.
Later, and shortly after the integration of the focusing NLS equation \cite{zsb},
the defocusing NLS equation was also shown \cite{zsd} to be completely integrable
by means of the Inverse Scattering Transform (IST) \cite{segur}; this way, single
as well as multiple dark soliton solutions of arbitrary amplitudes were found
analytically. The IST approach allowed for an understanding of the formation of
dark solitons \cite{gred1,gred2,zhaobour,zhao,KonVek,slavin}, the interaction and
{\it collision} between dark solitons \cite{zsd,Blow} (see also
Refs.~\cite{AA,gagnon,Thurston,snyder,lads} and \cite{foursa} for relevant theoretical
and experimental studies, respectively), and paved the way for the development of
perturbation methods for investigating their dynamics in the presence of perturbations
\cite{lads,uz,KY,vvkpert,jost1,jost2,lashkin}. From a physical standpoint, dark solitons
were mainly studied in the field of nonlinear optics --- from which the term ``dark'' was
coined. The first theoretical work in this context, namely the prediction of dark
solitons in nonlinear optical fibers at the normal dispersion regime \cite{Hasegawa},
was subsequently followed by extensive studies of {\it optical dark solitons}
\cite{kivpr,book}.

A new era for dark solitons started shortly after the realization of atomic BECs
\cite{bec1,bec2,bec3}; this achievement was awarded the Nobel prize in physics of 2001
\cite{becnl1,becnl2}, and has been recognized as one of the most fundamental recent
developments in quantum and atomic physics over the last decades (see, e.g., the books
\cite{book1,book2} for reviews). In an effort to understand the properties of this
exciting state of matter, there has been much interest in the macroscopic nonlinear
excitations of BECs (see reviews in Refs.~\cite{BECBOOK,revnonlin}). In that regard,
the so-called {\it matter-wave dark solitons}, were among the first purely nonlinear
states that were experimentally observed in BECs \cite{han1,nist,dutton,bpa,han2}.

The interest on matter-wave dark solitons is not surprising due to a series of reasons.
First of all, for harmonically confined BECs, these structures are the nonlinear
analogues of the excited states of a ``prototype'' quantum system \cite{gp1d_B,konotop1},
namely the quantum harmonic oscillator \cite{landau}. On the other hand, the topological
nature of matter-wave dark solitons (due to the phase jump at their density minimum)
renders them a ``degenerate'', one-dimensional (1D) analogue of {\it vortices}, which
are of paramount importance in diverse branches of physics \cite{pismen}. Additionally,
and perhaps more importantly, matter-wave dark solitons are --- similarly to vortices
\cite{kibble,zurek1,bpanat} --- quite fundamental structures arising spontaneously upon
crossing the BEC phase-transition \cite{zurek2,zurek3}, with properties which may be
used as diagnostic tools probing the rich physics of a purely quantum system (BEC)
at the mesoscale \cite{anglin}. Finally, as concerns applications, it has been proposed
that the dark soliton position can be used to monitor the phase acquired in an atomic
matter-wave interferometer in the nonlinear regime \cite{appl1,appl2} (see also
relevant experiments of Refs.~\cite{jo,appl3} devoted to atom-chip interferometry
of BECs).

The early matter-wave dark soliton experiments, as well as previous works on dark
solitons in optics, inspired many theoretical efforts towards a better understanding
of the stability, as well as the static and dynamical properties of matter-wave dark
solitons. Thus, it is probably not surprising that a new series of experimental results
from various groups have appeared
\cite{scottOL,ginsberg2005,engels,hamburg,hambcol,kip,technion,draft6},
while still other experiments --- not directly related to dark solitons ---
reported observation of these structures \cite{jo,appl3,enghoef}. These new, very recent,
experimental results were obtained with an unprecedented control over the condensate
and the solitons as compared to the earlier soliton experiments. Thus, these
``new age'' experiments were able not only to experimentally verify various theoretical
predictions, but also to open new exciting possibilities. Given this emerging interest,
and how new experiments in BEC physics inspire novel ideas --- both in theory and
in experiments --- new exciting results are expected to appear.

The present paper aims to provide an overview of the theoretical and experimental
progress on the study of dark solitons in atomic BECs. The fact that there are many
similarities between optical and matter-wave dark solitons \cite{analogies}, while
there exist excellent reviews on both types of dark solitons (see Ref.~\cite{kivpr}
for optical dark solitons and Ch.~4 in Ref.~\cite{BECBOOK} for matter-wave dark solitons),
provides some restrictions in the article: first, the space limitations of the article,
will not allow for an all-inclusive presentation; in that regard, important
entities --- relevant to dark solitons --- such as vortices
\cite{pismen,fetter,ourvortex} and vortex rings \cite{kominsolo,barenghi}
will only be discussed briefly. In fact, this review (which obviously entails
a ``personalized'' perspective on dark solitons) will cover the basic theory
emphasizing, in particular, to the connection between the theoretical results and
experimental observations; this way, in most cases, theoretical discussion will be
immediately followed by a presentation of pertinent experimental results. In that regard, it is also
relevant to note that our theoretical approach will basically be based on the mean-field theory:
as will be shown, the latter can be used as a basis of understanding of most
effects and experimental findings related to matter-wave dark solitons;
this way, thermal and quantum effects --- which may be particularly relevant
and important in certain cases --- will only be briefly covered.
Following the above limitations, the structure of the manuscript will be as follows.

Section \ref{sec2} is devoted to the mean-field description of BECs. Particularly,
we first present the GP equation and discuss its connection with the respective full
quantum many-body problem. Next, we present the ground state of the condensate and discuss how
its small-amplitude excitations can be studied by means of the Bogoliubov-de Gennes (BdG) equations.
Lower-dimensional versions of the GP model, pertinent to highly anisotropic trapping potentials,
are also discussed; this way, depending on the shape of the trap, we start from purely three-dimensional
(3D) BECs and introduce elongated (alias ``cigar-shaped'') BECs, quasi one-dimensional
(1D) BECs and quasi two-dimensional (2D) (alias ``disk-shaped'') ones, as well as discuss
cases relevant to the dimensionality crossover regimes. The topics of strongly-interacting
Bose gases, and their relevant mean-field description, are also briefly covered.

Section \ref{sec3} provides the theoretical basis for the study of matter-wave dark
solitons. Specifically, first we present the completely integrable 1D NLS equation,
its basic properties and the dark soliton solutions. Relevant mathematical tools, such
as Inverse Scattering Transform (IST), the renormalization of the integrals of motion
of dark solitons and the small-amplitude approximation --- leading to the connection
of matter-wave dark solitons to Korteweg-de Vries (KdV) solitons --- are discussed.
Furthermore, the generation of matter-wave dark solitons by means of the phase-,
density- and quantum-state-engineering methods are also presented. We also provide
the multiple-dark soliton solutions of the NLS equation, and discuss their interactions
and collisions.

Section \ref{sec4} deals with matter-wave dark solitons in quasi-1D Bose gases.
Particularly, we first discuss the adiabatic dynamics of dark solitons in the presence
of the harmonic trap by means of different analytical techniques; these include the
Hamiltonian and Lagrangian approaches of the perturbation theory, the Landau dynamics
and the small-amplitude approximation approaches. Next, a connection between the
stability, statics and dynamics of dark solitons is presented, relying on a study of the
Bogoliubov spectrum of single- and multiple-dark solitons and the role of the pertinent
anomalous modes. Non-adiabatic effects, namely emission of radiation of solitons in the
form of sound waves, as well as rigorous results concerning the persistence and stability
of matter-wave dark solitons, are also discussed.

Section \ref{sec5} studies matter-wave dark solitons in higher-dimensional settings.
Considering, at first, the case of purely 2D or 3D geometries, the transverse
(alias ``snaking'') instability of rectilinear dark solitons, and the concomitant
soliton decay into vortex pairs or vortex rings, is presented. The theme of matter-wave dark solitons
of radial symmetry, namely ring dark solitons and spherical shell solitons, is also
covered. Furthermore, we present results concerning the stability of dark solitons in
cigar-shaped (3D) BECs, and in BECs in the dimensionality crossover regime from 3D to 1D;
in the latter experimentally relevant setting, both single- and multiple- dark soliton
statics and dynamics are analyzed.

In Section \ref{exp}, we discuss various experimentally relevant settings and parameter regimes
for matter-wave dark solitons. In particular, we first present results concerning matter-wave dark solitons in multi-component (pseudo-spinor and spinor) BECs.
Next, we discuss how matter-wave interference and the breakdown of
BEC superfluidity are connected to the generation of matter-wave dark solitons.
We continue by referring to matter-wave dark solitons in periodic potentials,
namely optical lattices (OLs) and superlattices, and conclude this Section by discussing
the statics and dynamics of dark solitons at finite temperatures.

Finally, in Section \ref{conclusions} we briefly summarize our conclusions and discuss
future challenges.


\section{Mean-field description of Bose-Einstein condensates}
\label{sec2}


Bose-Einstein condensation of dilute atomic gases is an unambiguous
manifestation of a macroscopic quantum state in a many-body system.
As such, this phenomenon has triggered an enormous amount of experimental and
theoretical work \cite{book1,book2}. Importantly, this field is intimately
connected with branches of physics such as superfluidity, superconductivity,
lasers, coherent optics, nonlinear optics, and physics of nonlinear waves.
Many of the common elements between BEC physics and the above areas, and in
particular optics, rely on the existence of macroscopic coherence in the
many-body state of the system. From a theoretical standpoint, this can be
understood by the fact that many effects related to BEC physics can be
described by a {\it mean-field} model, namely the Gross-Pitaevskii (GP)
equation \cite{GP}. The latter is a partial differential equation (PDE)
of the NLS type, which plays a key role --- among other fields --- in
nonlinear optics \cite{book}. Thus, BEC physics is closely connected to
nonlinear optics (and the physics of nonlinear waves), with vortices
and solitons being perhaps the most prominent examples of common nonlinear
structures arising in these areas \cite{BECBOOK,revnonlin}.

Below we will briefly discuss the theoretical background for the description of BECs.
We emphasize, in particular, lowest-order mean-field theory, as this can be
used as a basis to understand the nonlinear dynamics of matter-wave
dark solitons. Interesting effects naturally arise beyond the GP mean-field,
both due to thermal and quantum fluctuations. Such effects become particularly relevant
in extremely tightly confining geometries, or when the Bose-Einstein
condensation transition region is approached.

\subsection{The Gross-Pitaevskii equation.}
\label{GPE}

In order to describe theoretically the statics and dynamics of BECs a
quantum many-body approach is required \cite{book1,book2} (see also
Ref.~\cite{rmpdalibard} for a recent review on the many-body aspects
of BECs). Particularly, a sufficiently dilute ultracold atomic gas,
composed by $N$ interacting bosons of mass $m$ confined by an external
potential $V_{\rm ext}({\bf r})$, can be described by the many-body
Hamiltonian; the latter can be expressed, in second quantization form,
through the boson annihilation and creation
field operators, ${\hat \Psi}({\bf r},t)$ and ${\hat \Psi}^{\dagger}({\bf r},t)$
(which create and annihilate a particle at the position ${\bf r}$) namely,
\begin{eqnarray}
{\hat H} &=&  \int  d{\bf r} {\hat \Psi}^{\dagger}({\bf r}, t) {\hat H}_{0}
{\hat \Psi}({\bf r}, t)
\nonumber \\
&+& \frac{1}{2} \int d{\bf r} d{\bf r'} {\hat \Psi}^{\dagger}({\bf r}, t)
{\hat \Psi}^{\dagger}({\bf r'}, t) V({\bf r}-{\bf r}'){\hat \Psi}({\bf r}', t) {\hat \Psi}({\bf r}, t),
\label{H}
\end{eqnarray}
where ${\hat H}_{0}= - (\hbar^{2}/2m) \nabla^2 + V_{\rm ext}({\bf r})$ is
the single-particle operator and $V({\bf r}-{\bf r}')$ is the two-body
interatomic potential. Apparently, the underlying full many-body problem is
very difficult to be treated analytically (or even numerically) as $N$
increases and, thus, for convenience, a mean-field approach can be adopted.
The mean-field approach is based on the separation of the condensate
contribution from the boson field operator as follows \cite{Bogoliubov}:
\begin{equation}
{\hat \Psi}({\bf r}, t) = \langle  {\hat \Psi}({\bf r}, t) \rangle + {\hat \Psi}'({\bf r}, t)
=\Psi ({\bf r}, t) + {\hat \Psi}'({\bf r}, t).
\label{bdec}
\end{equation}
In the above expression, the expectation value of the field operator,
$\langle  {\hat \Psi}({\bf r}, t) \rangle \equiv \Psi ({\bf r}, t)$,
is known as the {\it macroscopic wave function of the condensate},
while ${\hat \Psi}'({\bf r}, t)$ describes the non-condensate part,
which accounts for quantum and thermal fluctuations.
Considering the case of a dilute ultracold gas with binary collisions
at low energy, characterized by the $s$-wave scattering length $a$,
the interatomic potential can be replaced by an effective delta-function
interaction potential,
$
V({\bf r'}-{\bf r}) = g \delta({\bf r'}-{\bf r})
$
\cite{book1,book2}, with the coupling constant $g$ given by
$g = {4 \pi \hbar^2 a}/{m}$.
Under these assumptions, a nontrivial zeroth-order theory for the BEC wave function
can be obtained by means of the Heisenberg evolution equation
$i \hbar (\partial {\hat \Psi} /\partial t) = [ {\hat \Psi}, {\hat H}]$,
upon replacing the field operator ${\hat \Psi}$ with the classical field $\Psi$,
i.e., ignoring the quantum and thermal fluctuations described by
${\hat \Psi}'({\bf r'}, t)$. Such a consideration leads to
the Gross-Pitaevskii (GP) equation \cite{GP}, which has the form:
\begin{equation}
i \hbar \partial_t \Psi({\bf r}, t) = \left[ -\frac{{\hbar}^2}{2 m} \nabla^2
+ V_{{\rm ext}}({\bf r}) + g |\Psi({\bf r}, t)|^2 \right] \Psi({\bf r}, t).
\label{gpe}
\end{equation}
In the above equation, $\Psi({\bf r}, t)$ is normalized to the number of atoms $N$,
namely,
\begin{equation}
N = \int |\Psi({\bf r}, t)|^2 d{\bf r},
\label{Ngpe}
\end{equation}
and the nonlinearity (which is obviously introduced by interatomic
interactions) is characterized by the $s$-wave scattering length $a$, which
is $a>0$ or $a<0$ for repulsive or attractive interatomic
interactions, respectively. Notice that Eq.~(\ref{gpe}) can be
written in canonical form, $i \hbar \partial_t \Psi = \delta E
/\delta \Psi^{\ast}$ (with $\ast$ denoting complex conjugate), where
the dynamically conserved energy functional $E$ is given by
\begin{equation}
E = \int d{\bf r} \left[ \frac{\hbar^2}{2m} |\nabla \Psi|^2
+ V_{\rm ext} |\Psi|^2 + \frac{1}{2}g |\Psi|^4 \right],
\label{Egpe}
\end{equation}
with the three terms in the right-hand side representing,
respectively, the kinetic energy, the potential energy and the
interaction energy.

A time-independent version of the GP equation can be obtained
upon expressing the BEC wave function as
$\Psi({\bf r}, t)=\Psi_{0}({\bf r}) \exp(-i\mu  t/\hbar)$,
where $\mu = \partial E / \partial N$ is
the chemical potential. This way, Eq.~(\ref{gpe}) yields the following equation for
the stationary state $\Psi_{0}$:
\begin{equation}
\left[-\frac{\hbar^2}{2m} \nabla^2 + V_{\rm ext}({\bf r})
+g |\Psi_{0}|^{2}({\bf r})\right]\Psi_{0}({\bf r})=\mu\Psi_{0}({\bf r}).
\label{tigpe}
\end{equation}


\subsection{The mean-field approach vs.~the many-body quantum mechanical problem.}

Although the GP equation is known from the early 60's \cite{GP}, it was only
recently shown that it can be derived rigorously from a self-consistent
treatment of the respective many-body quantum mechanical problem
\cite{chap12:Lieb1}. In particular, in Ref.~\cite{chap12:Lieb1} --- which dealt
with the stationary GP Eq.~(\ref{tigpe}) --- it was proved that the GP energy
functional describes correctly the energy and the particle density of a
trapped Bose gas to the leading-order in the small parameter $\bar{n}|a|^{3}$
\footnotemark[1]
\footnotetext[1]{The condition $\bar{n}|a|^{3} \ll 1$, which is also required for
the derivation of the GP Eq.~(\ref{gpe}), implies that the Bose-gas is ``dilute''
or ``weakly-interacting''; typically, in BEC experiments,
$\bar{n} |a|^{3} < 10^{-3}$ \cite{book2}.},
where $\bar{n}$ is the average density of the gas. The above results were proved in
the limit where the number of particles $N \rightarrow \infty$ and the scattering
length $a \rightarrow 0$, such that the product $N a$ stays constant. Importantly,
although Ref.~\cite{chap12:Lieb1} referred to the full three-dimensional (3D) Bose
gas, extensions of this work for lower-dimensional settings were also reported
(see the review \cite{chap01:elieb} and references therein).

The starting point of the analysis of Ref.~\cite{chap12:Lieb1} is the effective
Hamiltonian of $N$ identical bosons, which can be expressed
(in units so that $\hbar=2m=1$) as follows:
\begin{equation}
\label{qmHam}
H = \sum_{j=1}^{{N}} \left[ - \nabla^2_j+ V_{\rm ext}({\bf r}_j) \right] +
\sum_{i < j} v(|{\bf r}_i - {\bf r}_j|),
\end{equation}
where $v(|{\bf r}|)$ is a general interaction potential assumed to be spherically
symmetric and decaying faster than $|{\bf r}|^{-3}$ at infinity.
Then, assuming that the quantum-mechanical ground-state energy of the
Hamiltonian (\ref{qmHam}) is $E_{\rm QM}(N, \tilde{a})$ (here $N$ is the
number of particles and $\tilde{a}$ is the dimensionless two-body scattering length),
the main theorem proved in Ref.~\cite{chap12:Lieb1} is the following.
The GP energy is the dilute limit of the quantum-mechanical energy:
\begin{equation}
\label{convergence-1} \forall \tilde{a}_1 > 0 : \quad \lim_{n \to \infty}
\frac{1}{{N}} E_{\rm QM}\left({N},\frac{\tilde{a}_1}{n}\right) = E_{\rm
GP}(1,\tilde{a}_1),
\end{equation}
where $E_{\rm GP}({N},\tilde{a})$ is the energy of a solution of the
stationary GP Eq.~(\ref{tigpe}) (in units such that $\hbar=2m=1$),
and the convergence is uniform on bounded intervals of $\tilde{a}_1$.

The above results (as well as the ones in Ref.~\cite{chap01:elieb}) were proved for stationary
solutions of the GP equation, and, in particular, for the ground state solution. More recently,
the time-dependent GP Eq.~(\ref{gpe}) was also analyzed within a similar asymptotic analysis
in Ref.~\cite{Erdos}. In this work, it was proved that the limit points of the $k$-particle
density matrices of $\Psi_{{N},t}$ (which is the solution of the ${N}$-particle Schr\"{o}dinger equation)
satisfy asymptotically the GP equation (and the associated hierarchy of equations)
with a coupling constant given by $\int v(x)dx$, where $v(x)$ describes the interaction potential.

These rigorous results, as well as a large number of experimental results related to the physics of BECs,
indicate that (under certain conditions) the GP equation is a good starting point for analyzing the statics
and dynamics of BECs.


\subsection{Ground state and excitations of the condensate.}
\label{gpebdg}
Let us now consider a condensate confined in a harmonic external potential,
namely,
\begin{equation}
V_{\rm ext}({\bf r}) = \frac{1}{2} m(\omega_x^2 x^2 + \omega_y^2 y^2 + \omega_z^2 z^2),
\label{hpot}
\end{equation}
where $\omega_x$, $\omega_y$, and $\omega_z$ are the (generally
different) trap frequencies along the three directions.
In this setting, and in the case of repulsive
interatomic interactions ($a>0$) and sufficiently
large number of atoms $N$, Eq.~(\ref{tigpe}) can be used to
determine analytically the {\it ground state} of the system.
In particular, in the asymptotic limit of
$N a/a_{\rm ho} \gg 1$ (where $a_{\rm ho}=\sqrt{\hbar/(m\omega_{\rm ho})}$ is
the harmonic oscillator length associated with the geometrical average
$\omega_{\rm ho}=(\omega_x \omega_y \omega_z)^{1/3}$ of the trap frequencies),
it is expected that the atoms are pushed
towards the rims of the condensate, resulting in slow spatial
variations of the density profile $n({\bf r}) \equiv |\Psi_{0}({\bf r})|^2$.
Thus, the latter can be obtained as an algebraic solution stemming from
Eq.~(\ref{tigpe}) when neglecting the kinetic energy term --- the
so-called {\it Thomas-Fermi (TF)} limit \cite{book1,book2,BECBOOK}:
\begin{equation}
n({\bf r})= g^{-1} \left[ \mu-V_{\rm ext}({\bf r}) \right],
\label{TF}
\end{equation}
in the region where $\mu>V_{\rm ext}({\bf r})$, and $n=0$ outside,
and the value of $\mu$ being determined by the normalization
condition [cf.~Eq.~(\ref{Ngpe})]. Notice that the TF approximation
becomes increasingly accurate for large values of $\mu$.

{\it Small-amplitude excitations} of the BEC can be studied upon
linearizing Eq. (\ref{tigpe}) around the ground state. Particularly,
we consider small perturbations of this state, i.e.,
%
\begin{equation}
\Psi({\bf r}, t)= e^{-i\mu t/\hbar} \left[ \Psi_{0}({\bf r})
+ \sum_{j} \left( u_{j}({\bf r}) e^{-i\omega_{j} t}
+\upsilon_{j}^{\ast}({\bf r}) e^{i \omega_{j} t} \right) \right],
\label{pert}
\end{equation}
where $u_j$, $\upsilon_j$ are the components of the linear response
of the BEC to the external perturbations that oscillate at
frequencies $\pm \omega_j$ [the latter are (generally complex)
eigenfrequencies]. Substituting Eq.~(\ref{pert}) into
Eq.~(\ref{tigpe}), and keeping only the linear terms in $u_j$ and
$\upsilon_j$, we obtain the so-called {\it Bogoliubov-de Gennes}
(BdG) equations:
\begin{eqnarray}
&&\left[\hat{H}_{0}-\mu+ 2g\, |\Psi_{0}|^{2}({\bf r})\right] u_{j}({\bf r})
+g\, \Psi_{0}^{2}({\bf r}) \upsilon_{j}({\bf r})
= \hbar\, \omega_{j}\, u_{j}({\bf r}),
\nonumber\\[1.0ex]
&&\left[\hat{H}_0-\mu+2g\, |\Psi_{0}|^{2}({\bf r})\right] \upsilon_{j}({\bf r})
+ g\, \Psi_{0}^{\ast 2} ({\bf r}) u_{j}({\bf r})
= -\hbar\, \omega_{j}\, \upsilon_{j}({\bf r}),
\label{BdG}
\end{eqnarray}
where $\hat{H}_0 \equiv -(\hbar^2 / 2m) \nabla^2 + V_{\rm ext}({\bf
r})$ is the single-particle Hamiltonian. Importantly, these equations
can also be used, apart from the ground state, for any other
stationary state (including, e.g., solitons) with the function
$\Psi_0$ being modified accordingly. In such a general context, the
BdG equations provide the eigenfrequencies $\omega \equiv
\omega_{r}+i \omega_{i}$ and the amplitudes $u_j$ and $\upsilon_j$
of the normal modes of the system. Note that due to the Hamiltonian
nature of the system, if $\omega$ is an eigenfrequency of the
{\it Bogoliubov spectrum}, so are $-\omega$,
$\omega^{\ast}$ and $-\omega^{\ast}$. In the case of stable
configurations with $\omega_i =0$, the solution of BdG equations
with frequency $\omega$ represent the same physical oscillation with
the solution with frequency $-\omega$ \cite{book2}.

In the case of a homogeneous gas ($V_{\rm ext}({\bf r})=0$) characterized by a constant
density $n_0 =|\Psi_0|^2$, the amplitudes $u_j$ and $\upsilon_j$ in the BdG equations
are plane waves, $\sim \exp(i {\bf k} \cdot {\bf r})$,
of wave vector $\bf k$. Then, Eqs.~(\ref{BdG}) lead to the dispersion relation,
\begin{equation}
(\hbar \omega)^2=\left(\frac{\hbar^2{\bf k}^2}{2m}\right)
\left(\frac{\hbar^2 {\bf k}^2}{2m}+2g n_0\right).
\label{Bogoliubov}
\end{equation}
In the case of repulsive interatomic interactions ($g>0$), Eq.~(\ref{Bogoliubov})
indicates that small-amplitude harmonic excitations of the stationary state
\begin{equation}
\Psi=\sqrt{n_0}\exp(-i\mu t/\hbar),
\label{pedestal}
\end{equation}
with $\mu = n_0$, are always stable since $\omega_i =0$ for every
$\bf k$. Thus, this state is not subject to the {\it modulational
instability} (see, e.g., Ref.~\cite{pandim} and references therein). This
fact is important, as the wave function of Eq. (\ref{pedestal}) can
serve as a stable background (alias ``pedestal''), on top of which
strongly nonlinear localized excitations may be formed; such
excitations may be, e.g., matter-wave dark solitons which are of
particular interest in this work. Notice that the above mentioned
small-amplitude harmonic excitations are in fact {\it sound waves},
characterized by the phonon dispersion relation $\omega= |{\bf k}|c_s$
[see Eq.~(\ref{Bogoliubov}) for small momenta $\hbar {\bf k}$],
where
\begin{equation}
c_s=\sqrt{g n_0/m},
\label{sound}
\end{equation}
is the {\it speed of sound}. We should note in passing that in the
case of attractive interatomic interactions ($g<0$), the speed of
sound becomes imaginary, which indicates that long wavelength
perturbations grow or decay exponentially in time. Thus, the
stationary state of Eq.~(\ref{pedestal}) is subject to the
modulational instability, which is responsible for the formation of
{\it matter-wave bright solitons} \cite{bright1,bright2,bright3} in
attractive BECs (see also the reviews
\cite{BECBOOK,revnonlin,pandim,rab} and references therein).

\subsection{Lower-dimensional condensates and relevant mean-field models.}
\label{lowerd}

Let us consider again a condensate confined in the harmonic trap of Eq.~(\ref{hpot}).
In this case, the trap frequencies set characteristic length scales for
the spatial size of the condensate through the harmonic oscillator
lengths $a_{j} \equiv (\hbar/m\omega_j)^{1/2}$ ($j\in\{x,y,z\}$).
Another important length scale, introduced by the effective
mean-field nonlinearity, is the so-called healing length defined as
$\xi = (8 \pi n_0 a)^{-1/2}$ (with $n_0$ being the maximum
condensate density). The healing length, being the scale over which
the BEC wave function ``heals'' over defects, sets the spatial
widths of nonlinear excitations, such as matter-wave dark solitons.

Based on the above, as well as the form of the ground state [cf. Eq.~(\ref{TF})],
it is clear that the shape of the BEC is controlled by the relative values of the
trap frequencies. For example, if
$\omega_x = \omega_y \equiv \omega_{\perp} \approx \omega_z$
(i.e., for an isotropic trap), the BEC is almost spherical, while for
$\omega_z < \omega_{\perp}$ (i.e., for an anisotropic trap) the BEC
is ``cigar shaped''. It is clear that such a cigar-shaped BEC
(a) may be a purely 3D object,
(b) acquire an almost 1D character
(for strongly anisotropic traps with
$\omega_z \ll \omega_{\perp}$ and $\mu \ll \hbar \omega_{\perp}$), or
(c) being in the so-called {\it dimensionality
crossover} regime from 3D to 1D. These regimes can be described by the
dimensionless parameter \cite{str},
\begin{equation}
d = N \Omega \frac{a}{a_{\perp}},
\label{dimparam}
\end{equation}
where $\Omega=\omega_z/\omega_{\perp}$ is the so-called ``aspect
ratio'' of the trap. Particularly, if the dimensionality parameter
is $d \gg 1$, the BEC locally retains its original 3D character
(although it may have an elongated, quasi-1D shape) and its ground
state can be described by the TF approximation in all directions. On
the other hand, if $d \ll 1$, excited states along the transverse
direction are not energetically accessible and the BEC is
effectively 1D. Apparently, this regime is extremely useful for an
analytical study of matter-wave dark solitons. Finally, if $d \approx 1$,
the BEC is in the crossover regime between 1D and 3D,
which is particularly relevant as recent matter-wave dark soliton
experiments have been conducted in this regime \cite{kip,draft6}.

Let us now discuss in more detail lower-dimensional mean-field
models describing cigar-shaped BECs. First, we consider the quasi-1D
regime ($d \ll 1$) characterized by an extremely tight transverse
confinement. In this case, following Refs.~\cite{gp1d_B,vpg,gp1d_A},
the BEC wave function is separated into transverse and longitudinal
components, namely $\Psi({\bf r}, t)= \Phi(r;t) \psi(z,t)$. Then,
the transverse component $\Phi(r;t)$ is described by the Gaussian
ground state of the transverse harmonic oscillator (and, thus, the
transverse width of the condensate is set by the transverse harmonic
oscillator length $a_\perp$), while the longitudinal wave function
$\psi(z,t)$ obeys the following effectively 1D GP equation:
\begin{equation}
i \hbar \partial_t \psi(z,t) = \left[ - \frac{\hbar^{2}}{2m} \partial_{z}^{2} + V(z)
+ g_{1D} |\psi(z,t)|^{2} \right] \psi(z,t),
\label{1dgpe}
\end{equation}
where the effective 1D coupling constant is given by $g_{1D} =
g/2\pi a_{\perp}^{2}=2 a \hbar \omega_{\perp}$ and $V(z)=(1/2)m
\omega_{z}^{2} z^{2}$. Notice that in the case under consideration,
if the additional condition $[(N/\sqrt{\Omega})(a/a_\perp)]^{1/3}
\gg 1$ is fulfilled, then the longitudinal condensate density
$n(z,t)\equiv |\psi(z,t)|^2$ can be described by the TF
approximation --- see Eq.~(\ref{TF}) with $\mu$ now being the 1D
chemical potential (and $g \rightarrow g_{1D}$) \cite{str}.
Following the terminology of Ref.~\cite{kip}, this regime will
hereafter be referred to as the TF-1D regime.

Next, let us consider the effect of the deviation from
one-dimensionality on the longitudinal condensate dynamics. In this
case, the wave function can be factorized as before, but with the
transverse component $\Phi$ assumed to depend also on the
longitudinal variable $z$ (and time $t$)
\cite{cqnls,npse,gerbier,delgado}. Physically speaking, it is
expected that the transverse direction will no longer be occupied by
the ground state, but $\Phi$ would still be approximated by a
Gaussian function with a width $w=w(z,t)$ that can be treated as a
variational parameter \cite{npse,gerbier,delgado}. This way, it is
possible to employ different variational approaches and derive the
following NLS equation for the longitudinal wave function,
\begin{equation}
i\hbar \frac{\partial \psi}{\partial t}=
\left[-\frac{\hbar^{2}}{2m}\frac{\partial^{2}}{\partial z^{2}}+ V(z) + f(n) \right]\psi.
\label{gengpe}
\end{equation}
The nonlinearity function $f(n)$ in Eq.~(\ref{gengpe}) depends on
the longitudinal density $n(z,t)$ and may take different forms.
Particularly, in Ref.~\cite{npse} (where variational equations
related to the minimization of the action functional were used)
$f(n)$ is found to be:
\begin{eqnarray}
f(n) =\frac{g}{2\pi a_{\perp}^{2}} \frac{n}{\sqrt{1+2a n}}
+\frac{\hbar \omega_{\perp}}{2}
\left(\frac{1}{\sqrt{1+2an}} +\sqrt{1+2an} \right),
\label{salnl}
\end{eqnarray}
and the respective NLS equation is known as the non-polynomial
Schr\"{o}dinger equation (NPSE). On the other hand, in
Refs.~\cite{gerbier,delgado} (where variational equations related to the
minimization of the transverse chemical potential were used) the result for $f(n)$ is:
\begin {equation}
f(n)=\hbar \omega_\perp\sqrt{1+4 a n}.
\label{gerbiernl}
\end {equation}
Since, as explained above, the derivation of the mean-field models with the nonlinearity functions in Eqs.~(\ref{salnl}) and (\ref{gerbiernl}) is based on different approaches, these nonlinearity functions are quite different. Nevertheless, they can be ``reconciled'' in the weakly-interacting limit of $a n \ll 1$: in this case, the width of the transverse wave function becomes $w=a_\perp$ and Eq.~(\ref{gengpe}) --- with either the nonlinearity function of Eq.~(\ref{salnl}) or that of Eq.~(\ref{gerbiernl}) --- is reduced to the 1D GP model of Eq.~(\ref{1dgpe}).

The above effective 1D models predict accurately ground state
properties of quasi-1D condensates, such as the chemical potential,
the axial density profile, the speed of sound, collective
oscillations, and others. Importantly, these models are particularly
useful in the dimensionality crossover regime, where they describe
the axial dynamics of cigar-shaped BECs in a very good approximation
to the 3D Gross-Pitaevskii (GP) equation (see, e.g., theoretical
work related to matter-wave dark solitons in Ref.~\cite{crossover}
and relevant experimental results in Ref.~\cite{kip}).

On the other hand, extremely weak deviations from
one-dimensionality can also be treated by means of a rather simple
non-cubic nonlinearity that can be obtained by Taylor expanding
$f(n)$, namely:
\begin{equation}
f(n)= g_{1} n - g_{2} n^{2},
\label{cqn}
\end{equation}
where $g_1 = g_{1D}$ and $g_2$ depends on the form of $f(n)$. In this case,
Eq.~(\ref{gengpe}) becomes a cubic-quintic NLS (cqNLS) equation. This model was derived
self-consistently in Ref.~\cite{cqnls}, where dynamics of matter-wave dark solitons in
elongated BECs was considered; there, the coefficient $g_{2}$ was found to be equal to
$g_{2} = 24 \ln(4/3)a^2 \hbar \omega_\perp$.


Here, it is worth mentioning that the quintic term in the cqNLS equation may have a
different physical interpretation, namely to describe {\it three-body interactions},
regardless of the dimensionality of the system. In this case, the coefficients $g_{1D}$
and $g_2$ in Eq.~(\ref{cqn}) are generally complex, with the imaginary parts describing
{\it inelastic} two- and three-body collisions, respectively \cite{mo}. As concerns the
rate of the three-body collision process, it is given by $(dn/dt) = - L n^3$ \cite{book1},
where $L$ is the loss rate (which is of order of $10^{-27}$--$10^{-30}$ cm$^{6}$s$^{-1}$
for various species of alkali atoms \cite{kooo}). Accordingly, the decrease of the density
is accounted for by the term $-(L/2)|\psi|^4 \psi$ in the time dependent GP equation,
i.e., to the quintic term in the cqNLS equation.


It is also relevant to note that the NLS Eq.~(\ref{gengpe}) has also been used as a
mean-field model describing strongly-interacting 1D Bose gases and, particularly,
the so-called {\it Tonks-Girardeau} gas of impenetrable bosons \cite{tg} (see also
Refs.~\cite{tgexp1,tgexp2} for recent experimental observations). In this case, the
function $f(n)$ takes the form \cite{kolom1},
\begin{equation}
f(n)= \frac{\pi^{2}\hbar^{2}}{2m} n^2,
\label{nltg}
\end{equation}
and, thus, Eq.~(\ref{gengpe}) becomes a quintic NLS equation.
Although the applicability of this equation has been criticized (as
in certain regimes it fails to predict correctly the coherence
properties of the strongly-interacting 1D Bose gases
\cite{girardeaubad}), the corresponding hydrodynamic equations for
the density $n$ and the phase $\varphi$ arising from the quintic NLS
equation under the Madelung transformation $\psi = \sqrt{n} \exp(i
\varphi)$ are well-documented in the context of the local density
approximation \cite{dunjko}. Additionally, it should be noticed that
the time-independent version of the quintic NLS equation has been
rigorously derived from the many-body Schr{\"o}dinger equation
\cite{liebprl}.

We finally mention that another lower-dimensional version of the fully 3D GP
equation can be derived for ``disk-shaped'' (alias ``pancake'')
condensates confined in strongly anisotropic traps with $\omega_{\perp} \ll \omega_z$
and $\mu \ll \hbar \omega_z$. In such a case, a procedure similar to the one used for
the derivation of Eq.~(\ref{1dgpe}) leads to the following $(2+1)$-dimensional NLS
equation:
\begin{equation}
i \hbar \partial_t \psi(x,y,t) = \left[ - \frac{\hbar^{2}}{2m} \nabla_{\perp}^{2} + V(r)
+ g_{2D} |\psi(x,y,t)|^{2} \right] \psi(x,y,t),
\label{2dgpe}
\end{equation}
where $r^2 = x^2 + y^2$, $\nabla_{\perp}^{2} = \partial_x^2+\partial_y^2$,
the effectively 2D coupling constant is given by
$g_{2D}=g/\sqrt{2\pi} a_{z} = 2\sqrt{2\pi} a a_{z} \hbar \omega_{z}$,
while the potential is given by $V(r)=(1/2)m \omega_{\perp}^{2}r^2$.
%
%
It should also be noticed that other effective 2D mean-field models (involving systems
of coupled 2D equations \cite{npse} or 2D GP equations with generalized nonlinearities
\cite{npse,delgado}) have also been proposed for the study of the transverse dynamics
of disk-shaped BECs.

The above
models will be
used below to investigate the static and dynamical properties of
matter-wave dark solitons arising in the respective settings.

\section{General background for the study of matter-wave dark solitons}
\label{sec3}

\subsection{NLS equation and dark soliton solutions.}
\label{nlsdark}


We start by considering the case of a quasi-1D condensate described by Eq.~(\ref{1dgpe}).
The latter, can be expressed in the following dimensionless form:
\begin{equation}
i \partial_t \psi(z,t) = \left[ -\frac{1}{2}\partial_z^{2} + V(z)
+ |\psi(z,t)|^{2} \right] \psi(z,t),
\label{dim1dgpe}
\end{equation}
where the density $|\psi|^2$, length, time and energy are respectively measured in units
of $2a$, $a_{\perp}$, $\omega_{\perp}^{-1}$ and $\hbar\omega_{\perp}$, while the
potential $V(z)$ is given by
\begin{equation}
V(z)=\frac{1}{2} \Omega^2 z^2.
\label{htrap}
\end{equation}
In the case under consideration, the normalized trap strength (aspect ratio)
is $\Omega \ll 1$ and, thus, as a first step in our analysis, the potential $V(z)$ is
ignored.
\footnotemark[1]
\footnotetext[1]{Note that in the limit of $z \rightarrow \pm \infty$ this
approximation always breaks down.}
In such a case, the condensate is homogeneous and can be described by
the {\it completely integrable} defocusing NLS equation
\cite{zsd} (see also the review \cite{kivpr}):
\begin{equation}
i \partial_t \psi(z,t) = \left[ -\frac{1}{2}\partial_z^{2}
+ |\psi(z,t)|^{2} \right] \psi(z,t).
\label{dim1dgpe2}
\end{equation}
This equation possesses an infinite number of conserved quantities (integrals of motion);
the lowest-order ones are the number of particles $N$, the momentum $P$, and the energy
$E$, respectively given by:
\begin{eqnarray}
N&=&\int_{-\infty}^{-\infty}|\psi|^{2}dz,
\label{NNLS} \\
P&=&\frac{i}{2}\int_{-\infty}^{-\infty}\left(\psi \partial_z \psi^{\ast}
-\psi^{\ast} \partial_z \psi \right)dz,
\label{PNLS} \\
E&=&\frac{1}{2}\int_{-\infty}^{-\infty} \left(|\partial_z \psi|^{2}+ |\psi|^{4}\right)dz.
\label{ENLS}
\end{eqnarray}
It is also noted that the NLS Eq.~(\ref{dim1dgpe2}) can be obtained by the
Euler-Lagrange equation
$\delta\mathcal{L}/\delta \psi^{\ast}=
\partial_t (\partial_{\partial_t \psi^{\ast}}\mathcal{L})
+\partial_z (\partial_{\partial_z \psi^{\ast}}\cal{L})
\partial_{\psi^{\ast}}\mathcal{L}$=$0$, where
the Lagrangian density $\mathcal{L}$ is given by:
\begin{equation}
\mathcal{L}=\frac{i}{2} \left(\psi \partial_t \psi^{\ast}
-\psi^{\ast} \partial_t \psi \right)
-\frac{1}{2}\left(|\partial_z \psi|^{2}+ |\psi|^{4}\right).
\label{L}
\end{equation}

The simplest nontrivial solution of Eq.~(\ref{dim1dgpe2})
is a plane wave of wave number $k$ and frequency $\omega$, namely,
\begin{equation}
\psi = \sqrt{n_{0}}\exp[i(k z-\omega t+\theta_{\rm o})], \qquad \omega=\frac{1}{2}k^{2} - \mu,
\label{cw}
\end{equation}
where the constant BEC density $n_{0}$ sets the chemical
potential, i.e., $n_0=\mu$ and $\theta_{\rm o}$ is an arbitrary constant phase.
This solution, which is reduced to the
stationary state of Eq.~(\ref{pedestal}) for $k=0$, is also
modulationally stable as can be confirmed by a simple stability
analysis (see, e.g., Refs.~\cite{kivpr,pandim}). For small densities,
$n_{0} \ll 1$, the above plane wave satisfies the linear
Schr\"{o}dinger equation, $i\partial_{t}\psi
+\frac{1}{2}\partial_{z}^2 \psi =0$, and the pertinent linear wave
solutions of the NLS equation
are characterized by the dispersion
relation $\omega=\frac{1}{2}k^{2}$. Notice that if the system is
characterized by a length $L$, then the integrals of motion for the
stationary solution in Eq. (\ref{cw}) take the values:
\begin{equation}
N=2 n_{0}L, \qquad P=k n_{0}L, \qquad  E=\frac{1}{2}(k^{2}-n_{0})n_{0}L.
\label{imcw}
\end{equation}
%


The NLS equation admits nontrivial solutions, in
the form of dark solitons, which can be regarded as strongly
nonlinear excitations of the plane wave solution (\ref{cw}). In the
most general case of a moving background [$k\ne 0$ in Eq.~(\ref{cw})],
a single dark soliton solution may be expressed as \cite{zsd},
\begin{equation}
\psi(z,t)=\sqrt{n_0} \left( B \tanh \zeta +iA \right)\exp[i(kz-\omega t +
\theta_{\rm o})],
\label{ds}
\end{equation}
where $\zeta \equiv \sqrt{n_0} B \left[z-z_0(t) \right]$; here,
$z_0(t) = vt+z_{\rm o}$ is the soliton center,
$z_{\rm o}$ is an arbitrary real constant representing the initial location
of the dark soliton,
$v$ is the relative velocity between the soliton and the background given
by $v=A \sqrt{n_{0}}+ k$, the frequency $\omega$ is provided by
the dispersion relation of the background plane wave, $\omega=(1/2)k^{2}+n_{0}$
[cf.~Eq. (\ref{cw})]
\footnotemark[1]
\footnotetext[1]{Here, this dispersion relation implies that $\omega>k^{2}$
and, thus, the allowable region in the $(k,\omega)$ plane for dark solitons
is located {\it above} the parabola $\omega=\frac{1}{2}k^{2}$ corresponding
to the linear waves.
},
and, finally, the parameters $A$ and $B$ are connected through the equation $A^2+B^2=1$.
In some cases it is convenient to use one parameter instead of two and, thus, one may
introduce
\begin{equation}
A = \sin \phi, \qquad B = \cos \phi,
\label{phi}
\end{equation}
where $\phi$ is the so-called ``soliton phase angle''
($|\phi |<\pi/2$). Notice that although the asymptotics of the dark soliton
solution (\ref{ds}) coincide with the ones of Eq.~(\ref{cw}), the
plane waves at $z \rightarrow \pm \infty$ have different phases; as
a result, there exists a nontrivial phase jump $\Delta \phi$ across
the dark soliton, given by:
\begin{equation}
\Delta \phi = 2\left[ \tan^{-1}\left(\frac{B}{A}\right) -\frac{\pi}{2}\right]
=-2\tan^{-1}\left(\frac{A}{B}\right).
\label{dphi}
\end{equation}
Note that, hereafter, we will consider the simpler case where the
background of matter-wave dark solitons is at rest, i.e., $k=0$;
then, the frequency $\omega$ actually plays the role of the
normalized chemical potential, namely $\omega = \mu = n_{0}$, which
is determined by the number of atoms of the condensate.

The soliton phase angle describes also the {\it darkness} of the soliton, namely,
\begin{equation}
|\psi|^{2}=n_0 (1-\cos^{2}\phi \,{\rm sech}^{2}\zeta).
\label{darkness}
\end{equation}
\begin{figure}[tbp]
\centering \includegraphics[width=9.0cm,height=4.1cm]{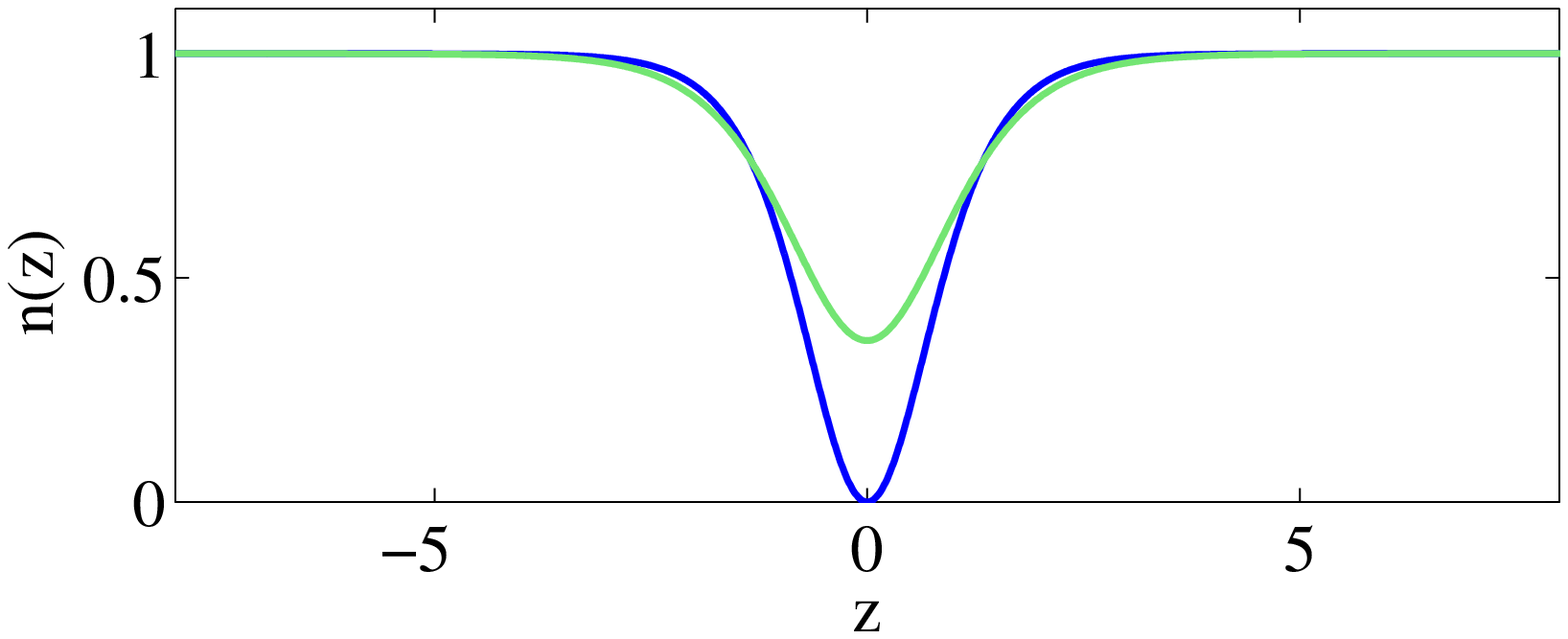}
\centering \includegraphics[width=9.0cm,height=4.1cm]{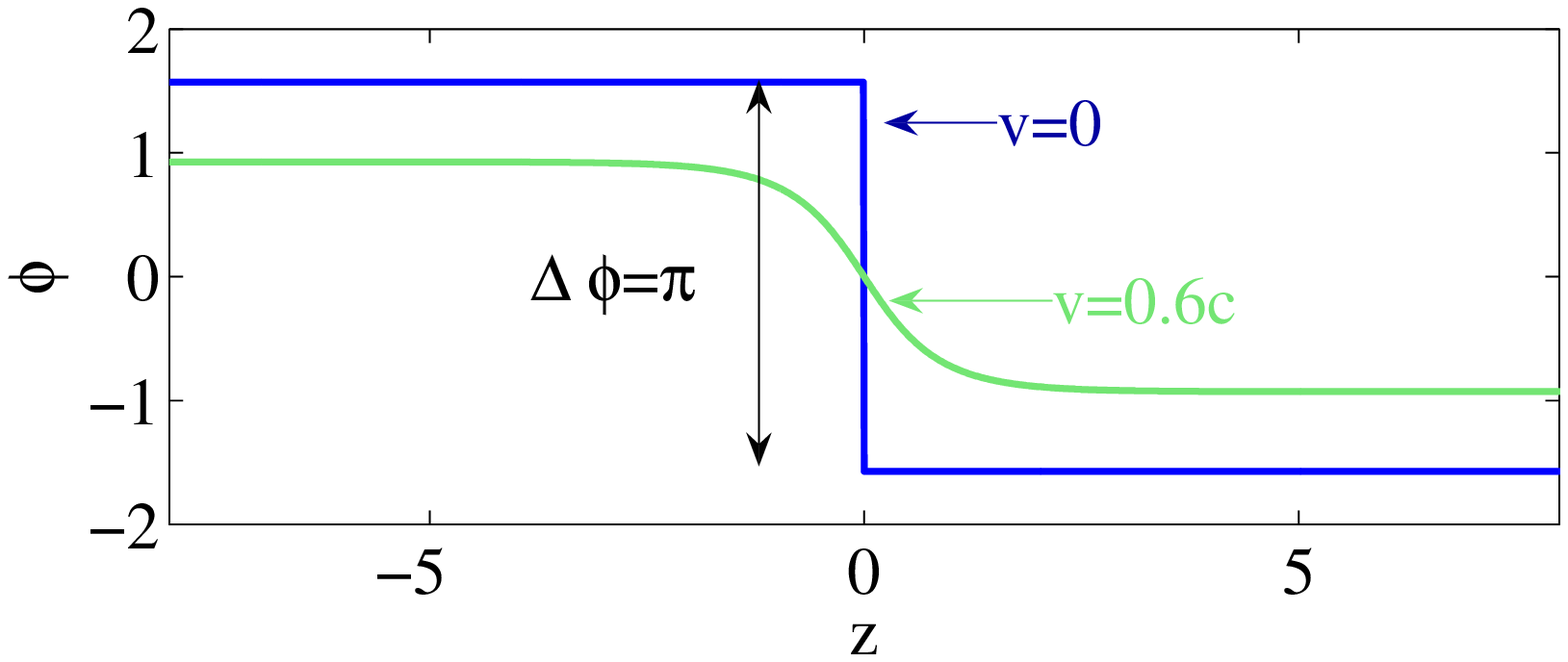}
\caption{ (color online) Examples of the density (top panel) and phase
(bottom panel) of a black (blue line) and a gray (green line) soliton on top of a
background with density $n_0=1$. The black soliton's parameters are $A=0$ and $B=1$,
i.e., $v=0$, $n_{\rm min}=0$ and $\Delta \phi = \pi$. The gray soliton's parameters
are $A=0.6$ and $B=0.8$, i.e., $v=0.6c_s$ (here $c_s=\sqrt{n_0}=1$),
$n_{\rm min}= n_0 (1-B^2)=n_0 A^2 = 0.36$, and $\Delta \phi = 0.31 \pi$.}
\label{figbg}
\end{figure}

This way, the cases $\phi=0$ and $0< \phi < \pi/2$ correspond to the so-called
{\it black} and {\it gray} solitons, respectively.
The amplitude and velocity of the dark soliton are given (for $k=0$)
by $\sqrt{n_0} \cos \phi$ and $\sqrt{n_0}\sin \phi$, respectively;
thus, the black soliton
\begin{equation}
\psi=\sqrt{n_0}\tanh(\sqrt{n_0}z)\exp(-i\mu t),
\label{black}
\end{equation}
is characterized by a zero velocity, $v=0$ (and, thus, it is also called
{\it stationary kink}), while the gray soliton moves with a finite velocity $v \ne 0$.
Examples of the forms of a black and a gray soliton are illustrated in Fig.~\ref{figbg}.

In the limiting case of a very shallow (small-amplitude) dark soliton with
$\cos\phi \ll 1$, the soliton velocity is close to the speed of sound which,
in our units, is given by:
\begin{equation}
c_s = \sqrt{n_0}.
\label{ssh}
\end{equation}
The speed of sound is, therefore, the maximum possible velocity of a dark soliton which,
generally, always travels with a velocity less than the speed of sound. We finally note
that the dark soliton solution (\ref{ds}) has two independent parameters (for $k=0$),
one for the background, $n_{0}$, and one for the soliton, $\phi$, while there is also
a freedom (translational invariance) in selecting the initial location of the dark
soliton $z_{\rm o}$
\footnotemark[1]
\footnotetext[1]{
Recall that the underlying model, namely the completely integrable NLS equation,
has infinitely many symmetries, including translational and Galilean invariance.
}.

In the case of a condensate confined in a harmonic trap [cf.~Eq.~(\ref{hpot})], the
background of the dark soliton is, in fact, of {\it finite extent}, being the ground
state of the BEC [which may be approximated by the Thomas-Fermi cloud,
cf.~Eq.~(\ref{TF})].
For example, in the quasi-1D setting of the 1D GP Eq.~(\ref{dim1dgpe}) with the harmonic
potential in Eq.~(\ref{htrap}), the ``composite'' wave function (describing both the
background and the soliton) can be approximated as
$\psi(z,t)=\Phi(z) \exp(-i \mu t) \psi_{\rm ds}(z,t)$,
where $\Phi(z)$ is the TF background and $\psi_{\rm ds}(z,t)$ is the dark soliton
wave function of Eq.~(\ref{ds}), which satisfies the 1D GP equation for $V(z)=0$.


\subsection{Dark solitons and the Inverse Scattering Transform.}
\label{IST}

The single dark soliton solution of the NLS Eq.~(\ref{dim1dgpe2}) presented in the
previous Section, as well as multiple dark soliton solutions (see Sec.~\ref{multiple}
below), can be derived by means of the Inverse Scattering Transform (IST) \cite{zsd}.
A basic step of this approach is the solution of the Zakharov-Shabat (ZS) eigenvalue
problem, with eigenvalue $\lambda$, for the auxiliary two-component eigenfunction
$U = (u_1, u_2)^T$, namely,
\begin{equation}
L U =
\left(
\begin{array}{cccc}
i\partial_z & \psi(z,0) \\
\psi^{\ast}(z,0) & -i\partial_z
\end{array}
\right)
\left(
\begin{array}{c}
u_1 \\ u_2
\end{array}
\right) = \lambda U,
\label{zakhshab}
\end{equation}
with the boundary conditions $\psi(z,0) \rightarrow \sqrt{n_0}$, for
$z \rightarrow +\infty$, and $\psi(z,0) \rightarrow \sqrt{n_0}\exp(i\theta)$, for
$z \rightarrow -\infty$. Here, $\sqrt{n_0}$
is the amplitude of the background wave function and $\theta$ is a constant phase.
Since the operator $L$ is self-adjoint, the ZS eigenvalue problem possesses real
discrete eigenvalues $\lambda_j$, with magnitudes $|\lambda_j|<\sqrt{n_0}$.
%
Importantly, each real discrete eigenvalue $\lambda_j = \sqrt{n_0}\sin\phi_j$ corresponds
to a dark soliton of depth $\sqrt{n_0}\cos\phi_j$ and velocity $\sqrt{n_0}\sin\phi_j$.
To make a connection to the dark soliton solutions of the NLS equation presented in
the previous Section, we note that the dark soliton of Eq.~(\ref{ds}) corresponds to a
{\it single} eigenvalue $\lambda = \sqrt{n_0}\sin\phi$.

Although the system of ZS Eqs.~(\ref{zakhshab}) is linear, its general solution for
arbitrary initial condition is not available. Thus, various methods have been developed
for the determination of the spectrum of the ZS problem, such as the so-called
quasi-classical method \cite{gred1,gred2} (see also Ref.~\cite{slavin}), the variational
approach \cite{fkhtsoy}, as well as other techniques that can be applied to the case of
dark soliton trains \cite{kamkra,brazkam}.
In any case, the {\it generation} of single- as well as multiple-dark solitons
(see Sec.~\ref{multiple} below) can be studied in the framework of the IST method,
and many useful results can be obtained.
In that regard, first we note that a pair of dark solitons --- corresponding to
a discrete eigenvalue pair in the associated scattering problem --- can always be
generated by an {\it arbitrary} small dip on a background of constant density
\cite{gred1} (see also Ref.~\cite{gred2}). This means that
the generation of dark solitons is a {\it thresholdless} process, contrary to the case
of bright solitons which are created when the number of atoms exceeds a certain
threshold \cite{yskbright}.
In another example, as dark solitons are characterized by a phase jump across them,
we
may assume that they can be generated by an anti-symmetric initial wave
function profile of the form,
\begin{equation}
\psi(z,0)=\sqrt{n_0}\tanh(\alpha z),
\label{tanh}
\end{equation}
characterized by a background density $n_0$ and a width $\alpha^{-1}$
(the ratio $\sqrt{n_0}/\alpha$ is assumed to be arbitrary). In such a case,
the ZS eigenvalue problem (\ref{zakhshab}) can be solved exactly
\cite{zhaobour,zhao,KonVek} and the resulting eigenvalues of the discrete
spectrum are given by $\lambda_1=0$ and
$\lambda_{2j}=-\lambda_{2j+1} = \sqrt{n_0-\mu_j^2}$, where positive
$\mu_j$ are defined as $\mu_j = \sqrt{n_0}-j\alpha$, $j=1,2, \cdots, N_0$,
and $N_0$ is the largest integer such that
$N_0 < \sqrt{n_0}/\alpha$. These results show that for arbitrary
$\sqrt{n_0}/\alpha$ the initial wave function profile of Eq.~(\ref{tanh})
will always produce a black soliton [cf.~Eq.~(\ref{black})] at $z=0$
(corresponding to the first, zero eigenvalue) and additional $N_0$ pairs of
symmetric gray solitons (corresponding to the even number of the secondary,
nonzero eigenvalues), propagating to the left and to the right of the primary
black soliton. Apparently, the total number of eigenvalues and, thus,
the total number of solitons, is $2N_0 +1$ and depends on the ratio $\sqrt{n_0}/\alpha$.
Apart from the above example, dark soliton generation was systematically studied in
Ref.~\cite{gred2} for a variety of initial conditions (such as box-like dark pulses,
phase steps, and others). Notice that, generally, initial wave function profiles with
odd symmetry will produce an odd number of dark solitons, while profiles with an even
symmetry (as, e.g., in the study of Ref.~\cite{gred1}) produce pairs of dark solitons;
this theoretical prediction was also confirmed in experiments with optical dark solitons
\cite{weinher}. Furthermore, the initial phase change across the wave function plays
a key role in dark soliton formation, while the number of dark solitons that are
formed can be changed by small variations of the phase.


\subsection{Integrals of motion and basic properties of dark solitons.}
\label{integrals}

Let us now proceed by considering the integrals of motion for dark solitons.
Taking into regard that Eqs.~(\ref{NNLS})--(\ref{ENLS})
refer to both the background and the soliton, one may follow
Refs.~\cite{uz,KY,igor,igor2}, and {\it renormalize} the integrals of
motion so as to extract the contribution of the background [see Eqs.~(\ref{imcw})].
This way, the {\it renormalized integrals of motion} become finite and, when calculated
for the dark soliton solution (\ref{ds}), provide the following results (for $k=0$).
The number of particles $N_{\rm ds}$ of the dark soliton reads:
\begin{equation}
N_{\rm ds}=\int_{-\infty}^{-\infty}(n_{0}-|\psi|^{2})dz = 2\sqrt{n_0}B.
\label{NDS}
\end{equation}
The momentum $P_{\rm ds}$ of the dark soliton is given by,
\begin{eqnarray}
P_{\rm ds}&=&\frac{i}{2}\int_{-\infty}^{-\infty}\left(\psi \partial_z \psi^{\ast}
-\psi^{\ast} \partial_z \psi \right)dz-n_{0}\Delta \phi
\nonumber \\
&=&\frac{i}{2}\int_{-\infty}^{-\infty}\left(\psi \partial_z \psi^{\ast}
-\psi^{\ast} \partial_z \psi \right)\left( 1-\frac{n_0}{|\psi|^2}\right) dz
\nonumber \\
&=&-2v(c_s^{2}-v^{2})^{1/2}+2c_s^{2} \tan^{-1}\left[ \frac{(c_s^{2}-v^{2})^{1/2}}{v} \right],
\label{PDS}
\end{eqnarray}
where $\Delta \phi $ is given by Eq.~(\ref{dphi}) and $c_s = \sqrt{n_0}$ is the speed of
sound. Furthermore, the energy $E_{\rm ds}$ of the dark soliton is given by,
\begin{equation}
E_{\rm ds}=\frac{1}{2}\int_{-\infty}^{-\infty}\left[|\partial_{z}\psi|^{2}
+\left(|\psi|^{2}-n_{0}\right)^{2}\right]dz
= \frac{4}{3}(c_s^{2}-v^{2})^{3/2},
\label{EDS}
\end{equation}
while the renormalized Lagrangian density takes the form \cite{lads}:

\begin{equation}
\mathcal{L_{\rm ds}}=\frac{i}{2} \left(\psi \partial_t \psi^{\ast}
-\psi^{\ast} \partial_t \psi \right)\left( 1-\frac{n_0}{|\psi|^2}\right)
-\frac{1}{2}\left[|\partial_z \psi|^{2}+ (|\psi|^{2}-n_0)^2 \right].
\label{LDS}
\end{equation}

The renormalized integrals of motion can now be used for a better understanding of
basic features of dark solitons. To be more specific, one may differentiate the
expressions (\ref{PDS}) and (\ref{EDS}) over the soliton velocity
$v \equiv A \sqrt{n_{0}}$ to obtain the result,
\begin{equation}
\frac{\partial E_{\rm ds}}{\partial P_{\rm ds}}= v,
\label{pnds}
\end{equation}
which shows that the dark soliton effectively behaves like a
{\it classical particle}, obeying a standard equation of classical mechanics.
Furthermore, it is also possible to associate an {\it effective mass} to the dark
soliton, according to the equation $m_{\rm ds} = \partial P_{\rm ds} / \partial v$.
This way, using Eq.~(\ref{PDS}), it can readily be found that
\begin{equation}
m_{\rm ds} = -4 \sqrt{n_0} B,
\label{mds}
\end{equation}
which shows the dark soliton is characterized by a {\it negative} effective mass.
The same result, but for almost black solitons ($B\approx 1$) with sufficiently
small soliton velocities ($v^{2} \ll c_s^{2}$), can also be obtained using
Eq.~(\ref{EDS}) \cite{fms}: in this case, the energy of the dark soliton can be
approximated as $E_{\rm ds}\approx (4/3)c_s^{3}-2c_s v^{2}$ or, equivalently,
\begin{equation}
E_{\rm ds} = E_{0}+\frac{1}{2}m_{\rm ds}v^{2},
\label{Ebs}
\end{equation}
where $E_{0} \equiv \frac{4}{3} c_s^{3}$, and the soliton's effective mass is
$m_{\rm ds}=-4\sqrt{n_{0}}$.


\subsection{Small-amplitude approximation: shallow dark solitons as KdV solitons.}
\label{kdvstuff}

As mentioned above, the case of $B^2 = \cos^2 \phi \ll 1$ corresponds to a
small-amplitude (shallow) dark soliton, which travels with a speed $v$ close to the
speed of sound, i.e., $v \approx c_s$. In this case, it is possible to apply the
{\it reductive perturbation method} \cite{rpm} and show that, in the small-amplitude
limit, the NLS dark soliton can be described by an effective KdV equation
(see, e.g., Ref.~\cite{ir} for various applications of the KdV model).
The basic idea of this, so-called, {\it small-amplitude approximation} can be
understood in terms of the similarity between the KdV soliton and the shallow dark
soliton's density profile: indeed, the KdV equation for a field $u(z,t)$ expressed as,
\begin{equation}
\partial_t u  +6 u \partial_z u + \partial_z^3 u = 0,
\label{kdveq}
\end{equation}
possesses a single soliton solution (see, e.g., Ref.~\cite{segur}):
\begin{equation}
u(z,t) = 2\kappa^2 {\rm sech}^{2}[\kappa(z-4\kappa^2 t)]
\label{kdvsol}
\end{equation}
(with $\kappa$ being an arbitrary constant), which shares the same functional form
with the density profile of the shallow dark soliton of the NLS equation
[see Eqs.~(\ref{kdvsol}) and (\ref{darkness})].
The reduction of the cubic NLS equation to the KdV equation was first presented in
Ref.~\cite{tsuzuki} and later the formal connection between several integrable
evolution equations was investigated in detail \cite{zk}. Importantly, such a connection
is still possible even in cases of strongly perturbed NLS models, a fact that triggered
various studies on dark soliton dynamics in the presence of perturbations (see, e.g.,
Refs.~\cite{bass,opt1,opt2} for studies in the context of optics, as well as the
recent review \cite{leblond} and references therein). Generally, the advantage of the
small-amplitude approximation is that it may predict approximate analytical dark
soliton solutions in models where exact analytical dark soliton solutions are not
available, or can only be found in an implicit form \cite{bass}.

Let us now consider a rather general case, and discuss small-amplitude dark solitons of
the generalized NLS Eq.~(\ref{gengpe}); in the absence of the potential ($V(z) = 0$),
this equation is expressed in dimensionless form as:
\begin{equation}
i \partial_t \psi = -\frac{1}{2} \partial_z^{2} \psi + f(n) \psi,
\label{dg}
\end{equation}
where the units are the same to the ones used for Eq. (\ref{dim1dgpe}). Then,
we use the Madelung transformation $\psi(z,t)=\sqrt{n(z,t)}\exp[i\varphi(z,t)]$
(with $n \equiv |\psi|^{2}$ and $\varphi$ representing the BEC density and phase,
respectively) to express Eq.~(\ref{dg}) in the hydrodynamic form:
\begin{eqnarray}
&&\partial_t \varphi +f(n)+\frac{1}{2} \left( \partial_z \varphi \right)^2
-\frac{1}{2} n^{-1/2} \partial_z^2 n^{1/2}=0,
\label{h1} \\
&&\partial_t n + \partial_z \left( n \partial_z \varphi \right) = 0.
\label{h2}
\end{eqnarray}
The simplest solution of Eqs.~(\ref{h1})--(\ref{h2}) is $n = n_0 \equiv |\psi_0|^2$ and
$\phi=-\mu t = -f_0 t$, where $f_0 \equiv f(n_0) = f(|\psi_0|^2)$. Note that in the model
of Eq.~(\ref{salnl}) one has $f_0 = \frac{2+3n_0}{2\sqrt{1+n_0}}$, for the model of
Eq.~(\ref{gerbiernl}), $f_0 = \sqrt{1+2n_0}$, and so on.
Next, assuming slow spatial and temporal variations, we define the slow variables
\begin{equation}
Z= \epsilon^{1/2} (z-ct), \,\,\,\,\, T=\epsilon^{3/2} t,
\label{slow}
\end{equation}
where $\epsilon$ is a formal small parameter ($0<\epsilon \ll 1$) connected with
the soliton amplitude.
Additionally, we introduce asymptotic expansions for the density and phase:
\begin{eqnarray}
n&=&n_{0}+ \epsilon n_{1}(Z,T)+\epsilon^{2} n_{2}(Z,T)+\cdots,
\label{ae1} \\
\varphi&=&-f_0 t+ \epsilon^{1/2} \varphi_{1}(Z,T)+\epsilon^{3/2} \varphi_{2}(Z,T)+\cdots.
\label{ae2}
\end{eqnarray}
Then, substituting Eqs.~(\ref{ae1})--(\ref{ae2}) into Eqs.~(\ref{h1})--(\ref{h2}),
and Taylor expanding the nonlinearity function $f(n)$ as
$f(n) = f_0 + \epsilon f_0' n_1 + \epsilon^2 [(1/2)f_0''n_1^2
+ f_0' n_2] + {\cal O}(\epsilon^3)$
(where $f_0'' \equiv \frac{d^2 f}{dn^2}\big|_{n=n_0}$), we obtain a
hierarchy of equations. In particular, Eqs.~(\ref{h1})--(\ref{h2}) lead,
respectively, at the order
${\cal O}(\epsilon)$ and ${\cal O}(\epsilon^{3/2})$, to the following linear system,
\begin{equation}
-c \partial_Z \varphi_1 + f_0' n_1 = 0,
\,\,\,\,\,\,\,\,\,\,\,
n_0 \partial_Z^2 \varphi_1 -c \partial_Z n_1 =0.
\label{lo}
\end{equation}
The compatibility condition of the above equations is the algebraic equation
$c^2 = f_0' n_0$, which shows that the velocity $c$ in Eq.~(\ref{slow})
is equal to the speed of sound, $c \equiv c_s$. Additionally, Eqs.~(\ref{lo}) connect
the phase $\varphi_1$ and the density $n_1$ through the equation:
\begin{equation}
\partial_Z \varphi_1 = \frac{c_s}{n_0} n_1.
\label{con}
\end{equation}
To the next order, viz. ${\cal O}(\epsilon^2)$ and ${\cal O}(\epsilon^{5/2})$,
Eqs.~(\ref{h1}) and (\ref{h2}), respectively, yield:
\begin{eqnarray}
&&\partial_T \varphi_1 - c_s \partial_Z \varphi_2
+ f_0' n_2 + \frac{1}{2}f_0'' n_1^2
+ \frac{1}{2} \left( \partial_Z \varphi_1 \right)^2 - \frac{1}{4} n_0^{-1} \partial_Z^2 n_1 =0,
\label{fo1} \\
&&\partial_T n_1 -c_s \partial_Z n_2
+ \partial_Z \left( n_1 \partial_Z \varphi_1 \right)
+n_0 \partial_Z^2 \varphi_2=0.
\label{fo2}
\end{eqnarray}
The compatibility conditions of Eqs.~(\ref{fo1})--(\ref{fo2}) are the algebraic equation
$c_s^2 = f_0' n_0$, along with a KdV equation [see Eq.~(\ref{kdveq})]
for the unknown density $n_1$:
\begin{equation}
2c_s \partial_T n_1 + (3f_0' +n_0 f_0'') n_1 \partial_Z n_1
- \frac{1}{4} \partial_Z^3 n_1 = 0.
\label{kdveqn}
\end{equation}
Thus, the density $n_1$ of the shallow dark soliton can be expressed as a
KdV soliton [see Eq.~(\ref{kdvsol})]. In terms of the original time and space
variables, $n_1$ is expressed as follows:
\begin{equation}
n_1(z,t) = -\frac{3\kappa^2}{2(3f_0' +n_0 f_0'')}
{\rm sech}^{2}\left[\epsilon^{1/2}\kappa \left(z - v t \right)\right],
\label{adso}
\end{equation}
where $\kappa$ is (as before) an arbitrary parameter [assumed to be of order
${\cal O}(1)$], while $v$ is the soliton velocity; the latter, is given by
\begin{equation}
v = c_s - \epsilon \frac{\kappa^2}{2c_s},
\label{ssolv}
\end{equation}
and, clearly, $v \lesssim c_s$. Apparently, Eq.~(\ref{adso}) describes a small-amplitude
{\it dip} [of order ${\cal O}(\epsilon)$ --- see Eq.~(\ref{ae1})] on the background
density of the condensate, with a phase $\varphi_1$ that can be found using
Eq.~(\ref{con}); in terms of the variables $z$ and $t$, the result is:
\begin{equation}
\varphi_1(z,t) = -\frac{3\kappa c_s}{2n_0(3f_0' +n_0 f_0'')}
{\rm tanh}\left[\epsilon^{1/2}\kappa \left(z - v t \right)\right].
\label{solph}
\end{equation}
The above expression shows that the density dip is accompanied by a ${\rm tanh}$-shaped
phase jump. Thus, the wave function characterized by the density $n_1$ in
Eq.~(\ref{adso}) and the phase $\varphi_1$ in Eq.~(\ref{solph}) is an approximate
{\it shallow dark soliton} solution of the GP Eq.~(\ref{dg}), obeying the effective
KdV Eq.~(\ref{kdveqn}).

Notice that the above analysis applies for $f(n)=n$ (i.e., for the cubic NLS model),
as well as for all forms of the nonlinearity function in Eqs.~(\ref{salnl})--(\ref{nltg}).
Furthermore, variants of the reductive perturbation method have also been applied for the
study of matter-wave dark solitons in higher-dimensional settings \cite{gxh1,gxh2},
multi-component condensates \cite{vectordark,bdspinor} (see also Sec.~\ref{multi})
and combinations thereof \cite{aguero}.


\subsection{On the generation of matter-wave dark solitons}
\label{gen}

Matter-wave dark solitons can be created in experiments by means of various methods,
namely the {\it phase-imprinting}, {\it density-engineering},
{\it quantum-state engineering} (which is a combination of phase-imprinting and
density engineering), the {\it matter-wave interference} method and by
{\it dragging an obstacle} sufficiently fast through a condensate.
In connection to Sec.~\ref{IST} --- and following the historical evolution
of the subject --- here we will discuss the phase-imprinting, density-engineering and
quantum-state engineering methods (the remaining two methods will be presented in
Secs.~\ref{interf} and \ref{drag} below).

\subsubsection{The phase-imprinting method.}
\label{pi}

The earlier results of Sec.~\ref{IST}, as well as more recent theoretical studies in
the BEC context \cite{DGLSBE,WLN} (see also Ref.~\cite{shukla}), paved the way for the
generation of matter-wave dark solitons by means of the phase-imprinting method. This
technique was used in the earlier \cite{han1,nist,han2} --- but also in recent
\cite{hamburg,hambcol} --- matter-wave dark soliton experiments. The phase-imprinting
method involves a manipulation of the BEC phase, without changing the BEC density,
which can be implemented experimentally by illuminating part of the condensate by a
short off-resonance laser beam (i.e., a laser beam with a frequency far from
the relevant atomic resonant frequency --- see details in the review \cite{chap01:odt2}).
This procedure can be described in the framework of
Eq.~(\ref{1dgpe}), by considering a time-dependent potential of the form
$V(z;t) \propto \phi(z) f(t)$, where $f(t)$ is the laser pulse envelope
and $\phi(z)$ is the imprinted phase, given by \cite{genburger},
\begin{equation}
\phi(z) = \frac{\Delta \phi}{2} \left[1+\tanh\left( \frac{z-z_{\ast}}{b W}\right)\right],
\label{impph}
\end{equation}
where $\Delta \phi$ is the phase gradient, while the width $W$ of the potential
edge sets the steepness of the phase gradient at $z_{\ast}$. Note that since
experimentally relevant values correspond to a $10$--$20$$\%$ absorption width of the phase
step, an empirical factor $b=0.45$ is also introduced in Eq.~(\ref{impph}) \cite{genburger}.

From a theoretical standpoint, phase-imprinting can be studied (in the absence of the
trapping potential) in the framework of the IST method, upon considering an initial
wave function of the form $\psi(z,0) = \exp[i\phi(z)]$; here the imprinted phase
$\phi(z)$ is assumed to increase from left to right and approach constants as
$z \rightarrow \pm \infty$ \cite{WLN} [as, e.g., in Eq.~(\ref{impph})]. The pertinent
ZS eigenvalue problem can be solved by mapping Eqs. (\ref{zakhshab}) to a damped
driven pendulum problem. This way, a formula for the number of both the even and
the odd number of generated dark solitons, traveling in both directions,
can be derived analytically.

In some experiments (see, e.g., Ref.~\cite{han1}), the generation of the
``dominant'' dark soliton is followed by the generation of a secondary wave packet
traveling in the opposite direction with a velocity near the speed of sound. This
effect can also be understood in the framework of IST: small perturbations of
the dark soliton produce shallow ``satellite'' dark solitons moving with
velocities $v \lesssim c_s$ \cite{KonVek}.

\subsubsection{The density-engineering method.}
\label{de}

The density-engineering method involves a direct manipulation of the BEC density,
without changing the BEC phase, such that local reductions of the density are created
which eventually evolve into dark solitons. This technique was used in the Harvard
experiments \cite{dutton,ginsberg2005}, where a compressed pulse of slow light was
used to create a defect on the condensate density. This defect induced the formation
of shock waves that shed dark solitons (or other higher-dimensional topological
structures, such as vortex rings \cite{ginsberg2005}). Notice that the use of a
compressed pulse of slow light is not really necessary or beneficial in order to create
dark solitons by means of the density engineering method: in fact, a local reduction of
the BEC density can also be created by modifying the (harmonic) trapping
potential with an additional barrier potential, which may be induced by an optical
dipole potential or a far-detuned laser beam; this barrier can then be switched
off non-adiabatically (while the harmonic trap is kept on),
creating the desired local reduction of the density \cite{genburger}.
This technique was employed in a recent experiment \cite{enghoef},
where such a dipole beam was used in different setups to induce merging and splitting
rubidium condensates; depending on the parameters, this process leads to the formation
of dark soliton trains, or a high density bulge and dispersive shock waves.

As in the case of phase-imprinting, the density-engineering technique can be studied by
means of the IST method (in the absence of the trapping potential). In fact, earlier works
\cite{gred1,gred2} (see also Ref.~\cite{brazkam}) have already addressed the problem
of dark soliton generation induced by initial change of the density: for example,
in the case of a box-like initial condition, namely $\psi(z,0)=\sqrt{n_0}$ for $|z|>z_0$
and $\psi(z,0)=\sqrt{n_1}$ for $|z|<z_0$ (with $n_1<n_0$), the ZS spectral problem admits
an explicit solution, as it can be solved exactly on the intervals $|z|<z_0$ and
$|z|>z_0$. In particular, it can be shown that there appear two discrete eigenvalues
$\lambda_{1,2} = \pm 2\sqrt{n_0}\left[1-2z_0^2 (\sqrt{n_0}-\sqrt{n_1})^2 \right]$ (for
$\sqrt{n_0}-\sqrt{n_1} \ll \sqrt{n_0}$) and, thus, two small-amplitude dark solitons
are generated.

\subsubsection{The quantum-state engineering method.}
\label{qse}

A combination of the phase-imprinting and density-engineering methods is also possible,
leading to the so-called {\it quantum-state engineering} technique \cite{genburger,CBBS}.
This method, which involves manipulation of both the BEC density and phase, has been used
in experiments at JILA \cite{bpa} and Hamburg \cite{hamburg} with a {\it two-component}
$^{87}$Rb BEC (see Sec.~\ref{multi} below): in the one component, a so-called ``filled''
dark soliton was created, with the hole in this component being filled by the other
component. Depending on the trap geometry, the created filled dark soliton was found
to be either unstable or stable. Particularly, in the JILA experiment \cite{bpa}, the
dark soliton evolved in a quasi-spherical trap (after the filling from the other
component was selectively removed) and, due to the onset of the so-called
{\it snaking instability}, the soliton was found to decay into {\it vortex rings}
(see Sec.~\ref{snaking} below). On the other hand, in the Hamburg experiment
\cite{hamburg}, the filled dark soliton in the one component was allowed to evolve
(in the presence of the other component) in an elongated cigar-shaped trap; this way,
a so-called {\it dark-bright soliton} pair was created (see Sec.~\ref{multi}), which was
found to be stable, performing slow oscillations in the trap as predicted in
theory \cite{BA01}.

Notice that a similar two-component engineering technique was also used for the
creation of vortices \cite{Matthews1}, while earlier experimental results from the
JILA group could be interpreted as a formation of a stack of filled dark solitons in
a single BEC \cite{Matthews2}.

\subsection{Multiple dark solitons and dark soliton interactions}
\label{multiple}

\subsubsection{The two-soliton state and dark soliton collisions.}

Apart from the single dark soliton solution, the NLS Eq.~(\ref{dim1dgpe2}) possesses
exact analytical {\it multiple} dark soliton solutions, which can be found by means of
the IST \cite{zsd,Blow} (see also Refs.~\cite{AA,gagnon}). Such solutions describe the
elastic collision between dark solitons as, in the asymptotic limit of
$t\rightarrow \pm\infty$, the multiple-soliton solution can be expressed as a linear
superposition of individual single-soliton solutions, which remain unaffected by the
collision apart from a collision-induced {\it phase-shift}. To be more specific, let us
consider the two-soliton wave function $\psi=\psi(z,t)$, which can be asymptotically
expressed as:
\begin{eqnarray}
&&\psi \rightarrow \psi(z-\sqrt{n_{0}} A_1 t, A_1, z_{1}^{+})
+ \psi(z-\sqrt{n_{0}} A_2 t, A_2, z_{2}^{+}),
\,\,\,
t \rightarrow +\infty,
\label{as2s1} \\
&&\psi \rightarrow \psi(z-\sqrt{n_{0}} A_1 t, A_1, z_{1}^{-})
+ \psi(z-\sqrt{n_{0}} A_2 t, A_2, z_{2}^{-}),
\,\,\,
t \rightarrow -\infty,
\label{as2s2}
\end{eqnarray}
where $z_{1,2}^{\pm}$ denote the position of each individual soliton
(in the above expressions, the parameters $A_{j}$ and $B_{j}$ ($j=1,2$),
with $A_{j}^2 + B_{j}^2=1$, characterize the velocity and depth of the soliton $j$).
Apparently the shape and the parameters of each soliton are preserved, while the
phase-shift of each soliton is given by:
\begin{eqnarray}
\Delta z_1 &\equiv& z_{1}^{+}-z_{1}^{-} = \frac{1}{2B_1} \ln\left[\frac{(A_1-A_2)^2
+ (B_1+B_2)^2}{(A_1-A_2)^2 + (B_1-B_2)^2} \right],
\label{dz1} \\
\Delta z_2 &\equiv& z_{2}^{+}-z_{2}^{-} = -\frac{1}{2B_2} \ln\left[\frac{(A_1-A_2)^2
+ (B_1+B_2)^2}{(A_1-A_2)^2 + (B_1-B_2)^2} \right].
\label{dz2}
\end{eqnarray}
Note that if the soliton velocities are equal, i.e., $A_1= - A_2 =A$ (hence, $B_1=B_2=B$),
then the phase-shift is equal for both solitons and is given by
$\Delta z = (2B)^{-1}\ln(1+B^2/A^2)$.

\begin{figure}[tbp]
\centering
\includegraphics[width=8.4cm,height=3.8cm]{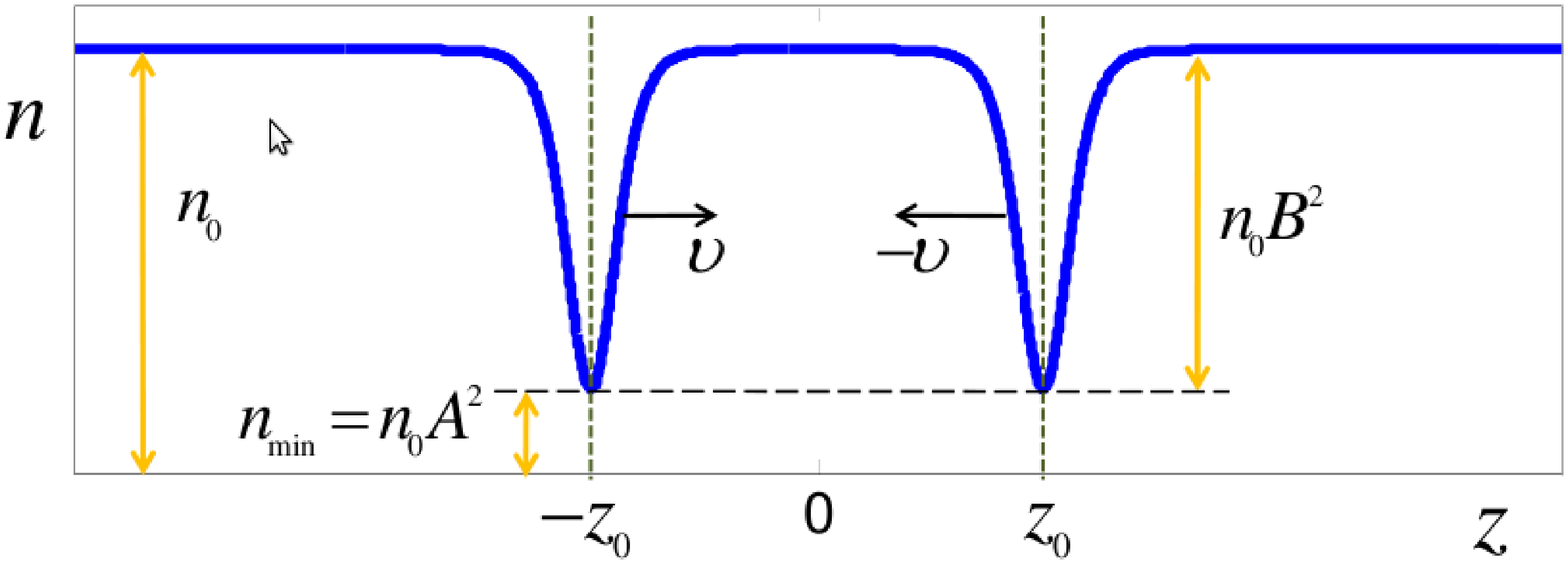}
\includegraphics[width=12cm,height=2.5cm]{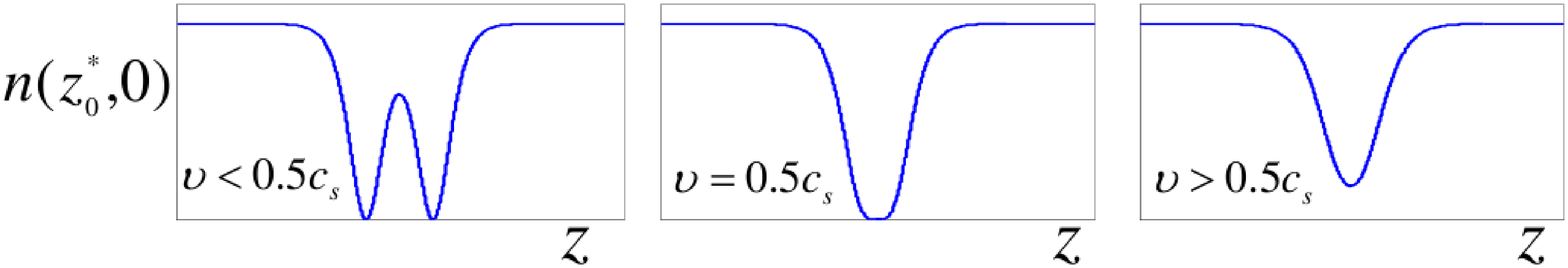}
\caption{(color online) The top panel shows the density profile of the two-soliton solution
in Eq.~(\ref{double}). The bottom panels show the density profile of two dark solitons at
their collision point corresponding to $z=z^{*}_{0}$ and $t=0$. The density of low-speed
solitons, $v < v_c$, is characterized by two distinguishable minima (bottom left panel),
while in the case of high-speed solitons, $v > v_c$, the density exhibits a single minimum
(bottom right panel); in the critical case, $v = v_c$, the density has a flat single minimum
(bottom middle panel).
}
\label{fig2}
\end{figure}

Equations (\ref{dz1})--(\ref{dz2}) show that the spatial shift of each soliton trajectory is
in the same direction as the velocity of each individual soliton and, thus, the dark
solitons always {\it repel} each other. Here it should be mentioned, however, that
this important result (as well as the collision dynamics {\it near} the collision point)
can  better be understood upon studying the explicit form of the two-soliton wave
function rather than its asymptotic limit considered above. To do so, we consider
again the case of a two-soliton solution, assuming for simplicity that the two
solitons are moving with equal velocities (i.e., $A_1=-A_2=A$).
In such a case, the two-soliton wave function is given by \cite{AA,gagnon}:
\begin{equation}
\psi(z,t)= \frac{F(z,t)}{G(z,t)} \exp(-i\mu t),
\label{double}
\end{equation}
where
\begin{eqnarray}
F(z,t)&=&2(n_0-2n_{min})\cosh(2n_0 AB t)
\nonumber \\
&-&2n_0 A \cosh(2\sqrt{n_0} B z)+i\sinh(2n_0 AB t),
\label{fzt} \\
G(z,t)&=&2\sqrt{n_0}\cosh(2n_0 AB t)+2 \sqrt{n_{min}}\cosh(2\sqrt{n_0} B z),
\label{gzt}
\end{eqnarray}
while $n_{min} =n_0 -n_0 B^2= n_0 A^2$ is the minimum density (i.e., the
density at the center of each soliton). The density profile
of the two-soliton solution in Eq.~(\ref{double}) is sketched in the top panel
of Fig.~\ref{fig2}.

To study analytically the interaction and collision between dark solitons,
we follow the approach of Ref. \cite{draft6} and find, at first, the trajectory of
the soliton coordinate $z_{0}$ as a function of time: using the
auxiliary equation $\partial_z |\psi|^2 = 0$
\footnotemark[1]
\footnotetext[1]{Recall that the dark soliton coordinate $z_{0}$ is the location of
the minimum density (see Fig.~\ref{fig2}).}
[where the density $|\psi|^2$ is determined by Eq.~(\ref{double})], the following result
is obtained:
\begin{equation}
\cosh(2\sqrt{n_0} B z_{0})=\sqrt{\frac{n_0}{n_{min}}}\cosh(2n_0 AB t)
-2\sqrt{\frac{n_{min}}{n_0}}\frac{1}{\cosh(2n_0 AB t)}.
\label{traj}
\end{equation}
Then, Eq.~(\ref{traj}) determines the distance $2z^{*}_0$
between the two solitons at the point of their closest proximity, i.e., the
{\it collision point} corresponding to $t=0$:
\begin{equation}
z^{*}_{0}=\frac{1}{2\sqrt{n_0-n_{min}}}\cosh^{-1}\left(\sqrt{\frac{n_0}{n_{min}}}
-2\sqrt{\frac{n_{min}}{n_0}}\right).
\label{closest}
\end{equation}
This equation [which holds for $n_{min}/n_0 = \nu^2 \le 1/4$, otherwise Eq.~(\ref{closest})
provides a complex (unphysical) value for $z^{*}_{0}$] shows that
$z^{*}_{0}=0$ for $n_{min}/n_0 = A^2 = 1/4$. Thus, it is clear that there exists
a critical value of the soliton velocity, namely
$v_c = \frac{1}{2} \sqrt{n_0} \equiv \frac{1}{2} c_s$, which defines two types of dark
solitons, exhibiting different behavior during their collision: ``low-speed'' solitons,
with $v < v_c$, which are {\it reflected} by each other, and ``high-speed'' solitons,
with $v > v_c$, which are {\it transmitted} through each other. In fact, as shown in the
bottom panels of Fig.~\ref{fig2}, the density profile of the low-speed (high-speed)
two-soliton state exhibits two separate minima (a single non-zero minimum) at the collision
point, namely $n(z^{*}_{0},0)=0$ ($n(z^{*}_{0},0) \ne 0$)
\footnotemark[1]
\footnotetext[1]{
In the case of solitons moving with the critical velocity, $v= v_c = \frac{1}{2}c_s$,
the two-soliton density exhibits a ``flat'' single zero minimum at the collision point
(see bottom-middle panel of Fig.~\ref{fig2}).
}.
In other words, low-speed solitons are, in fact, {\it well-separated} solitons, which can
always be characterized by two individual density minima --- even at the collision point ---
while high-speed solitons completely overlap at the collision point. According to the
nomenclature of Ref.~\cite{Huang}, the collision between slow-speed (high-speed) solitons is
called ``black collision'' (``gray collision''), since the dark solitons become black
(remain gray) at $t=0$. Notice that the case of gray collision can effectively be
described --- in the small-amplitude approximation --- by the collision dynamics of the
KdV equation \cite{Huang}.

\subsubsection{The repulsive interaction between slow dark solitons.}

Let us now investigate in more detail the case of well-separated solitons, which are always
reflected by each other, with their interaction resembling the one of
{\it hard-sphere-like particles}. In particular, we consider the limiting case of extremely
slow solitons, i.e., $n_0/n_{min}=A^2 \ll \frac{1}{4}$, for which the soliton separation is
large for every time (i.e., $z^{*}_{0} \gg 0$); in this case, the second term in the
right-hand side of Eq.~(\ref{traj}) is much smaller than the first one and can be ignored.
This way, the soliton coordinate is expressed as:
\begin{equation}
z_{0}=\frac{1}{2\sqrt{n_0} B} \cosh^{-1}\left[A^{-1}\cosh(2n_0 A B t)\right].
\label{x0}
\end{equation}
The above equation yields the soliton velocities:
\begin{equation}
\frac{dz_{0}}{dt}=\frac{\sqrt{n_0} \sinh(2n_0\nu B t)}{\sqrt{A^{-1}\cosh^{2}(2n_0\nu B t)-1}},
\label{dx00}
\end{equation}
which, in the limit $t\rightarrow 0$, become $dz_0/dt=0$. Thus, as the dark solitons
approach each other, their depth (velocity) is increased (decreased), and become
black at the collision point ($t=0$), while remaining at some distance away from each
other. Afterwards, the dark solitons are reflected by each other and continue their motion
in opposite directions, with their velocities approaching the asymptotic values
$dz_0/dt = \pm \sqrt{n_{0}} A$ for $t\rightarrow \pm\infty$ [see Eq.~(\ref{dx00})],
i.e., the velocity values of each individual soliton.

Next, differentiating Eq.~(\ref{x0}) twice with respect to time, and using
Eq.~(\ref{traj}) (without the second term which is negligible for well-separated
solitons), one may derive an equation of motion for the soliton coordinate in the form
$d^2 z_0/dt^2 =-\partial V_{\rm int}(z_0)/\partial z_0$,
where the interaction potential $V_{\rm int}(z_0)$ is given by:
\begin{equation}
V_{\rm int}(z_{0})=\frac{1}{2}\frac{n_0 B^2}{\sinh^2(2\sqrt{n_0}Bz_{0})}.
\label{dx01}
\end{equation}
It is clear that $V_{\rm int}$ is a {\it repulsive} potential, indicating that the dark
solitons {\it repel each other}. If the separation between the dark solitons is sufficiently
large (i.e., $2z_0 \gg 1$) then the hyperbolic $\sinh$ function in Eq.~(\ref{dx01}) can be
approximated by its exponential asymptote, and the potential in Eq.~(\ref{dx01})
can be simplified as:
\begin{equation}
V_{\rm int}(z_{0}) \approx 2n_0 B^2 \exp(-4\sqrt{n_0}Bz_{0}).
\label{dx01ap}
\end{equation}
The latter expression can also be derived by means of a Lagrangian approach \cite{lads}.
Importantly, although the above result refers to a symmetric two-soliton collision, the
results of Ref.~\cite{draft6} show that it is possible to use the repulsive potential
(\ref{dx01}) in the cases of {\it non-symmetric} collisions --- using an ``average depth''
of the two solitons --- and {\it multiple} dark solitons --- with each soliton interacting
with its neighbors (see also relevant discussion in Sec.~\ref{crossover}).

\subsubsection{Experiments on multiple dark solitons.}

Multiple dark solitons were first created in a $^{23}$Na BEC in the NIST experiment
\cite{nist} by the phase-imprinting method (see Sec.~\ref{pi}), while the interaction and
collision between two dark solitons in a $^{87}$Rb BEC was first studied in the Hannover
experiment of Ref.~\cite{han2}. Nevertheless, in this early experiment the outcome of the
collision was not sufficiently clear due to the presence of dissipation caused by the
interaction of the condensate with the thermal cloud. In the more recent Hamburg experiment
\cite{hambcol}, the phase-imprinting method was also used to create two dark solitons in
a $^{87}$Rb BEC with slightly different depths. These solitons propagated to opposite sides
of the condensate, reflected near the edges of the BEC, and subsequently underwent a
single ``gray'' collision near the center of the trap. In addition, in the recent Heidelberg
experiment \cite{kip} two dark solitons were created in a $^{87}$Rb BEC by the so-called
{\it interference method} (see Sec.~\ref{interf} below). The solitons observed in this
experiment, which were ``well-separated'' ones, propagated to opposite directions, reflected
and then underwent multiple genuine elastic ``black'' collisions, from which the solitons
emerged essentially unscathed. Notice that the experimentally observed dynamics of the
oscillating and interacting dark-soliton pair of Ref.~\cite{kip}, as well as the one of
multiple dark solitons in another Heidelberg experiment \cite{draft6}, was in a very good
agreement with theoretical predictions based on the effective particle-like picture for
dark solitons (see Sec.~\ref{adiabatic} and Sec.~\ref{crossover} below) and the interaction
potential of Eq.~(\ref{dx01}).

\subsubsection{Stationary dark solitons in the trap.}

At this point, it is relevant to briefly discuss the case where multiple dark solitons are
considered in a trapped condensate. In this case, both the single dark soliton and all other
multiple dark soliton states can be obtained in a {\it stationary} form from the
non-interacting (linear) limit of Eq.~(\ref{dim1dgpe}), i.e., in the absence of the nonlinear term.
In this case, Eq.~(\ref{dim1dgpe}) is reduced to a linear Schr{\"o}dinger
equation for a confined single-particle state. For the harmonic potential of
Eq.~(\ref{hpot}), this Schr{\"o}dinger equation describes the quantum harmonic oscillator,
characterized by discrete energies and corresponding localized eigenmodes in the form of
Hermite-Gauss polynomials \cite{landau}. As shown in Refs.~\cite{gp1d_B,konotop1}, all these
eigenmodes exist also in the fully nonlinear problem, and describe an analytical continuation
of the above mentioned linear modes to a set of nonlinear stationary states. Additionally,
analytical and numerical results of the recent work \cite{AZ} suggest that in the case of
a harmonic trapping potential there are no solutions of the 1D GP Eq.~(\ref{dim1dgpe})
without a linear counterpart. This actually means that interatomic interactions (i.e.,
the effective mean-field nonlinearity in the GP model) transforms all higher-order stationary
modes into a sequence of stationary dark solitons confined in the harmonic trap
\cite{gp1d_B,konotop1}; note that as concerns its structure, this chain of, say $n$, stationary dark solitons shares the same spatial profile with the linear eigenmode of quantum number $n$.
From a physical point of view, multiple stationary dark soliton states exist due to the
fact that the repulsion between dark solitons is counter-balanced by the restoring force
induced by the trapping potential.


\section{Matter-wave dark solitons in quasi-1D Bose gases}
\label{sec4}

\subsection{General comments.}
\label{comments}

We consider again the quasi-1D setup of Eq.~(\ref{dim1dgpe}), but now incorporating the
external potential $V(z)$. In this setting, the dynamics of matter-wave dark solitons can
be studied analytically by means of various perturbation methods, assuming that the trapping
potential $V(z)$ is smooth and slowly-varying on the soliton scale. This means that in
the case, e.g., of the conventional harmonic trap [cf.~Eq.~(\ref{htrap})], the normalized
trap strength is taken to be $\Omega \sim \epsilon$, where $\epsilon \ll 1$ is a formal
small (perturbation) parameter. In such a case, Eq.~(\ref{dim1dgpe}) can be expressed as
a perturbed NLS equation, namely,
\begin{eqnarray}
i \partial_t \psi +\frac{1}{2} \partial_z^{2} \psi - |\psi|^{2} \psi = R(\psi) \equiv V(z) \psi.
\label{gpe1d_u}
\end{eqnarray}
Then, according to the perturbation theory for solitons \cite{pertkm}, one may assume that a
perturbed soliton solution of Eq.~(\ref{gpe1d_u}) can be expressed in the following general
form,
\begin{equation}
\psi(z,t)=\psi_{s}(z,t) + \epsilon \psi_{r}(z,t).
\label{pes}
\end{equation}
Here, $\psi_{s}(z,t)$ has the functional form of the dark soliton solution (\ref{ds}),
but with the soliton parameters depending on time, and $\psi_{r}$ is the radiation ---
in the form of sound waves ---  emitted by the soliton.
Generally, the latter term is strong only for sufficiently strong perturbations (see,
e.g., Refs.~\cite{parker1,parker2}, as well as Ref.~\cite{analogies} and discussion in
Sec.~\ref{radiation}). Thus, the simplest possible approximation for a study of matter-wave
dark solitons in a trap corresponds to the so-called {\it adiabatic approximation} of the
perturbation theory for solitons \cite{pertkm}, namely $\psi(z,t) \approx \psi_{s}(z,t)$.
In any case, the study of matter-wave dark solitons in a trap should take into regard that
the trap changes the boundary conditions for the wave function, and BEC density, namely
$n \rightarrow 0$ [instead of $n \rightarrow n_0$ in the homogeneous case --- see, e.g.,
Eq.~(\ref{darkness})] as $z \rightarrow \pm \infty$. From a physical viewpoint, and based on
the particle-like nature of dark solitons (see Sec.~\ref{integrals}), one should expect
that dark solitons could be reflected from the trapping potential; apparently, such a
mechanism should then result in an {\it oscillatory motion} of dark solitons in the trap.

There exist many theoretical works devoted to the oscillations of dark solitons in trapped
BECs. The earlier works on this subject reported that solitons oscillate in a condensate
confined in a harmonic trap of strength $\Omega$, and provided estimates for the oscillation
frequency. In particular, in Ref.~\cite{motion1} soliton oscillations were observed in
simulations and a soliton's equation of motion was presented without derivation; in the same
work, it was stated that the solitons oscillate with frequency $\Omega$ (rather than the
correct result which is $\Omega/\sqrt{2}$ --- see below). The same result was derived in
Ref.~\cite{motion2}, considering the dipole mode of the condensate supporting the dark soliton.
Other works \cite{motion3a,motion3b,motion3c} also considered oscillations of dark solitons
in trapped BECs. An analytical description of the dark-soliton motion, and the correct result
for the soliton oscillation frequency, $\Omega/\sqrt{2}$, were first presented in
Ref.~\cite{fr2} by means of a multiple-time-scale boundary-layer theory (this approach is
commonly used for vortices \cite{pismen}). The same result was obtained in
Refs.~\cite{fms,Muryshev} by solving the BdG equations (for almost black solitons performing
small-amplitude oscillations around the trap center --- see Sec.~\ref{BdGanalysis} below),
using a time-independent version of the boundary-layer theory. Furthermore, in Ref.~\cite{fms}
a kinetic-equation approach was used to describe dissipative dynamics of the dark soliton due
to the interaction of the BEC with the thermal cloud.

Matter-wave dark soliton dynamics in trapped BECs was also analyzed in other works by means
of different techniques that were originally developed for optical dark solitons \cite{kivpr}.
In particular, in Ref.~\cite{fr1} the problem was analyzed by means of the adiabatic
perturbation theory for dark solitons devised in Ref.~\cite{KY}, in Ref.~\cite{huang1} by
means of the small-amplitude approximation (see Sec.~\ref{kdvstuff}), while in
Ref.~\cite{braz1} by means of the perturbation theory of Ref.~\cite{vvkpert}. Later, in
Refs.~\cite{fr3,bkp} the so-called ``Landau dynamics'' approach was developed, based on the
use of the renormalized soliton energy [cf. Eq.~(\ref{EDS})], along with a local density
approximation. Models relevant to the dynamics of matter-wave dark solitons in 1D
{\it strongly-interacting} Bose gases, were also considered and analyzed by means of the
small-amplitude approximation \cite{tonks1,tonks2} (see also work for dark solitons
in this setting in Refs.~\cite{marvin1,marvin2,kavoulakis,baizakov,marvin3}). In other works,
a Lagrangian approach for matter-wave dark solitons was presented \cite{fr4} (see also
Ref.~\cite{BilPav1}), and an asymptotic multi-scale perturbation method was used to
describe dark soliton oscillations and the inhomogeneity-induced emission of radiation
\cite{fr5}. Recently, the motion of dark solitons was rigorously analyzed in
Ref.~\cite{pelpan} (where a wider class of traps was considered), while in
Ref.~\cite{salernokam} the same problem was studied in the framework of a
generalized NLS model.

Finally, as far as experiments are concerned, the oscillations of dark solitons were only
recently observed in the Hamburg \cite{hamburg,hambcol} and Heidelberg \cite{kip,draft6}
experiments. In these works, the experimentally determined soliton oscillation frequencies
were found to deviate from the theoretically predicted value $\Omega/\sqrt{2}$. This deviation
was explained in Refs.~\cite{hamburg,hambcol} by the anharmonicity of the trap, while in
Refs.~\cite{kip,draft6} by the dimensionality of the system and the soliton interactions
(see also Sec.~\ref{crossover} below).

\subsection{Adiabatic dynamics of matter-wave dark solitons}
\label{adiabatic}

\subsubsection{The perturbed NLS equation.}

The adiabatic dynamics of dark matter-wave solitons may be studied analytically by means of
the Hamiltonian \cite{uz,KY} or the Lagrangian \cite{lads} approach of the perturbation theory
for dark solitons, which were originally developed for the case of a constant background.
These approaches were later modified (see Ref. \cite{fr1} for the Hamiltonian approach and
Refs.~\cite{fr4,BilPav1} for the Lagrangian approach) to take into regard that, in the
context of BECs, the background is inhomogeneous  due to the presence of the external
potential. The basic steps of these perturbation methods are: (a) determine the background
wave function carrying the dark soliton, (b) derive from Eq.~(\ref{gpe1d_u}) a perturbed NLS
equation for the dark soliton wave function, and (c) determine the evolution of the dark
soliton parameters by means of the renormalized Hamiltonian [cf. Eq.~(\ref{EDS})] or the
renormalized Lagrangian [cf. Eq.~(\ref{LDS})] of the dark soliton. Here, we will present the
first two steps of the above approach and, in the following two subsections, we will
describe the adiabatic soliton dynamics in the framework of the Hamiltonian and Lagrangian
approaches.

We consider again Eq.~(\ref{gpe1d_u}) and seek the background wave function in the form,
\begin{equation}
\psi(z,t) = \Phi(z) \exp(-i\mu t + i\theta_{\rm o}),
\label{backgr}
\end{equation}
where $\mu$ is the normalized chemical potential, $\theta_{\rm o}$ is an arbitrary phase,
while the unknown real function $\Phi(z)$ satisfies the following equation,
\begin{equation}
\mu\Phi+\frac{1}{2}\frac{d^{2} \Phi}{dz^2}-\Phi^{3}=V(z)\Phi.
\label{ub11}
\end{equation}
Then, we seek for a dark soliton solution of Eq.~(\ref{gpe1d_u}) on top of the inhomogeneous
background satisfying Eq.~(\ref{ub11}), namely,
$
\psi = \Phi(z) \exp(-i\mu t+ i\theta_{\rm o}) \psi_{s}(z,t),
$
where the unknown wave function $\psi_{s}(z,t)$ represents a dark soliton. This way, employing
Eq.~(\ref{ub11}), the following evolution equation for the dark soliton wave function is
readily obtained:
\begin{equation}
i \partial_t \psi_{s} +\frac{1}{2} \partial_z^{2} \psi_{s}
- \Phi^{2}(|\psi_{s}|^{2}-1)\psi_{s} = -\frac{d}{dz}\ln(\Phi) \partial_z \psi_{s}.
\label{upsa}
\end{equation}
It is clear that if the trapping potential $V(z)$ is smooth and slowly-varying on the soliton
scale, then the right-hand-side, and also part of the nonlinear terms of Eq.~(\ref{upsa}),
can be treated as a perturbation. To obtain this perturbation in an explicit form, we
use the TF approximation to express the background wave function as $\Phi(z)=\sqrt{1-V(z)}$
[see Eq.~(\ref{TF}) for $g=1$ and $\mu=1$
\footnotemark[1]
\footnotetext[1]{It can easily be shown that the main result of the analysis [cf. Eq. (\ref{eqmd})]
can be generalized for every value of $\mu$ such that the system is in the TF-1D regime.}
] and approximate the logarithmic derivative of $\Phi$ as
\begin{equation}
-\frac{d}{dz} \ln \Phi \approx \frac{1}{2} \frac{dV}{dz} \left(1+V+V^2\right).
\label{approxPhi}
\end{equation}
This way, Eq.~(\ref{upsa}) leads to the following perturbed NLS equation,
\begin{equation}
i\frac{\partial \psi_{s}}{\partial t}+\frac{1}{2}\frac{\partial^{2} \psi_{s}}{\partial z^2}
-(|\psi_{s}|^{2}-1)\psi_{s}=Q(\psi_{s}),
\label{pnlsd}
\end{equation}
where the perturbation $Q(\psi_{s})$ is given by:
\begin{eqnarray}
Q(\psi_{s})= \left( 1-|\psi_{s}|^{2}\right) \psi_{s} V
+ \frac{1}{2}\partial_z \psi_{s} \frac{dV}{dz} (1+V+V^2).
\label{pertQ}
\end{eqnarray}

\subsubsection{Hamiltonian approach of the perturbation theory.}
\label{hamap}

\begin{figure}[tbp]
\centering
\includegraphics[width=10cm]{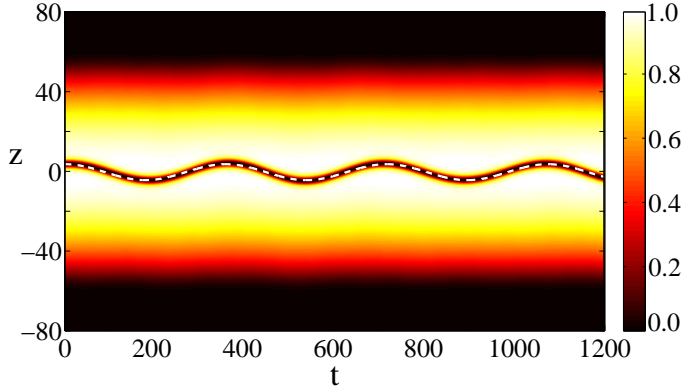}
\caption{(Color online) Contour plot showing the evolution of the density of a harmonically
confined  BEC, as obtained by direct numerical integration of the GP Eq. (\ref{gpe1d_u}).
The initial condition is $\psi = [\mu-(1/2)\Omega^2 z^2]\tanh(z-z_0(0))$, i.e., a TF cloud,
characterized by a chemical potential $\mu$, and carrying a dark soliton initially placed at
$z=z_0(0)$. The parameter values are $\mu=1$, $\Omega=0.025$ and $z_0(0)=4$. The dark soliton
oscillates with frequency $\omega_{\rm osc}=\Omega/\sqrt{2} \approx 0.018$. The dotted line
across the soliton trajectory corresponds to the analytical prediction of Eq.~(\ref{eqmd}).
}
\label{figosc}
\end{figure}

First we note that in the absence of the perturbation (\ref{pertQ}), Eq.~(\ref{pnlsd}) has a
dark soliton solution of the form:
\begin{equation}
\psi_{s}(z,t)=\cos \phi \tanh \zeta +i \sin \phi,
\label{ansatzHam}
\end{equation}
where $\zeta \equiv \cos \phi \left[ x-(\sin \phi)t \right]$ [see Eq.~(\ref{ds})].
Then, considering an adiabatic evolution of the dark soliton, we assume that in the
presence of the perturbation the dark soliton parameters become slowly-varying unknown
functions of $t$ \cite{uz,KY,fr1}. Thus, the soliton phase angle becomes
$\phi \rightarrow \phi(t)$ and, as a result, the soliton coordinate becomes
$\zeta \rightarrow \zeta(t) =\cos\phi(t) \left[z-z_{0}(t) \right]$.
In the latter expression, the dark soliton center $z_{0}(t)$ is connected to the soliton
phase angle through the following equation:
\begin{equation}
\frac{d z_{0}(t)}{dt} = \sin\phi(t).
\label{dscenter}
\end{equation}
The evolution of the soliton phase angle can be found by means of the evolution of the
{\it renormalized soliton energy}. In particular, employing Eq.~(\ref{EDS}) (for $\mu=1$),
it is readily found that
$dE_{\rm ds}/dt=-4\cos^2\phi \sin\phi (d\phi/dt)$.
On the other hand, using Eq.~(\ref{pnlsd}) and its complex conjugate, it can be found that
the evolution of the renormalized soliton energy is given by
$dE_{\rm ds}/dt=-\int_{-\infty}^{+\infty}
\left[Q(\psi_{s})\partial_t \psi_s^{\ast}+Q^{\ast}(\psi_{s})\partial_t \psi_s \right]dz$.
Then, the above expressions for $dE_{\rm ds}/dt$ yield the evolution of $\phi$,
namely \cite{KY}:
\begin{equation}
\frac{d\phi}{dt}=\frac{1}{2\cos ^{2}\phi \sin \phi} {\rm Re}
\left[ \int_{-\infty}^{+\infty}Q(\psi_{s})\partial_t \psi_{s}^{\ast} dz \right].
\label{evphi}
\end{equation}
Next, we Taylor expand the potential $V(z)$ around the soliton center $z_0$, and assume that
the dark soliton is moving in the vicinity of the trap center, i.e., $\mu \equiv 1 \gg V$,
which means that the last two terms in the right-hand side of Eq.~(\ref{pertQ}) can be
neglected. This way, one may further simplify the expression for the perturbation in
Eq.~(\ref{pertQ}) which, when inserted into Eq.~(\ref{evphi}), yield the following result:
\begin{eqnarray}
\frac{d \phi}{dt}=- \frac{1}{2} \cos\phi \frac{\partial V}{\partial z_{0}}.
\label{dark_par2}
\end{eqnarray}
To this end, combining Eq.~(\ref{dark_par2}) with Eq.~(\ref{dscenter}), we obtain the
following equation of motion for nearly stationary (black) solitons with $\cos \phi \approx 1$,
\begin{eqnarray}
\frac{d^2 z_{0}}{dt^2}=-\frac{1}{2} \frac{\partial V }{\partial z_{0}}.
\label{eqmd}
\end{eqnarray}
The above result indicates that the dark soliton center can be regarded as a Newtonian
particle: Eq.~(\ref{eqmd}) has the form of a Newtonian equation of motion of a classical
particle, of an effective mass $M_{\rm eff} = 2$, in the presence of the external potential
$V$. In the case of the harmonic potential [cf. Eq.~(\ref{htrap})], Eq.~(\ref{eqmd})
becomes the equation of motion of the classical linear harmonic oscillator,
$d^2 z_0/dt^2 = -(1/2)\Omega^2 z_0$, and shows that the dark soliton oscillates with
frequency
\begin{equation}
\omega_{\rm osc} = \frac{\Omega}{\sqrt{2}},
\label{soloscfreq}
\end{equation}
or, in physical units, with $\omega_z/\sqrt{2}$. An example of an oscillating matter-wave
dark soliton is shown in Fig.~\ref{figosc}.

At this point, it is relevant to follow the considerations of Ref.~\cite{fms}
(see also \cite{fr2,Muryshev}) and estimate the energy $E_{\rm ds}$ of this almost
dark soliton in the trap. Taking into regard that in the case of a homogeneous BEC
this energy is given by Eq.~(\ref{Ebs}), one may use a {\it local density approximation}
and use in Eq.~(\ref{EDS}) the local speed of sound, $c(z) = \sqrt{n_0(z)}$ \cite{Bogoliubov}
(here, $n_0(z)$ is the density of the ground state of the BEC), rather than the constant
value $c_s = \sqrt{n_0}$ [cf. Eq.~(\ref{ssh})]. Then, in the TF limit, the density is
expressed as $n_0(z) = \mu-\frac{1}{2}\Omega^2 z^2 = c_{s}^2 - \frac{1}{2}\Omega^2 z^2$ and,
thus, one may follow the lines used for the derivation of Eq.~(\ref{Ebs}) (for sufficiently
slow solitons and weak trap strengths) and obtain the result:
\begin{equation}
E_{\rm ds} = E_0 +\frac{1}{2}m_{\rm ds}v^{2} +\frac{1}{4}m_{\rm ds}\Omega^2 z^2,
\label{Ebsinh}
\end{equation}
where $E_{0} \equiv \frac{4}{3} c_s^{3}$ and $m_{\rm ds}=-4\sqrt{n_{0}}$ as in Eq.~(\ref{Ebs}).
The above equation shows that the incorporation of the harmonic trap results in a decrease of
the energy of the dark soliton by the potential energy term
$\frac{1}{4}|m_{\rm ds}|\Omega^2 z^2$.
Moreover, the ratio of the soliton mass over this potential energy is given by
$(\Omega^2 z^2/4)^{-1}$, which is exactly two times the ratio of the atomic mass
(which is equal to $m=1$ in our units) over the external potential, namely
$(\Omega^2 z^2/2)^{-1}$. This is another interpretation of the result that the effective mass
of the dark soliton center is $M_{\rm eff} = 2$.

\subsubsection{Lagrangian approach for matter-wave dark solitons.}

The perturbed NLS Eq. (\ref{pnlsd}), with the perturbation of Eq.~(\ref{pertQ}), can also be
treated by means of a variational approach as discussed in the beginning of
Sec.~\ref{adiabatic}. First, we assume that the solution of Eq.~(\ref{pnlsd}) is expressed as
[see Eqs.~(\ref{ds}) and (\ref{phi})]:
\begin{equation}
\psi_s(z,t)=B\tanh\zeta +iA.
\label{ansatzLagr}
\end{equation}
Here, $A$ and $B$ are unknown slowly-varying functions of time (with $A^2+B^2=1$)
representing, respectively, the velocity and amplitude of the dark soliton (which
become time-dependent due to the presence of the perturbation), while
$\zeta \equiv B(t) \left[ z-z_0(t) \right]$,
where $z_0(t)$ is the dark soliton center. Note that in the unperturbed case,
$dz_{0}/dt \equiv A$, but in the perturbed case under consideration, this simple relationship
may not be valid (see below). Next, the evolution of the unknown soliton parameters
$\alpha_{j}(t)$ (which is a generic name for $z_{0}(t)$ and $A(t)$) are obtained via the
Euler-Lagrange equations \cite{lads,fr4}:
\begin{equation}
\frac{\partial L_{\rm ds}}{\partial \alpha_{j}}
-\frac{d}{dt}\left(\frac{\partial L_{\rm ds}}{\partial \dot{\alpha}_{j}}\right)
=2 {\rm Re}\left\{\int_{-\infty}^{+\infty} Q^{\ast}(\psi_{s})
\frac{\partial \psi_s }{\partial \alpha_{j}}dz \right\},
\label{EL}
\end{equation}
where $\dot{\alpha}_{j} \equiv d \alpha_{j}/dt$ and
$L_{\rm ds}=\int_{-\infty}^{+\infty} dz{\cal L}_{\rm ds}\{\psi_{s}\}$
represents the averaged Lagrangian of the dark soliton of the unperturbed NLS equation
(namely for $Q(\psi_{s})=0$), with the Lagrangian density ${\cal L}_{\rm ds}$ being given
by Eq.~(\ref{LDS}) (for $n_0=1$). The averaged Lagrangian can readily be obtained by
substituting the ansatz (\ref{ansatzLagr}) into Eq.~(\ref{LDS}):
\begin{equation}
L_{\rm ds}=2\frac{dz_{0}}{dt}\left[-AB+\tan^{-1}\left(\frac{B}{A}\right)\right]-\frac{4}{3}B^3.
\label{Lds}
\end{equation}
Therefore, substituting Eqs.~(\ref{Lds}) and (\ref{pertQ}) into Eq.~(\ref{EL}), it is
straightforward to derive evolution equations for the soliton parameters. For completeness,
we will follow Ref.~\cite{fr4} and present the final result taking also into account the
last two terms in the right-hand side of Eq.~(\ref{pertQ}) --- which were omitted in the
previous subsection --- so as to describe the motion of shallower solitons as well.
This way, and employing a Taylor expansion of the potential around the soliton center
(as in the previous subsection), we obtain the following evolution equations for $z_0(t)$
and $A(t)$:
\begin{eqnarray}
&\frac{dz_{0}}{dt}= A \left[1-\frac{1}{2}V(z_{0})\right]
- \frac{A}{4B^2} \left( \frac{5}{3}-\frac{\pi^2}{9} \right)
\left(\frac{\partial V}{\partial z_{0}}\right)^{2}
\left[1- 2V(z_{0})\right],
\label{x0t}
\end{eqnarray}

\begin{eqnarray}
\frac{dA}{dt}=&-&\frac{1}{2} B^{2} \frac{\partial V}{\partial z_{0}}
-\frac{1}{3} B^{2} V(z_{0}) \frac{\partial V}{\partial z_{0}}
\nonumber \\
&-&B^{2} \frac{\partial V} {\partial z_{0}} \left[ \frac{1}{3}V^{2}(z_{0})+
\frac{1}{4}\left(\frac{2}{3}
-\frac{\pi^2}{9}\right) \left(\frac{\partial V}{\partial z_{0}}\right)^{2} \right].
\label{At}
\end{eqnarray}
Equations (\ref{x0t})-(\ref{At}) describe the dark soliton dynamics in the trap, in both
cases of nearly black solitons ($A \approx 0$ or $B \approx 1$) and gray ones (with arbitrary
$A$ or $B$). In the former case, and neglecting the higher-order corrections arising from the
inclusion of the last two terms in the right-hand side of Eq.~(\ref{pertQ}), the result of
Eq.~(\ref{eqmd}) is recovered: nearly black solitons oscillate near the trap center with the
characteristic frequency given in Eq.~(\ref{soloscfreq}). On the other hand, numerical
simulations in Ref.~\cite{fr4} have shown that the full system of
Eqs.~(\ref{x0t})--(\ref{At}) predicts that shallow solitons oscillate in the trap with
the {\it same} characteristic oscillation frequency. Therefore,
there is a clear indication that the oscillation frequency of Eq.~(\ref{soloscfreq}) does
{\it not} depend on the dark soliton amplitude. This result is rigorously proved by means
of the Landau dynamics approach that will be discussed below.

\subsubsection{Landau dynamics of dark solitons.}

The oscillations of dark solitons of arbitrary amplitudes in a trap can also be studied by
means of the so-called {\it Landau dynamics} approach devised in Refs.~\cite{fr2,bkp}. This
approach, which further highlights the particle-like nature of the matter-wave dark solitons,
relies on a clear physical picture: when a dark soliton moves in a weakly inhomogeneous
background, its local energy stays constant. Hence, one may employ the local density
approximation, and rewrite the energy conservation law of Eq.~(\ref{EDS}) as
$c^2 (z_0) - v^2 = (3E_{\rm ds}/4)^{2/3}$, where $c^2(z_0)$ is the local speed of sound
evaluated at the dark soliton center $z_0$. Then, in the TF limit, one has
$c^2(z_0)= c_{s}^2 - \frac{1}{2}\Omega^2 z_0^2$ (as before), and taking into regard that
the soliton velocity is $v=dz_0/dt$, the following equation for the energy of the dark
soliton is readily obtained:
\begin{equation}
\frac{1}{2}M_{\rm eff} \left( \frac{dz_0}{dt}\right)^2
+ \frac{1}{2} \Omega^2 z_0^2 = \tilde{E}_{\rm ds},
\label{EdsL}
\end{equation}
where $\tilde{E}_{\rm ds} = c_s^2 - (3E_{\rm ds}/4)^{2/3}$ and the effective mass of the dark
soliton center is again found to be $M_{\rm eff} = 2$. It is readily observed that
Eq.~(\ref{EdsL}) can be reduced to Eq.~(\ref{eqmd}) and, thus, it leads to the oscillation
frequency of Eq.~(\ref{soloscfreq}). Nevertheless, the result obtained in the framework of
the Landau dynamics approach is more general, as it actually refers to dark solitons of
arbitrary amplitudes. Moreover, the same approach can be used also in the case of more general
models, including, e.g., the cases of non-harmonic traps and/or more general nonlinearity
models, such as the physically relevant ones described by Eqs.~(\ref{salnl})--(\ref{nltg})
\cite{bkp}. Nevertheless, it should be noted that in such more general cases the problem can
be treated analytically for almost black solitons ($v \ll c$), performing small-amplitude
oscillations. In this case, the conservation law $E_{\rm ds}(c(z_0), v) = E_0$ can be Taylor
expanded around $z_0=0$ and $v=0$, leading to expressions for the soliton's effective
soliton mass and oscillation frequency \cite{bkp}.

\subsubsection{The small-amplitude approximation.}

Next, we discuss the adiabatic dynamics of {\it small-amplitude} dark solitons in trapped
1D Bose gases. In this case, one may formally reduce the more general GP model of
Eq.~(\ref{dg}) (including the potential term $V \psi$) to a KdV equation with
{\it variable coefficients} --- see Ref.~\cite{revnonlin} for details and Ref.~\cite{asano}
for applications of this KdV model. The main result of such an analysis is that the density
and the phase of the approximate shallow dark soliton solution of Eq.~(\ref{dg}) have,
to the leading-order of approximation, the functional form of their counterparts in
Eqs.~(\ref{adso}) and (\ref{solph}), but with the soliton parameter $\kappa$ depending on
a slow variable, say $Z$ (see earlier work for a calculation of $\kappa(Z)$ in
Refs.~\cite{ko,karpman}). This way, approximate analytical shallow soliton solutions
have been found in various works \cite{Huang,huang1,tonks1,tonks2} for different forms
of the nonlinearity function $f(n)$. Nevertheless, there are some subtle issues concerning
the validity of this approximation, as discussed in Refs.~\cite{braz1,bkp,fr5}, which is,
strictly speaking, valid {\it away from the turning points} (where the soliton velocity
vanishes). On the other hand, numerical results (see, e.g., Ref.~\cite{tonks2}), illustrate
that the range of validity of the above results is, in fact, wider than what may be expected
based on the limitations of this approach.

The small-amplitude approximation, along with a local-density approximation, has also been
used to estimate the shallow soliton's oscillation frequency: the shallow soliton's velocity
$v$, which in the homogeneous problem was found to be close to the speed of sound, namely
$v \approx c_s = \sqrt{f_0' n_0}$  [see Eq.~(\ref{ssolv})], can be approximated in the
inhomogeneous system as follows:
\begin{equation}
\frac{dZ}{dt} \approx c_s(Z) = \sqrt{f_0' n_0(Z)}.
\label{sc}
\end{equation}
In some cases, Eq.~(\ref{sc}) can be used for the derivation of physically relevant results.
For example, following Ref.~\cite{bkp}, we assume that $f(n) = n^{\alpha}$, where
$\alpha =1$ for weakly-interacting BECs, or $\alpha =2$ for strongly-interacting Tonks
gases \cite{kolom1}. Then, in the TF limit, $n_0(Z) = [\mu - V(Z)]^{1/\alpha}$, and
Eq.~(\ref{sc}) is reduced to the form $dZ / \sqrt{\mu-V(Z)} = \sqrt{\alpha}dt$. The latter
is integrated and yields [for $V(Z) = (1/2)\Omega^2 Z^2$] the soliton trajectory:
\begin{equation}
Z = L \sin[(\Omega \sqrt{\alpha/2})t],
\label{trajs}
\end{equation}
where $L=\sqrt{2\mu}/\Omega$ is the length of the TF cloud. Equation~(\ref{trajs}) predicts
that the shallow dark soliton will perform oscillations approximately in the entire spatial
region occupied by the gas, with an oscillation frequency which takes the following values
(in physical units): for $\alpha=1$, i.e., for quasi-1D BECs described by the cubic GP
equation, $\omega_{\rm osc}=\omega_z/\sqrt{2}$, while for $\alpha=2$, i.e., for the Tonks
gas described by a quintic NLS equation, $\omega_{\rm osc}=\omega_z$. Note that the latter
result was first obtained via a many-body calculation \cite{bh}, and later was derived by
means of the KdV approximation \cite{tonks1}.


\subsection{Bogoliubov-de Gennes analysis of stationary dark solitons}
\label{BdGanalysis}

\subsubsection{The single-soliton state.}

\begin{figure}[tbp]
\centering
\includegraphics[width=12cm]{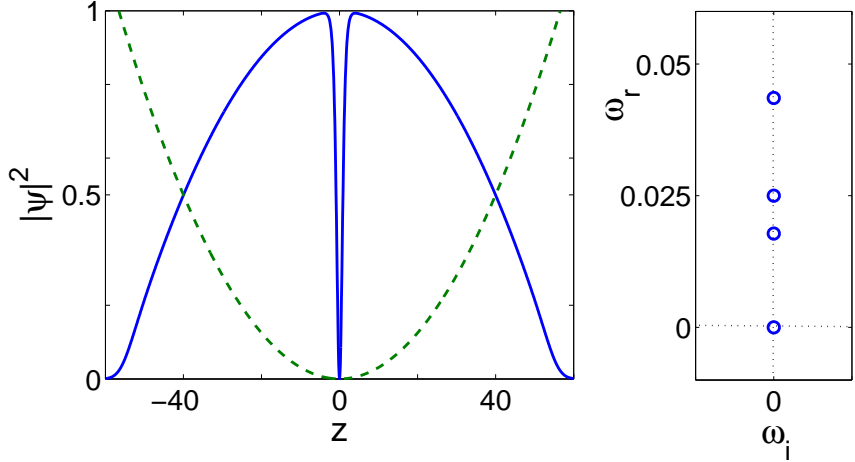}
\centering
\caption{(Color online)
Left panel: The density of a condensate carrying a stationary (black) soliton located at $z=0$.
The normalized chemical potential is $\mu=1$. The (green) dashed line shows the trapping
potential with normalized trap strength $\Omega=0.025$. Right panel: The lowest
characteristic eigenfrequencies of the Bogoliubov excitation spectrum: The eigenfrequency
located at the origin corresponds to the Goldstone mode, the one at $\Omega = 0.025$ to the
Kohn mode, and the one at $\sqrt{3}\Omega$ to the quadrupole mode. Finally,
there exist a anomalous mode with $\omega_{A} = \Omega/\sqrt{2}$.}
\label{figspectrum}
\end{figure}

The above result, namely the fact that almost black solitons perform oscillations around the
trap center with the frequency given in Eq.~(\ref{soloscfreq}), can also be derived by means
of a BdG analysis as was first demonstrated in Ref.~\cite{Muryshev} (see also results in
Refs.~\cite{mprizolas,brand,dz}). Such an analysis can be done in the TF limit for the
{\it stationary} dark soliton state, namely the black soliton $\psi_0$ [see Eq.~(\ref{black})]
located at the trap center, i.e., $z_{0}=0$, which is actually the first excited state of the
condensate. Then, following the discussion in Sec.~\ref{gpebdg}, the excitation spectrum can
be found as follows: using the ansatz
$\psi(z,t)=\left[\psi_{0}(z)+u(z)e^{-i\omega t}
+\upsilon^{\ast}(z)e^{i\omega t}\right]e^{-i \mu t}$
[where $\omega=\omega_r+i\omega_i$ is a (generally complex)
eigenfrequency and $(u, v)$ are perturbation eigenmodes],
we derive from Eq.~(\ref{dim1dgpe}) the following BdG equations:
\begin{eqnarray}
&&[\hat{H} - \mu + \psi_{0}^{2}] u + \psi_{0}^{2}\upsilon = \omega u,
\label{BdG2a} \\
&&[\hat{H} - \mu + \psi_{0}^{2}] \upsilon + \psi_{0}^{*2}u = -\omega \upsilon,
\label{BdG2b}
\end{eqnarray}
where $\hat{H}= -(1/2)\partial_{z}^{2}+(1/2)\Omega^2 z^2$ is the single particle operator.
A typical example showing the initial configuration, i.e., the condensate and the stationary
dark soliton (which can be found, e.g., by a Newton-Raphson method), as well as the
corresponding spectral plane $(\omega_r, \omega_i)$, are shown in Fig.~\ref{figspectrum}.

The BdG analysis reveals that all the eigenfrequencies of the spectrum are real,
which indicates that the stationary dark soliton is dynamically stable. The four smallest
magnitude eigenfrequency pairs
\footnotemark[1]
\footnotetext[1]{Recall that due to the Hamiltonian nature of the system, the eigenfrequencies
$\pm \omega_r$ correspond to the same physical oscillation.}
and their corresponding eigenmodes have the following physical significance (see, e.g.,
Ref.~\cite{book2}). First, there exists a zero eigenfrequency, located at the origin of the
spectral plane $(\omega_{r}, \omega_{i})$, which reflects the phase invariance of the 1D GP
equation. The respective eigenfunction is the so-called Goldstone mode and does not result in
any physical excitation (oscillation) of the system. The solutions with eigenfrequencies
$\omega_r= \pm \Omega$ correspond to the so-called {\it dipole mode} (or {\it Kohn mode}),
which is relevant to the motion of the center of mass of the system; note that as the system
is harmonically confined, the center of mass oscillates with the frequency of the harmonic
trap \cite{kohn}. The solutions with eigenfrequencies having the next larger magnitude
eigenfrequencies correspond to the {\it quadrupole mode}, with the location of the
eigenfrequencies at $\omega_r= \pm \sqrt{3}\Omega$, being particular to the one-dimensionality
of the system \cite{str}. Note that the excitation of the quadrupole mode (induced, e.g., by
a time-modulation of the trap strength) results in a breathing behavior of the condensate,
with its width oscillating with the above-mentioned frequency.

Of particular interest are the solutions with eigenfrequencies
$\omega_r =\omega_{A} \equiv \Omega/\sqrt{2}$, which correspond to the so-called
{\it anomalous mode}. This mode appears in the Bogoliubov analysis only when topological
excitations of the condensate are involved, namely dark solitons or vortices \cite{fetter}.
A characteristic property of the anomalous mode is that the integral of the
$norm~\times~energy$ product, $\int (|u|^2 -|v|^2)\omega dz $ (in our units), is negative
rather than positive as is the case for all the positive frequency modes associated with
the ground state of the system \cite{book2}. Note that, from a mathematical viewpoint,
the anomalous mode possesses a topological property of the so-called
{\it negative Krein signature} \cite{MacKay}, namely
$K \equiv {\rm sign}\{\int (|u|^2-|v|^2) \omega dz \} < 0$
(for positive eigenfrequencies $\omega$). Practically, this means that the
anomalous mode becomes structurally unstable (i.e., it becomes complex) upon collision with other
eigenvalues, as is the case when dissipation is present \cite{sand}.
In our case, {\it finite temperature} automatically implies
the presence of dissipation which, as discussed in more detail
in Sec.~\ref{finiteT} below, may be described --- in the simplest possible approach ---
by including a phenomenological temperature-induced damping term in the GP model.

\begin{figure}
\centering
\includegraphics[width=14.5cm]{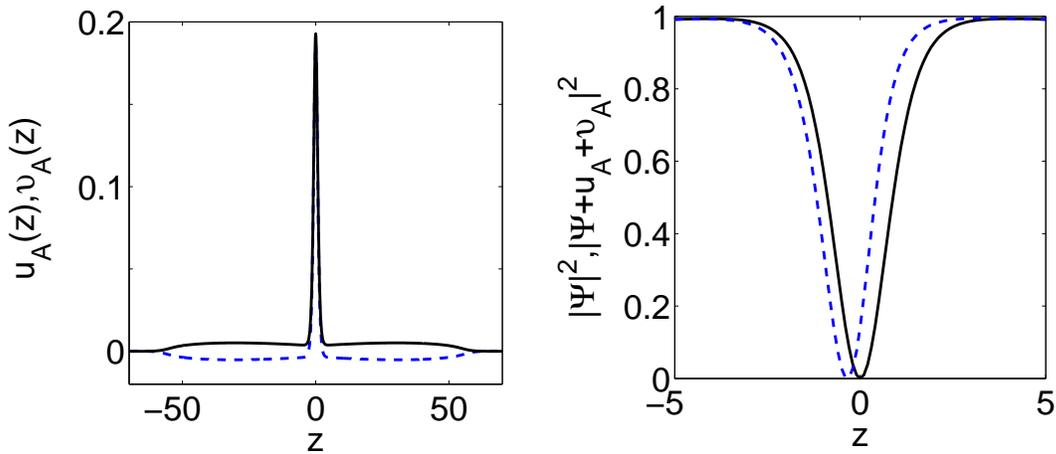}
\caption{(Color online)
Left panel: The eigenfunctions $u_A$ (solid line) and $\upsilon_A$ (dotted line) of the
anomalous mode. Right panel: The solid (black) line shows the density of the stationary
dark soliton in a region around the trap center. The dashed (blue) line shows the density
of the dark soliton when excited by the anomalous mode. The parameter values are the same
to the ones of Fig.~\ref{figspectrum}.}
\label{anomalous}
\end{figure}

In order to better clarify the above, and discuss in more detail the stability of the excitation
corresponding to the anomalous mode, namely of the dark soliton, we note the following.
At temperatures $T \rightarrow 0$ (as is the case under consideration), the negative energy of
the dark soliton does not imply any instability (e.g., a decay process) and, thus, the soliton is
dynamically stable. Nevertheless, at finite temperatures, i.e., in the presence of a thermal cloud,
the above mentioned properties of the anomalous mode indicate that the soliton will become unstable:
in this case, the presence of the temperature-dependent damping results in the decay of the soliton
(see discussion in Refs.~\cite{fms,Muryshev} as well as in Sec.~\ref{finiteT} below). From
a physical point of view, the decay mechanism resembles the one of the low-energy excitations
of trapped BECs \cite{fsw} and originates from the scattering of thermal particles on the dark soliton. Thus, according to these arguments, matter-wave dark solitons can be regarded as
thermodynamically unstable excitations as, in the presence of the temperature-induced dissipation,
the system will be driven towards configurations with lower energy; in other words, the dark solitons
will decay to the ground state. This scenario is also often referred to as {\it energetic instability}
\cite{wuniu}.

As mentioned above, the eigenfrequency of the anomalous mode is equal to the oscillation
frequency of a dark soliton around the center of the harmonic trap in the TF limit. On the
other hand, the eigenfunctions $u_A$ and $\upsilon_A$ of the anomalous mode, shown in
Fig.~\ref{anomalous}, are localized within the notch of the dark soliton \cite{Muryshev,dz},
and their sum can be approximated as $u_A+\upsilon_{A}\propto {\rm sech}^2(\sqrt{n_0}z)$.
Notice that in the case of a uniform condensate (i.e., in the absence of the trap), there
exists a translational mode with zero frequency which has the same functional form, namely
$\partial_z \psi_{0}$, due to the translational invariance of the dark soliton solution. When
the trap is present, however, this symmetry is broken, which suggests that the anomalous mode
can be regarded as the ``ghost'' of the broken translational invariance of the dark soliton
solution. We finally mention that the direct connection of the anomalous mode to the
oscillation of the dark soliton can be better explained by the fact that an excitation of
the stationary black soliton $\psi_0$ by the anomalous mode results in a displacement of the
soliton from the trap center. In other words, the GP Eq.~(\ref{dim1dgpe}) with the initial
condition $\psi(z;t=0)=\psi_{0}(z)+ u_A(z)+\upsilon_A^{\ast}(z)$, will naturally lead to dark
soliton oscillations studied in Sec.~\ref{adiabatic}.

\subsubsection{The multi-soliton state.}

\begin{figure}[tbp]
\centering
\includegraphics[width=12cm]{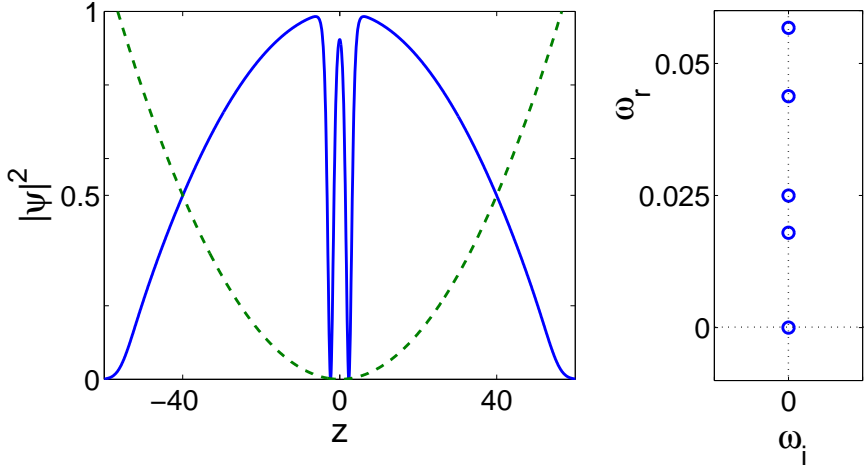}
\centering
\caption{(Color online)
Left panel: The density of a condensate carrying a stationary two-dark-soliton state
(parameter values are $\mu=1$ and $\Omega=0.025$, as in Fig.~\ref{figspectrum}). The
solitons are located at $z =\pm 2.3$. Right panel: The lowest characteristic eigenfrequencies
of the Bogoliubov excitation spectrum: The eigenfrequency located at the origin corresponds
to the Goldstone mode, the one at $\Omega = 0.025$ to the Kohn mode, and the one at
$\sqrt{3}\Omega=0.043$ to the quadrupole mode. Finally, there exist two anomalous modes
with eigenfrequencies $\omega_{1}=0.0179$ and $\omega_2= 0.0566$.}
\label{2dsgp}
\end{figure}

The BdG analysis can also be performed in the more general case of multi-soliton states
\cite{law}, which may be found in a {\it stationary} form (as explained in Sec.~\ref{multiple}).
In this case, starting from the non-interacting limit, it can be found that the Bogoliubov
spectrum of the $n$-th excited state consists of one zero eigenvalue (corresponding to the
Goldstone mode), $n$ double eigenvalues (accounted for by the presence of the harmonic trap),
and infinitely many simple ones. Then, in the nonlinear regime, one of the eigenvalues of each
double pair becomes an anomalous mode of the system (characterized by a negative valued
integral of the $norm~\times~energy$ product) and, thus, the number of anomalous modes in the
excitation spectrum equals to the number of dark solitons \cite{law}. This is in agreement
with the fact that the number of eigenvalues with negative Krein signature equals to the
number of the nodes of the stationary state \cite{Kapitula}. Notice that in the framework of
the 1D GP Eq.~(\ref{dim1dgpe}) --- i.e., in the TF-1D regime --- the first anomalous mode
coincides with the oscillation frequency $\omega_{\rm osc}=\Omega/\sqrt{2}$ of the single dark
soliton. An example of a condensate with a stationary two-dark soliton state, as well as the
pertinent spectral plane, are shown in Fig.~\ref{2dsgp}.

\begin{figure}
\centering
\includegraphics[width=9cm]{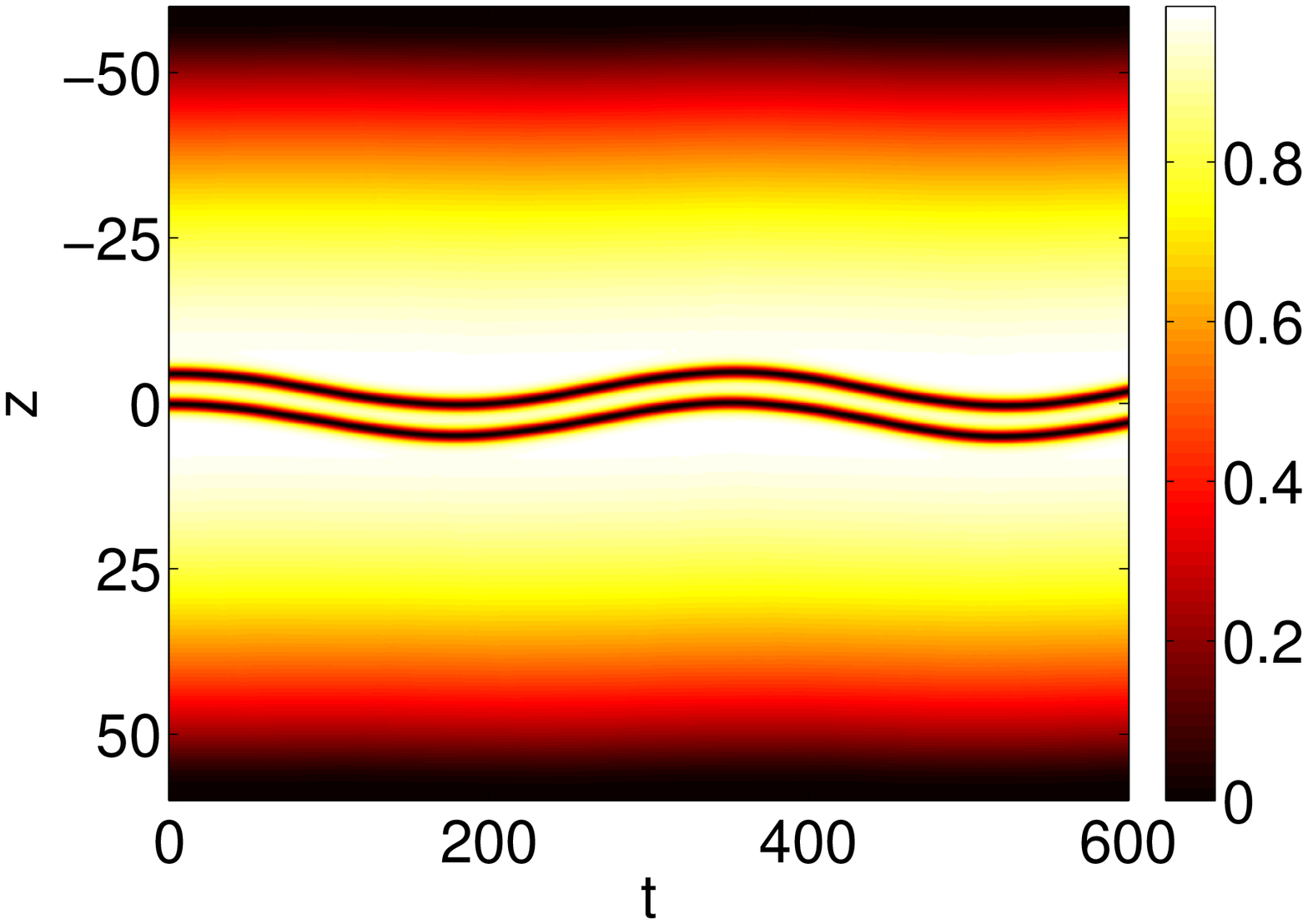}
\includegraphics[width=9cm]{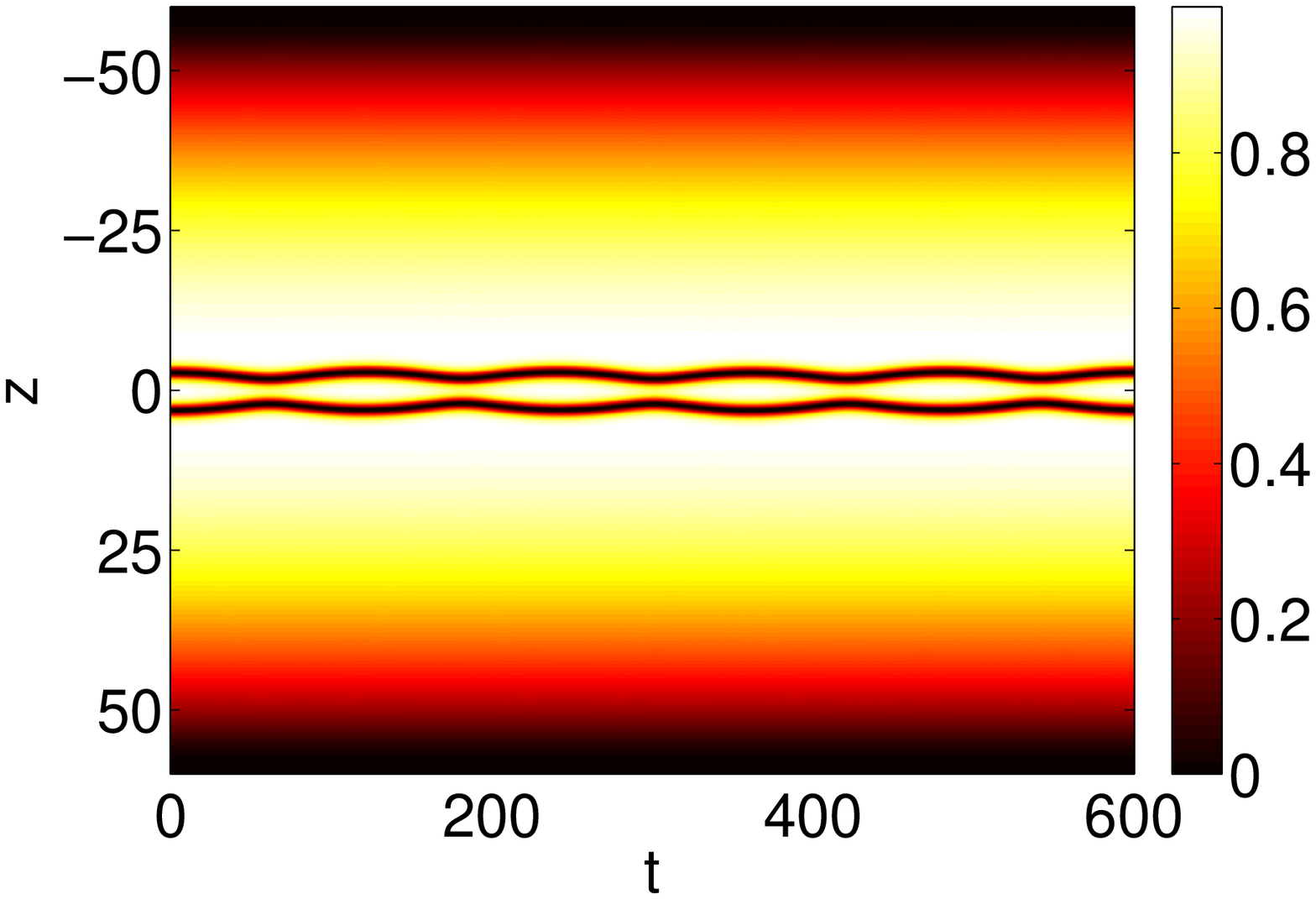}
\caption{(Color online) Spatio-temporal contour plot of the condensate density,
for parameter values $\mu=1$ and $\Omega=0.025$ (as in Fig.~\ref{2dsgp}).
Top panel: The dark solitons, initially placed at $z_1=0$ and $z_2=5$,
oscillate {\it in-phase} with a frequency
$\omega_{\rm osc}=0.018 \approx \omega_{1}=0.0179 \approx \Omega/\sqrt{2} = 0.0176$.
Bottom panel: The dark solitons, initially placed at $z=\pm 3$,
oscillate {\it out-of-phase} with a frequency
$\omega_{\rm osc}=0.057 \approx \omega_2 = 0.0566$.
Here, $\omega_{1,2}$ are the eigenfrequencies of the first and second anomalous mode,
respectively.
}
\label{inoutgp}
\end{figure}

The physical significance of the $n$-anomalous modes has been discussed in
Refs.~\cite{draft6,law}: for example, in the case of a two-dark soliton state, the
smallest of the anomalous modes corresponds to an {\it in-phase} oscillation (i.e.,
when the two dark solitons oscillate together without changing their relative spatial
separation), the largest anomalous mode corresponds to an {\it out-of-phase} oscillation
(i.e., when the two dark solitons move to opposite directions with the same velocity and
undergo head-on collision), and so on. An example of the in-phase and the out-of-phase
oscillation of the two-dark soliton state shown in the left panel of Fig.~\ref{2dsgp}
(when the solitons are properly displaced from their equilibrium positions) is illustrated
in Fig.~\ref{inoutgp}.
Here it should be noticed that since the starting point of the above considerations is the
non-interacting limit, a similar analysis can also be performed in the case of other
mean-field models with {\it non-cubic nonlinearity}, as the ones describing cigar-shaped BECs
in the dimensionality crossover regime from 3D to 1D (see Sec.~\ref{lowerd} and
Sec.~\ref{crossover} below).

Finally, as concerns the stability of nonlinear modes (see relevant investigations in
Refs.~\cite{draft6,Carr,konotop1,ZA}), all higher-order nonlinear modes are unstable near
the non-interacting limit \cite{draft6,ZA} (but can be stabilized by using anharmonic traps
\cite{ZA}); nevertheless, the instability ceases to exist sufficiently deep inside the
nonlinear regime (i.e., for sufficiently large BECs, with large $N$) \cite{draft6}.


\subsection{Radiation effects: inhomogeneity-induced sound emission by the soliton.}
\label{radiation}

As indicated in Sec.~\ref{comments}, the dark soliton experiences a background density gradient
in the presence of the perturbation $R(\psi)$ in Eq.~(\ref{gpe1d_u}) and, thus, it continuously
emits energy in the form of sound waves. Here, we will study this effect in more detail,
considering the case of $R(\psi)= V(z)\psi$ with $V(z)$ being a harmonic potential. A particular
feature of this setup, is that the emitted sound energy remains confined within the spatial
region of the trap and, hence, the soliton continuously re-interacts with the emitted sound
waves \cite{parker1}; in fact, this process is such that, on average, the dark soliton reabsorbs
the radiation it emits. Thus, in the case of harmonic traps, an investigation of the
inhomogeneity-induced sound emission, as well as an estimation of the rate of emission
of energy, is relevant for short timescales, i.e.,
$0\le t \le T_{\rm osc} \equiv 2\pi/\omega_{\rm osc}$
[where $\omega_{\rm osc}$ is given by Eq.~(\ref{soloscfreq}),
with $\Omega \ll 1$]. On the other hand, if dark solitons evolve in the presence of
{\it non-harmonic} potentials (such as localized barriers
\cite{analogies,parker2,fr1,BilPav1,radou2}, disordered potentials \cite{BilPav2,sach},
anharmonic traps \cite{recnpp}, or other ``properly designed'' potentials --- see below),
the problem  may be easier -- at least in terms of a numerical investigation: in fact, as is
explained below, it is possible to consider suitable setups that either damp off the emitted
sound density or cause the emitted sound to dephase.

Various such setups were proposed and analyzed in the past; the most prominent example is,
perhaps, a tight inner ``dimple'' trap, confining a dark soliton, located within a much weaker
outer harmonic potential (such a configuration can be realized by focusing an off-resonant laser
beam within a harmonic trap) \cite{parker1}. In this case, if the depth of the dimple trap is
sufficiently small, the sound waves can escape (to the outer trap), while the soliton can remain
confined in this region. In this limit, sound energy is damped off for short enough timescales,
until it bounces off the weaker outer trap and thus becomes forced to re-interact with the
soliton. An alternative setup considered in Refs.~\cite{parker3,gt1} consists of a harmonic trap
perturbed by an {\it optical lattice} potential (see Sec.~\ref{OLs}). In this case, the optical
lattice can confine a soliton within a single or a few lattice sites, with the sound (again
for short enough times) escaping to neighboring sites. Although in this case the sound
re-interacts with the soliton on faster timescales than in the case of the dimple trap mentioned
above, the presence of the lattice dephases the emitted sound waves, and hence accelerates the
soliton decay.

The radiation-induced dissipation of matter-wave dark solitons in harmonic traps was studied
analytically in Ref.~\cite{fr5} by means of an asymptotic multi-scale expansion method.
Particularly, assuming that $R(\psi) = \epsilon^2 z^2 \psi$ (with $\epsilon$ being a formal
small parameter defined by the aspect ratio $\Omega$), the following results were obtained.
In the limit $\epsilon \rightarrow 0$, the dark soliton evolves adiabatically so that the dark
soliton center $z_0(t) = vt+z_{\rm o} \rightarrow s(T)/\epsilon$, i.e., it becomes a function
of the slow timescale $T=\epsilon t$, while the soliton velocity is given by
$v(T)=\dot{s} \equiv ds/dT$. The adiabatic dynamics is followed by generation of sound waves,
which can be taken into regard as per Eq.~(\ref{pes}). In fact, in the decomposition of the wave
function $\psi$ into an inner and an outer asymptotic scale, the leading-order radiative effects
are taken into account when the complex phase $\theta_{\rm o}$ [see Eq.~(\ref{backgr})] depends
also on $T = \epsilon t$, i.e., $\theta_{\rm o} \equiv \theta(T)$,
and the first-order corrections to the dark soliton (\ref{ds}) grow linearly in $z$.
Neglecting reflections from the trap, the extended dynamical equation for the position
$s(T)$ of the dark soliton (\ref{ds}) takes the form:
\begin{equation}
\label{pumped-oscillator-main} \ddot{s} + s = \frac{\epsilon
\dot{s}}{2 \sqrt{(1-s^2)^3} \sqrt{1 - s^2 - \dot{s}^2}} + {\rm O}(\epsilon^2).
\end{equation}
The left-hand-side of Eq.~(\ref{pumped-oscillator-main}) represents the leading-order adiabatic
dynamics of the dark soliton [see also Eq.~(\ref{eqmd})] oscillating on the ground state of
the trap, namely a harmonic oscillator with the obvious solution $s(T) = s_0 \cos(T + \delta_0)$
(with $s_0$ and $\delta_0$ being arbitrary parameters representing the initial position and
phase of the soliton).
As long as $s^2 + \dot{s}^2 < 1$, the adiabatic dynamics approximation remains valid for
large values of the position $s(0)$ and speed $\dot{s}(0)$ of the dark soliton. In other words,
dark solitons oscillate in the trap with a {\it uniform} oscillation frequency for solitons of
{\it all} amplitudes and velocities, in agreement with the prediction of the Landau dynamics
approach (see Sec.~\ref{adiabatic}). Apparently, in the limiting case of
$s^2 + \dot{s}^2 \rightarrow 1$ (i.e., for extremely shallow dark solitons)
Eq.~(\ref{pumped-oscillator-main}) is not applicable.

Next, letting $E = \frac{1}{2} \left( \dot{s}^2 + s^2 \right)$ be the energy of the harmonic
oscillator, one may employ Eq. (\ref{pumped-oscillator-main}) to calculate the rate of change
of $E$ due to the leading-order radiative effects appearing in the right-hand side
of Eq.~(\ref{pumped-oscillator-main}). The result is:
\begin{equation}
\label{energy-increase} \dot{E} = \frac{\epsilon \dot{s}^2}{2
\sqrt{(1-s^2)^3} \sqrt{1 - s^2 - \dot{s}^2}} + {\rm O}(\epsilon^2)
> 0,
\end{equation}
and shows that due to the energy pumping (\ref{energy-increase}), the amplitude of the harmonic
oscillator increases in time. In the limit $s^2 + \dot{s}^2 \to 0$,
Eqs.~(\ref{pumped-oscillator-main}) and (\ref{energy-increase}) can be simplified. First, the
energy of the dark soliton oscillations
accelerates by the squared law $\dot{E} = \epsilon \dot{s}^{2}/2$, which was confirmed
in numerical simulations in the setup of Ref.~\cite{parker1}. Second, the nonlinear equation
(\ref{pumped-oscillator-main}) is linearized as follows:
\begin{equation}
\ddot{s} + s - \frac{\epsilon}{2} \dot{s} = {\rm O}(\epsilon^2, s^3),
\label{linearized}
\end{equation}
Equation~(\ref{linearized}) includes an {\it anti-damping} term accounting for the emission
of radiation, indicating that the center point $(0,0)$ becomes an unstable spiral
point on the plane $(s,\dot{s})$. Apparently, the leading-order solution reads
$s(T) = s_0 e^{\epsilon T/4} \cos(T + \delta_0)$; thus, the amplitude of oscillations
of a dark soliton increases while its own amplitude decreases.

The above results of the asymptotic analysis were confirmed by the numerical simulations of
Ref.~\cite{fr5}, but also by numerical findings reported in other works: the radiation-induced
effects were also observed for dark solitons oscillating between two Gaussian humps
\cite{parker2}, or for dark solitons that are parametrically driven by a pair of two
periodically-modulated Gaussian barriers, oscillating in anti-phase at a frequency close
to the soliton frequency \cite{npppd}. It is worth noticing that the mechanism proposed
in Ref.~\cite{npppd} pumps energy into the dark soliton, which may compensate the
inhomogeneity-induced emission of radiation, as well as the damping due to the presence
of the thermal cloud \cite{fms} (see Sec.~\ref{finiteT} below).


\subsection{Persistence and stability of dark solitons.}
\label{pers}

As was highlighted in this Section, there exist many alternative approaches for the study of
the statics and dynamics of matter-wave dark solitons in the quasi-1D setting. Nevertheless,
rigorous results concerning the persistence and stability of dark solitons in a generalized
GP-like model [cf.~Eq.~(\ref{gengpe})] were obtained only recently \cite{pelpan,Menza}.
Particularly, in Ref.~\cite{Menza}, the existence and stability of a black soliton of
Eq.~(\ref{dg}) were studied in the absence of the potential term, while in Ref.~\cite{pelpan}
a more general model, incorporating the potential term, was considered. More specifically,
the model used in Ref.~\cite{pelpan} was of the following form,
\begin{equation}
i \partial_t \psi = -\frac{1}{2} \partial_z^{2} \psi + f(n) \psi+ \varepsilon V(z) \psi,
\label{dgpot}
\end{equation}
where $\varepsilon$ is a formal small parameter setting the strength of the potential. The
results obtained in Ref.~\cite{pelpan} for bounded and exponentially decaying potentials
(as, e.g., ones corresponding to red-detuned laser beams --- see, e.g., the experiment of
Ref.~\cite{onofrio2}) can be summarized as follows.

Let us consider that, in the absence of the potential, Eq.~(\ref{dgpot}) admits a black
soliton solution of the form $\psi(z,t)= q(z-s)\exp[-if(n_0)t+i\theta]$ (here, $s$ is the
soliton center and $\theta$ an arbitrary constant phase), with boundary conditions
$q_0 \rightarrow \pm \sqrt{n_0}$ as $z \rightarrow \pm \infty$. Then, the above solution
persists in the presence of the perturbation induced by the potential term in Eq.~(\ref{dgpot})
provided that the function
\begin{equation}
M'(s)= \int_{-\infty}^{+\infty} V'(z)[n_0 - q^2(z-s)]dz,
\label{Mprime}
\end{equation}
possesses at least one single root, say $s_0$. Then, the stability of the dark soliton solution
depends on the sign of the first derivative of the function in Eq.~(\ref{Mprime}), evaluated at
$s_0$: an instability occurs, with one imaginary eigenfrequency pair for
$\varepsilon M''(s_0)<0$, and with exactly one complex eigenfrequency quartet for
$\varepsilon M''(s_0)>0$. In fact, this instability is dictated by the translational eigenvalue,
which bifurcates from the origin as soon as the perturbation is present.
For $\varepsilon M''(s_0)<0$, the relevant eigenfrequency pair moves along the
imaginary axis, leading to an instability associated with exponential growth of a
perturbation along the relevant eigendirection. On the other hand, for
$\varepsilon M''(s_0)>0$, although the eigenfrequency should move along
the real axis, it can not do so because the latter is filled with continuous spectrum;
thus, since the translation mode and the eigenmodes of the continuous spectrum have opposite
Krein signature, the collision of the eigenfrequency of the translational mode with the
continuous spectrum results in a complex eigenvalue quartet, signalling the presence of
an oscillatory instability. The relevant eigenfrequencies can be determined by a
quadratic characteristic equation which, in the case of the cubic GP model (\ref{dgpot})
with $f(n)=n$ and $n_0=1$, takes the form \cite{pelpan},
\begin{eqnarray}
\lambda^{2}+\frac{\varepsilon}{4}M''(s_{0})\left(1-\frac{\lambda}{2}\right)=O(\varepsilon^{2}),
\label{chareqpelpan}
\end{eqnarray}
and the eigenvalues $\lambda$ are related to the eigenfrequencies $\omega$ through
$\lambda^2=-\omega^2$. For sufficiently small $\varepsilon>0$, this equation has only
one real root $\lambda(\varepsilon)>0$ for $M''(s_0)<0$ and two complex-conjugate roots,
with ${\rm Re}\{\lambda(\varepsilon)\}>0$ for $M''(s_0)>0$.

It is interesting to observe that if the characteristic equation (\ref{chareqpelpan}) is formally
applied to the cubic GP model (\ref{dgpot}) with $f(n)=n$, $n_0=1$ and $V(z)=z^2$, one obtains
$M''(s_0)=2\int_{-\infty}^{+\infty}{\rm sech}^2(z)dz=4$ and, thus, Eq.~(\ref{chareqpelpan})
takes the form $\lambda^{2} - (\varepsilon/2)\lambda +\varepsilon = O(\varepsilon^{2})$. Using
appropriate rescalings, it can easily be shown that the latter characteristic equation can be
derived from Eq.~(\ref{linearized}) of Sec.~\ref{radiation}. Although the validity of the
radiative boundary conditions for $V(z)=z^2$ cannot be verified by the analysis of
Ref.~\cite{pelpan}, the above observation leads to the following conjecture \cite{pelpan}:
in the most general GP model [cf.~Eq.(\ref{dgpot})], the two complex-conjugate eigenvalues
with positive real part for $M''(s_0)>0$ result from the following Newton's particle equation
of motion for the soliton center $s(t)$:
\begin{equation}
\mu_0 \ddot{s} -\varepsilon \lambda_0 M''(s) \dot{s} = -\varepsilon M'(s),
\label{eqmotpelpan}
\end{equation}
where $M(s)$ is the effective potential implied by Eq.~(\ref{Mprime}), while the constants
$\mu_0$ and $\lambda_0$ represent, respectively, the soliton's mass and anti-damping --- as per
the discussion of Sec.~\ref{radiation}.

The validity of Eq.~(\ref{eqmotpelpan}), as well as the other theoretical predictions presented
in this Section, were tested against numerical simulations in Ref.~\cite{pelpan} for small
decaying potentials, and the agreement between the analytical and numerical results was found
to be very good. Notice that although the above results of Ref.~\cite{pelpan} can only be
rigorously applied to the case of small, bounded and exponentially decaying potentials, the
basic qualitative features may formally persist for other types of external potentials as well.
A pertinent example is the work of Ref.~\cite{constgofx}, where the the persistence and stability
of matter-wave black solitons were studied in a condensate characterized by a periodic,
piecewise-constant scattering length
\footnotemark[1]
\footnotetext[1]{BECs with spatially-varying coupling constant $g$,
so-called {\it collisionally inhomogeneous condensates} \cite{gofx}, have attracted
much attention, as they provide a variety of interesting phenomena
\cite{gofxo1,gofxo2,gofxo3,gofxo4,gofxo5,gofxo6,gofxo7,gofxo8,gofxo9}.}:
as shown in Ref.~\cite{constgofx}, a formal application of the predictions of
Ref.~\cite{pelpan} concerning the persistence and stability of dark solitons in this setting,
was found to be in very good agreement with relevant numerical findings. Nevertheless, an
analysis similar to the one presented in Ref.~\cite{pelpan}, but for other types of potentials
(such as confining and periodic ones) is still missing.


\section{Matter-wave dark solitons in higher-dimensional settings}
\label{sec5}

\begin{figure}[tbp]
\centering
\hskip+1cm
\includegraphics[width=12.0cm]{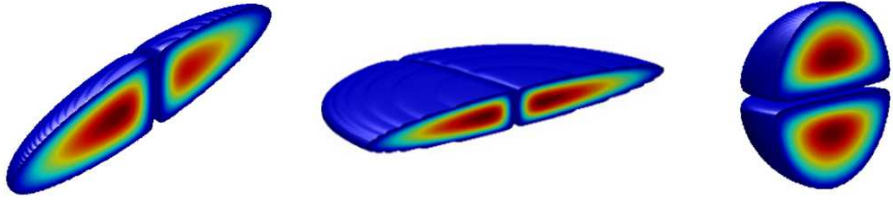}
\caption{(color online) Examples of the 3D densities of condensates, confined in various
types of anisotropic harmonic traps and carrying quasi-1D dark solitons. Shown are
(from left to right) the longitudinal cuts of a cigar-shaped BEC, a disk-shaped BEC,
and a spherical BEC.
}
\label{fignice}
\end{figure}

Quasi-1D matter-wave dark solitons may naturally exist in higher-dimensional settings.
For example, in the experimentally relevant case of a {\it cigar-shaped} trap, the actual
dimensionality of the BEC density is 3D rather than 1D, despite of the fact that the BEC can
be treated as an effectively 1D object using the NLS Eq.~(\ref{gengpe}) with the generalized
nonlinearities of Eqs.~(\ref{salnl})--(\ref{cqn}) --- see Sec.~\ref{lowerd}. The density of
such a cigar-shaped BEC with a dark soliton on top of it is illustrated in the left panel of
Fig.~\ref{fignice}. Furthermore, quasi-1D dark solitons may also exist in {\it disk-shaped}
BECs (see Sec.~\ref{lowerd}), which are described by the $(2+1)$-dimensional GP
Eq.~(\ref{2dgpe}). The latter, can be expressed in the following dimensionless form,
\begin{equation}
i \partial_t \psi  = \left[ - \frac{1}{2} \nabla^2 + V(r)
+ |\psi|^{2} \right] \psi,
\label{2dgpedime}
\end{equation}
where $\nabla^{2} = \partial_x^2 +\partial_y^2 $, the density $|\psi|^2$, length, time and
energy are respectively measured in units of $2\sqrt{2\pi}a a_z$, $a_z$, $\omega_z^{-1}$ and
$\hbar\omega_z$, while the potential $V(r)$ is given by
\begin{equation}
V(r)=\frac{1}{2} \Omega^2 r^2,
\label{htrap2dd}
\end{equation}
with the aspect ratio being $\Omega = \omega_{\perp}/\omega_z \ll 1$. In this case, the soliton
of Eq.~(\ref{ds}), with the variable $z$ being replaced by $x$, is an exact analytical solution
of Eq. (\ref{2dgpedime}) for $V(z) =0$. This ``rectilinear'' soliton has the form of a
dark ``stripe'' on top of a 2D TF cloud, and the BEC wave function can be expressed
(similarly to the 1D case) as $\psi=\psi_{\mathrm{TF}}(r) \exp(-i \mu t) \psi_{\rm ds}(x,t)$.
It is also natural to consider the full $(3+1)$-dimensional version of Eq.~(\ref{2dgpedime}),
with $\nabla^{2} = \partial_x^2 +\partial_y^2 + \partial_z^2$, where quasi-1D dark soliton
stripes exist as well. In this case, the potential and the 3D TF cloud are modified according
to the relative values of the confining frequencies in the three directions. Examples of the
densities of a disk-shaped BEC and a spherical BEC carrying a rectilinear dark soliton are shown,
respectively, in the middle and right panels of Fig.~\ref{fignice}.

Apart from the quasi-1D dark solitons, purely 2D dark solitons have also been predicted to
occur in theory (but they have not been observed so far in experiments). Such dark soliton
solutions of the GP Eq.~(\ref{2dgpedime}), which have been derived in the framework of the
small-amplitude approximation (see Sec.~\ref{kdvstuff}), may have the form of {\it lumps}
satisfying an effective Kadomtsev-Petviashvili (KP) equation \cite{gxh1} or {\it dromions}
satisfying an effective Davey-Stewartson system \cite{gxh2}; quasi-1D and 2D dark solitons
of the dromion type, have also been predicted to occur in disk-shaped multi-component
condensates \cite{aguero}.

\subsection{Snaking instability of rectilinear dark solitons.}
\label{snaking}

\subsubsection{Basic phenomenology and results.}

An important issue arising in higher-dimensional settings is the stability of dark solitons
which, for simplicity, will be studied at first in the $(2+1)$-dimensional geometry (relevant
to disk-shaped BECs) and in the absence of the potential $V(r)$. The stability of the 1D dark
soliton stripe (lying, say, along the $x$-direction) in such a 2D setting was first studied in
Ref.~\cite{kuz} (see also Refs.~\cite{kuz2,kuz3}). In this work, it was shown that the soliton
is prone to {\it transverse modulational instability}, i.e., it is unstable against
long-wavelength transverse periodic perturbations $\sim \cos(Qy)$, where $Q$ is the
wave number of the perturbation. In particular, the instability band is defined by
$Q < Q_{\rm cr}$, where the critical value of the perturbation wave number is given by
(for $\mu=1$):
%
%
\begin{equation}
Q_{\rm cr}^2 \equiv \cos^2 \phi -2 +2 \sqrt{ \cos^4 \phi + \sin^2 \phi }.
\label{Qcr}
\end{equation}
The above expression is a result of a linear stability analysis, which indicates that the
amplitude of the rectilinear dark soliton will grow exponentially in the transverse direction.
Nonlinear regimes of this instability were also studied analytically by means of asymptotic
expansion techniques (see, e.g., Ref.~\cite{pelkiv} and references therein). This instability
was extensively studied in the context of nonlinear optics, both theoretically \cite{kivpr}
and experimentally \cite{tikho,mamaev}, and was found to be responsible for a possible
decay of a plane dark soliton into a chain of vortices of opposite topological charges
(vortex--anti-vortex pairs). Particularly, when the transverse modulational instability
sets in, a plane {\it black} soliton undergoes a transverse ``snake'' deformation (hence the
name ``snaking instability'') \cite{kivpr,pelkiv}, causing the nodal plane to decay into
vortex pairs. On the other hand, unstable {\it gray} solitons may not decay into vortices,
but rather perform long lived oscillations accompanied by emission of radiation in the form
of sound waves.

\begin{figure}[tbp]
\centering
\hskip+1cm
\includegraphics[width=12.0cm]{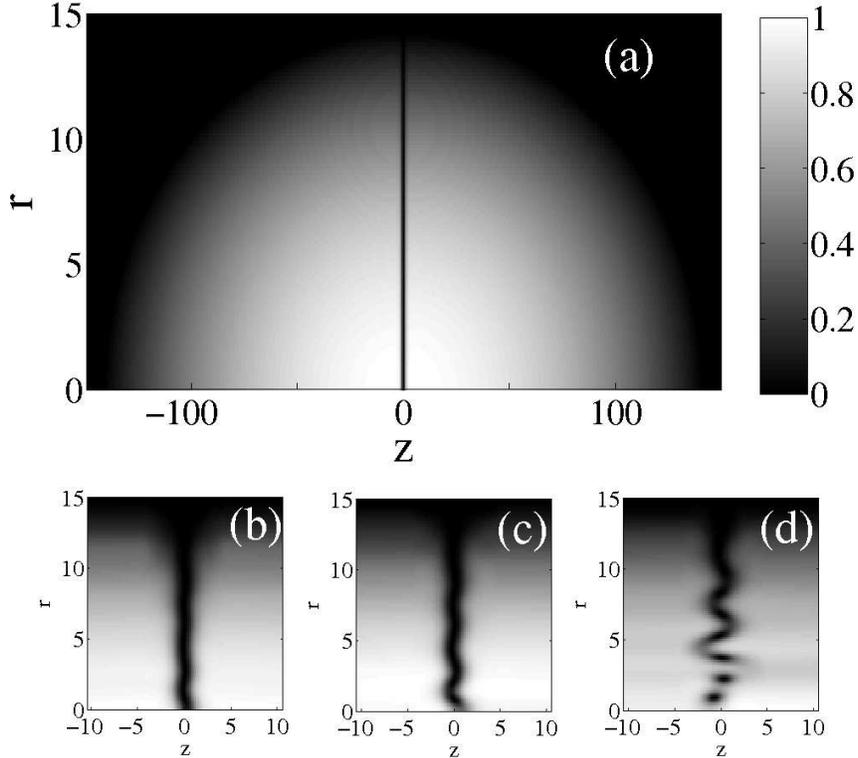}
\caption{Contour plots showing the evolution of the density of a cigar-shaped BEC confined
in the trap $V(z,r)=(1/2)(\Omega_r^2 r^2+\Omega_z^2 z^2)$ (parameter values are
$\Omega_r=10\Omega_z=0.1$ and $\mu=1$) and carrying a rectilinear dark soliton.
Panel (a) shows the initial condition,
$\psi(r,z,0)=\sqrt{1-V(r,z)}\tanh(z)$ and panels
(b), (c) and (d) are close-up snapshots --- at $t=97$, $t=101$ and $t=110$,
respectively --- showing the onset of the snaking instability and the decay of the soliton into vortex rings.
}
\label{snaking2}
\end{figure}

The basic phenomenology and results described above, persist in the case of other
higher-dimensional setups as well. For example, in Fig.~\ref{snaking2}, we show the onset of
the snaking instability of a rectilinear dark soliton on top of a cigar-shaped BEC
(see also Sec.~\ref{stabilitycigar} below), confined in a trap of the form
$V(z,r)=(1/2)(\Omega_r^2 r^2+\Omega_z^2 z^2)$ (with $r^2=x^2+y^2$). In such
higher-dimensional settings, the unstable soliton collapses into more stable
vortex structures, namely vortex rings. From the viewpoint of experimental observations,
the snaking instability and the decay of matter-wave dark solitons into vortex rings
was first observed in a JILA experiment \cite{bpa} with a two-component $^{87}$Rb BEC
(see Sec.~\ref{multi}). In particular, a quasi-1D dark soliton created in one component
(see Sec.~\ref{qse}) evolved in a quasi-spherical trap and, thus, the onset of the
snaking instability caused the soliton to decay into vortex rings --- as predicted in
theory \cite{mprizolas}.

\subsubsection{Avoiding the snaking instability.}

As is known from the context of nonlinear optics \cite{kivpr}, the snaking instability can
be avoided by using finite-sized background optical beams (see, e.g., relevant experimental
results in Ref.~\cite{spatial1}). Thus, one should expect that the suppression of the snaking
instability may also be possible in the case of a trapped (disk-shaped) condensate, which also
constitutes a background of a finite extent. Indeed, in such a case, a simple criterion for
the suppression of the snaking instability can be found by means of scale competition
arguments \cite{lwi}. In particular, if the characteristic length scale of the condensate
$L_{\rm BEC} \equiv 2\sqrt{2}/\Omega$ (i.e., the TF diameter for $\mu=1$) is below the
critical length $L_c \equiv 2\pi/Q_{\rm cr}$ stemming from Eq.~(\ref{Qcr}), then the snaking
instability will not manifest itself. Considering the case of a black (stationary) soliton
with $\sin\phi=0$, Eq.~(\ref{Qcr}) yields $Q_{\rm cr} =1$ and, thus, $L_c =2\pi$; in such
a case, the above scale competition argument, $L_{\rm BEC} <L_c$, leads to the prediction
that a use of a sufficiently strong trap, such that
$\Omega > \Omega_c \equiv \sqrt{2}/\pi \simeq 0.45$,
can suppress the snaking instability. The above criterion was tested against direct
numerical simulations \cite{lwi}, and it was found that the critical value of
the trap strength is less than the theoretically predicted, namely $\Omega_c \approx 0.31$.
This discrepancy can be understood by the fact that for small BECs (i.e., for tight traps)
the presence of the dark soliton significantly modifies the maximum density which is less
than $\mu$ by a ``rescaling'' factor $f$, found to be $f \approx 0.5$.

On the other hand, it was recently predicted \cite{ddids} that stable 3D stationary dark
solitons may exist in {\it dipolar condensates} (for this type of BECs see, e.g., the recent
review \cite{pfau} and references therein). In particular, the special feature of dipolar
condensates, namely the dipole-dipole interaction, together with the use of a sufficiently
deep optical lattice in the soliton's nodal plane, allows for the existence of dark solitons
of arbitrarily large transversal sizes, which are not prone to the snaking instability. In
this case, the underlying reason for the suppression of the snaking instability is that
the dipole-dipole interaction is {\it long-range} (it decays like $r^{-3}$, where $r$
is the inter-particle distance), which means that the respective GP equation incorporates
a {\it nonlocal nonlinear} term. Generally, such a nonlocal response may arrest collapse
and stabilize solitons in higher-dimensions, as was shown in the context of optics
(see, e.g., Ref.~\cite{nonlocalNLS}, as well as Ref.~\cite{strillo} for a relevant
recent work on dark solitons).

We also note that a more ``exotic'' dark soliton configuration in the 2D setting, which is
not subject to the snaking instability, was presented in Ref.~\cite{crosses} (see also
Ref.~\cite{crosses2}). This configuration, which refers to a two-component BEC (see
Sec.~\ref{multi}), consists of a ``cross'' formed by the intersection of two rectilinear
domain walls, with the wave functions of the same species filling each pair of opposite
quadrants having a $\pi$ phase difference. This way, a  dark soliton configuration is formed,
which was found to be stable for long times --- and even in the presence of rotation of
the trap --- in a large parametric region.

\subsection{Matter-wave dark solitons of radial symmetry}
\label{RDS_SSS}

\subsubsection{Ring dark solitons (RDS) and spherical shell solitons (SSS).}

\begin{figure}[tbp]
\centering
\hskip+1cm
\includegraphics[width=12.0cm]{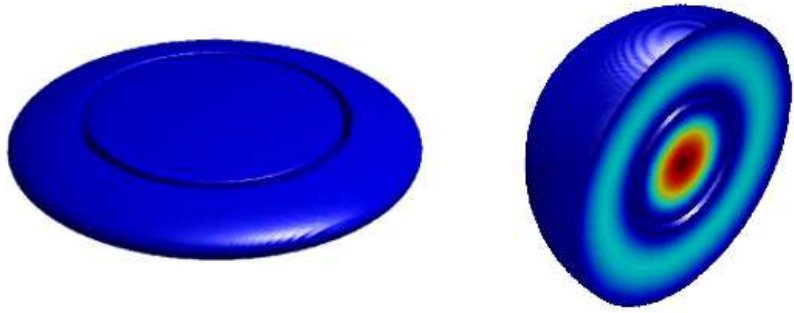}
\caption{(color online) Examples of the 3D densities of condensates, confined in
a disk-shaped (left panel) and a spherical (right panel) traps, and carrying a ring dark
soliton and a spherical shell soliton, respectively.
}
\label{fignice2}
\end{figure}

A special class of dark solitons in higher-dimensional settings consists of dark solitons
exhibiting a {\it radial symmetry}, that can be realized by wrapping the nodal plane around on
itself. Such structures were first introduced in the context of nonlinear optics \cite{kyang},
with the motivation being that these dark solitons may not be prone to the snaking
instability: indeed, if a dark stripe is bent so as to form a dark ring of length
$L < 2\pi/Q_{\rm cr}$ (in the $(2+1)$-dimensional geometry), then the snaking instability
will be suppressed. Such {\it ring dark solitons} (RDSs) were studied in nonlinear optics
both theoretically \cite{framalpla,nistaz01} and experimentally \cite{dreis96,ords,dreis02},
while later were also predicted to occur in BECs \cite{rds}. In this context, and in the
2D setting (i.e., in a disk-shaped BEC), the RDS has the form of an annular ``trough''.
On the other hand, in a 3D setting (i.e., in a spherical BEC), the radially symmetric dark
soliton is called {\it spherical dark soliton} \cite{fr4}, or {\it spherical shell soliton}
(SSS) ---  according to the nomenclature of Ref.~\cite{carr2006d} --- and has the form of
a nodal spherical ``shell''. Examples of the 3D densities of a disk-shaped and a spherical
BEC, carrying a RDS and a SSS, are shown in Fig.~\ref{fignice2}.

As was originally proposed in Ref.~\cite{rds}, RDS may be generated in BECs by means of
phase-engineering techniques (i.e., by a proper phase-imprinting method  --- see Sec.~\ref{pi}
below), similar to the ones used for the generation of optical RDS \cite{dreis96,ords,dreis02}.
Another technique that has been proposed for the generation of RDS in BEC is the matter-wave
interference method (see Sec.~\ref{interf}): if the condensate is initially trapped in a narrow
cylindrical box-like potential, and then is allowed to coherently expand in the presence of a
wider cylindrical impenetrable hard-wall potential, it is reflected from the boundary, and the
self-interference pattern has the form of a sequence of non-stationary concentric RDS
\cite{genrds1,genrds2,genrds3} (see also relevant work in Ref.~\cite{nate}).

\subsubsection{Dynamics and stability of RDS and SSS.}

From a mathematical standpoint, matter-wave dark solitons of radial symmetry can be considered
as quasi-1D objects and, accordingly, be analyzed by means of a quasi-1D GP equation.
In particular, either RDS or SSS can be described by Eq.~(\ref{2dgpedime}), with the Laplacian
being in the form,
\begin{equation}
\nabla^{2}= \partial_r^2+\frac{(D-1)}{r} \partial_r,
\label{laplacian}
\end{equation}
with $D=1,2,3$. In this setup, the simplest case of $D=1$ reduces Eq.~(\ref{2dgpedime}) to
the 1D GP Eq.~(\ref{dim1dgpe}) describing a quasi-1D BEC (here, $r\equiv z$). The
higher-dimensional setups correspond to the cases of $D=2$ and $D=3$: in the former case,
Eq.~(\ref{2dgpedime}) describes a disk-shaped BEC in the $(x,y)$ plane (with $r$ given
by $r=\sqrt{x^{2}+y^{2}}$), while in the latter case Eq.~(\ref{2dgpedime}) describes a
spherical BEC (with $r$ given by $r=\sqrt{x^{2}+y^{2}+z^{2}}$).

In such a quasi-1D setup, the Hamiltonian or the Lagrangian perturbation theory for dark
solitons (see Sec.~\ref{adiabatic}) may also be applied for the study of the dynamics of
RDS and SSS. In particular, Eq.~(\ref{2dgpedime}), with the Laplacian of Eq.~(\ref{laplacian}),
can be treated as a perturbed 1D NLS equation [similar to Eq.~(\ref{gpe1d_u})] provided that
both the potential term and the term $\sim r^{-1}$ can be considered as small perturbations.
The latter assumption is physically relevant for radially symmetric solitons of large
radius $r_{0}$. Then, approximating the functional form of the RDS or SSS as
[cf. Eq.~(\ref{ansatzHam})],
\begin{equation}
\psi_{s}(r,t)=\cos \phi(t) \tanh \zeta +i \sin \phi(t), \,\,\,\,
\zeta =\cos\phi(t) \left[r-r_{0}(t) \right],
\label{ansatzrds}
\end{equation}
where $r_0(t)$ is the soliton radius, it can be found \cite{rds} (see also Ref.~\cite{fr4})
that $r_0$ is governed by the following Newtonian equation of motion:
\begin{equation}
\frac{d^{2}r_{0}}{dt^{2}}=- \frac{1}{2} \frac{\partial V_{\rm eff}}{\partial r_{0}}.
\label{sem}
\end{equation}
Here, the effective potential is given by
\begin{equation}
V_{\rm eff}(r_{0})=V(r_0)- \ln r_{0}^{2(D-1)/3},
\label{effpotrds}
\end{equation}
and $V(r_0) =(1/2)\Omega^{2}r_{0}^{2}$ is the trapping potential evaluated at the soliton
radius $r_0$. In the $1$D limit of $D=1$, the last term in the right-hand side of
Eq.~(\ref{effpotrds}) vanishes and Eq.~(\ref{sem}) is reduced to Eq.~(\ref{eqmd}).
In the higher-dimensional cases of $D=2$ or $D=3$, the equation of the soliton motion
(\ref{sem}) is clearly nonlinear (even for nearly black RDS or SSS) due to the presence
of the repulsive curvature-induced logarithmic potential.

\begin{figure}[tbp]
\centering
\includegraphics[width=12.0cm]{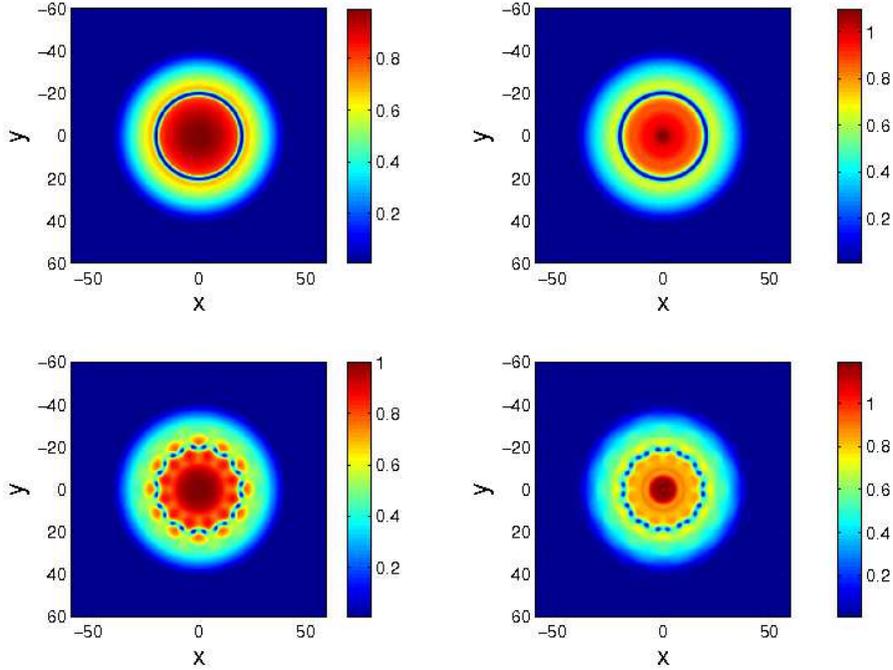}
\caption{(color online) Contour plots of the density of a disk-shaped condensate carrying a
stationary RDS, which develops the snaking instability. The initial condition used for the
integration of the GP Eq.~(\ref{2dgpedime}) is
$\psi(r,0)=\sqrt{\mu - (1/2)\Omega^2 r^2} \tanh(r-r_0(0))$,
with $\mu=1$, $\Omega=0.035$ and initial soliton radius
$r_0(0)=(\sqrt{2}\Omega)^{-1}=20.2$ (for a discussion concerning the value
of $r_0(0)$ for stationary RDS see Ref.~\cite{fr4}). The top left (top right) panels show,
respectively, the initial condition ($t=0$) and a snapshot at $t=40$, while the bottom left
(bottom right) right panels show, respectively, the onset of the snaking instability
($t = 80$) and the formation of the vortex necklace ($t=100$).
}
\label{grds}
\end{figure}

Equation~(\ref{sem}) predicts the existence of both oscillating (grey) and stationary (black)
RDS or SSS: the former, perform oscillations on top of the TF cloud, changing their radii between
a minimum and a maximum value, while the latter correspond to the minimum of the effective
potential Eq.~(\ref{effpotrds}) (such stationary states do not exist in the case of a uniform
ground state --- as in the context of nonlinear optics \cite{kyang}). As was shown in
Ref.~\cite{rds} (see also Ref.~\cite{fr4}), such a particle-like approach can describe
quite effectively the above generic scenarios and the RDS dynamics up to a certain time
(before the development of instabilities --- see below). Furthermore, in
Refs.~\cite{xue1,xue2} it was shown that the dynamics of small-amplitude RDS,
as well as the collisions between them, can be described in the framework of an
effective {\it cylindrical KdV equation} \cite{ir} (see also
Refs.~\cite{kyang,framalpla,nistaz01} for similar findings in optics).
On the other hand, numerical simulations of Ref.~\cite{rds} revealed that RDS are generally
unstable, as they either decay to radiation (the small-amplitude ones) or are subject to the
snaking instability (the moderate- and large-amplitude ones). Interestingly, as shown in
Fig.~\ref{grds}, the snaking instability of the RDS results in the formation of
vortex--anti-vortex pairs in multiples of four, which are initially set along a ring,
forming a so-called {\it vortex necklace}. Eventually, this pattern relaxes to a set of four
pairs located on a ring, which  oscillates in the radial direction between the same limits
which confined the oscillations of the original RDS; simultaneously, the pairs perform
oscillatory motion along the ring \cite{rds}.

Matter-wave dark solitons of radial symmetry were also analyzed by means of other approaches.
For example, in Ref.~\cite{carr2006d} (see also Ch.~7 in Ref.~\cite{BECBOOK}) RDS and SSS were
considered as {\it nonlinear Bessel functions}, namely solutions of the equation,
\begin{equation}
q'' + \frac{1}{r}q' - \frac{S^2}{r^2}q +2\mu q -2 q^3 = 0,
\label{bessel}
\end{equation}
resulting from Eq.~(\ref{2dgpedime}) (with $V(r)=0$) when the ansatz
$\psi = q(r)\exp(-i\mu t +i S \phi)$ is introduced (in the latter expression, $S$ is the
topological charge of a central vortex). In this setting (i.e., in the absence of the trap),
the solutions of Eq.~(\ref{bessel}) include, apart from the ground state, singly- and
multiply-charged vortices, as well as infinitely many RDS; the nodes of the latter
correspond to the nodes of the nonlinear Bessel function governed by Eq.~(\ref{bessel}).
On the other hand, if an external harmonic potential is incorporated in Eq.~(\ref{bessel}),
then it is possible to find infinite branches of nonlinear bound states, with each branch
stemming from the respective mode of the underlying linear problem (the radially-symmetric
2D quantum harmonic oscillator) \cite{lcarr}.

A stability analysis performed in Refs.~\cite{carr2006d,lcarr} also revealed that the
radially-symmetric dark solitons are typically unstable but, in agreement to the findings
of Ref.~\cite{rds}, their life-times may be considerably long. Additionally, as shown in
a recent work \cite{stabrds}, the life-time of RDS may be extended employing the so-called
\cite{frm1} {\it Feshbach Resonance Management (FRM)} technique, which is based on the use
of external fields to periodically modulate in time the $s$-wave scattering length
\cite{frm2,frm3,frm4,frm5,frm6}. In any case, the theoretical investigations indicate
that RDS and SSS have a good chance to be experimentally observed. In fact, structures
similar to stationary SSS have already been observed as transients in the Harvard experiment
of Ref.~\cite{ginsberg2005} (see also Ref.~\cite{komineas:110401} were SSS are predicted to
occur as a result of collisions of vortex rings).


\subsection{Stability of dark solitons in cigar-shaped condensates.}
\label{stabilitycigar}


The transverse (in)stability of dark solitons confined in a purely 3D setting, namely in
a cylindrical trap of the form $V(z,r)=(1/2)m[\omega_z^2 z^2 + \omega_{\perp}^2 (x^2+y^2)]$,
was first studied in Ref.~\cite{Muryshev}. In that work, it was shown that dynamical stability
of a {\it black soliton (stationary kink)}, say $\psi_0$, with a nodal plane perpendicular to
the axis of the cylindrical trap (see left panel of Fig.~\ref{fignice}), requires a strong
radial confinement [as in the $(2+1)$-dimensional case]. Particularly, it was shown that the
instability can be suppressed if the transverse (radial) condensate component is {\it not}
in the TF regime, which is guaranteed as long as $\hbar \omega_{\perp} > \mu$ (where $\mu$ is
the 3D chemical potential). This criterion can physically be understood as follows:
if the condensate is confined in a highly anisotropic (cigar-shaped) trap, then the
energy of the lowest possible radial excitation, $\hbar \omega_{\perp}$, must exceed the
kink-related kinetic energy, $K_0 = -(1/2)\psi_0 \partial_z^2 \psi_0 \sim \mu$, so that
the latter can not be transferred to the BEC's unstable transverse modes by the inter-atomic
interaction. Moreover, a systematic study in Ref.~\cite{Muryshev} revealed a criterion of
dynamical stability for the black soliton in terms of the ratio $\omega_{\perp}/\omega_z$,
namely,
\begin{equation}
\gamma \equiv \frac{\mu}{\hbar \omega_{\perp}} < \gamma_c.
\label{gama}
\end{equation}
The critical value $\gamma_c$ was calculated for various values of $\omega_{\perp}/\omega_z$
and it was found that a characteristic value of $\gamma_c$, pertinent to the limiting case of
$\omega_{\perp} \gg \omega_z$, is $\gamma_c \approx 2.4$.

In a more quantitative picture, a detailed study of the BdG equations in Ref.~\cite{Muryshev}
(see also relevant work in Refs.~\cite{mprizolas,brand}) revealed the emergence of complex
eigenvalues in the excitation spectrum and their connection to oscillatory dynamical
instabilities, including the snaking instability. Additionally, in Ref.~\cite{kominpapa}
it was found that the emergence of complex eigenvalues is directly connected to bifurcations
of the rectilinear black solitons to other stationary states that may exist in BECs confined
in cylindrical traps. In particular, an investigation of the dependence of the excitation
energy on the dimensionality parameter $d$ [cf. Eq.~(\ref{dimparam})] led to the following
results: for sufficiently low excitation energies, the black soliton may bifurcate to a
{\it solitonic vortex} or to an axisymmetric vortex ring (see also Ch.~7 in \cite{BECBOOK}
and references therein), with the corresponding bifurcation points occurring at a low and a
higher value of $d$. It was also found that the emergence of the first (second) complex
eigenvalue in the Bogoliubov spectrum coincides with the above mentioned bifurcation points.
Therefore, according to the above results, it can be concluded that the emergence of the
complex eigenvalues in the excitation spectrum (a) denote the onset of dynamical instabilities
of black solitons and (b) indicate the excitation of lower energy topological states. Since
these states are energetically preferable, the onset of the dynamical instability will result
in decay of the ``high-energy'' black soliton to these ``low-energy'' states carrying vorticity.

On the other hand, the stability of {\it moving (gray)} solitons was analyzed in
Ref.~\cite{cqnls}. According to this work, and following the arguments of Ref.~\cite{Muryshev},
a criterion for {\it not} being in the radial TF regime (which is required for dynamical
stability of the gray soliton) is $w \lesssim R \sim \xi$, where $R$ is the radial size of the
BEC and $w$ is the soliton width ($\xi$ is the healing length). Recalling that the soliton
width $w$ and velocity $v$ depend on the soliton phase angle as $w \sim 1/\cos\phi$
and $v \sim \sin\phi$, it is clear that as the soliton is moving towards the boundaries
of the BEC, its width (velocity) is increased (decreased). Thus, the instability border
$R \sim w$ for the gray soliton is reached for larger values of the parameter $\gamma$
[see Eq.~(\ref{gama})] than the ones pertaining to the black soliton. In the case of strongly
anisotropic traps, the critical value of the chemical potential required for the dynamical
stability of the gray soliton is proportional to the soliton velocity. In other words, the
stability domain of gray solitons is wider than the one of black solitons. In fact, the
shallower the soliton it is, it becomes more stable, similarly to the case of homogeneous
systems: see Eq.~(\ref{Qcr}), which indicates that the instability band vanishes for shallow
solitons with $\cos\phi \rightarrow 0$.

Numerical simulations of Ref.~\cite{cqnls} have also revealed that for $\gamma >10$, a
phase-imprinted dark soliton --- with a $\pi$-phase jump --- always decays in a cigar-shaped
BEC (on a time scale of order of $\omega_{\perp}^{-1}$), while for $\gamma \lesssim 5$ it
transforms into a dark soliton characterized by a flat notch region and $r$-independent
velocity. Here, it is relevant to mention that in the recent Technion experiment \cite{technion},
an interesting nonlinear excitation that evolves periodically between a dark soliton and
a vortex-ring was observed in a $^{87}$Rb BEC, for $\gamma =4.95$ (see also relevant
theoretical work in Refs.~\cite{s_th_2,komineas02}).

\subsection{Matter-wave dark solitons in the dimensionality crossover from 3D to 1D}
\label{crossover}

\subsubsection{The single-soliton state.}

As discussed in Sec.~\ref{lowerd}, if the dimensionality parameter [cf. Eq.~(\ref{dimparam})]
takes values $d \approx 1$, then a cigar-shaped condensate is in the so-called dimensionality
crossover regime from 3D to 1D. However, in such an experimentally relevant regime
\footnotemark[1]
\footnotetext[1]{Note that the recent Heidelberg experiments \cite{kip,draft6} were conducted in
this regime.
},
exact analytical dark soliton solutions (of arbitrary amplitudes) of the pertinent
effectively 1D mean-field models (see Sec.~\ref{lowerd}) are not available. As a result,
the analytical techniques presented in Sec.~\ref{adiabatic} cannot be applied for the study of
matter-wave dark soliton dynamics in this regime. Nevertheless, the results of
Secs.~\ref{adiabatic} and \ref{stabilitycigar} indicate that the evolution of dark solitons
should be similar to the one pertaining to the TF-1D regime, while the solitons would
not be prone to the snaking instability, as in the case of the purely higher-dimensional
setups.

The statics and dynamics of matter-wave dark solitons in the crossover regime between
1D and 3D were studied in Ref.~\cite{crossover}. Particularly, numerical simulations of
the 3D GP equation revealed that matter-wave dark solitons are indeed dynamically stable,
and perform harmonic oscillations in the harmonic trap. Importantly, in the case of
small-amplitude oscillations, the oscillation frequency $\omega_{\rm osc}$ resulting
from the 3D GP equation was found to be equal to the eigenfrequency $\omega_{A}$ of the
anomalous mode of the effectively-1D NPSE model [cf. Eqs.~(\ref{gengpe}) and (\ref{salnl})];
note that the latter, can be expressed in the dimensionless form:
\begin{equation}
i\partial_t \psi = \left[-\frac{1}{2} \partial_z^2 + \frac{1}{2} \Omega^2 z^2 +
\frac{3|\psi|^2+2}{2(1+|\psi|^2)^{1/2}}\right]\psi,
\label{1dNPSE}
\end{equation}
where units are the same to the ones used for Eq.~(\ref{dim1dgpe}). The above finding
leads, in turn, to the following conclusion: an equation of motion for the center $z_0$
of a dark soliton in a condensate in the dimensionality crossover regime can be expressed
as follows:
\begin{equation}
\frac{d^2 z_{0}}{dt^2}=-\frac{\partial V_{\rm eff} }{\partial z_{0}},
\,\,\,\,\,\,\
V_{\rm eff} = \frac{1}{2}\omega_{\rm osc}^2 z_0^2,
\,\,\,\,\,\,\
\omega_{\rm osc} \equiv \omega_A.
\label{eqmocr}
\end{equation}
As shown in Ref.~\cite{crossover}, the soliton oscillation frequency is a decreasing function
of the dimensionality parameter $d$, taking values ranging from
$\omega_{\rm osc} = \Omega$ (corresponding to the non-interacting limit of
$d \rightarrow 0$), to $\omega_{\rm osc}=\Omega/\sqrt{2}$ [corresponding to
the TF-1D regime of $d \ll 1$ --- cf. Eq.~(\ref{soloscfreq})], with $\Omega$
being the normalized strength of the harmonic trap. In any case, the oscillation
frequency is {\it up-shifted} from its TF-1D value and, as a result, the effective
trapping potential felt by the dark soliton during its motion [see Eq.~(\ref{eqmd})]
will effectively become steeper. In that regard, it is relevant to mention that
substantial shifts of the oscillation frequency (which may be of order of $10\%$) have
been predicted in Ref.~\cite{crossover} and later confirmed in the Heidelberg experiment
of Ref.~\cite{kip} (see discussion in Sec.~\ref{large}). It should also be noticed that
the soliton oscillation frequency is also a decreasing function of the soliton amplitude
(or, in other words, of the oscillation amplitude), contrary to the result corresponding
to the TF-1D regime
\footnotemark[1]
\footnotetext[1]{Recall that the results of Sec.~\ref{adiabatic} indicate that the
soliton oscillation frequency does not depend on the soliton amplitude in the TF-1D regime.}
\cite{kip,draft6}.

Results similar to the ones obtained in the framework of the NPSE model \cite{crossover}
can also be obtained in the framework of the generalized NLS Eq.~(\ref{gengpe}) with the
nonlinearity function of Eq.~(\ref{gerbiernl}) (see, e.g., the analysis of Ref.~\cite{draft6}).
Furthermore, we should mention that the oscillations of dark solitons were also analyzed in
the framework of a GP model with generalized nonlinearities, and specific results in the
physically relevant case of a cubic–quintic nonlinearity (modeling two- and three-body
interactions --- see Sec.~\ref{lowerd}) were presented \cite{salernokam}. The same model
was also studied in Ref.~\cite{kinshukla}, were various stationary states, including
dark solitons, were found and analyzed in detail.

\subsubsection{The multiple-soliton state.}

Apart from the single-dark soliton state, the case of a multiple-dark soliton state can also
be analyzed in the dimensionality crossover regime using, as a guideline, the methodology
exposed in Secs.~\ref{multiple}, \ref{adiabatic} and \ref{BdGanalysis}. In particular,
in the case of well-separated and symmetrically interacting dark solitons in a harmonic trap,
one may follow the analysis of Refs.~\cite{kip,draft6} and analyze this problem by adopting
a simple physical picture: each soliton in the multi-soliton state follows the evolution of
the single-soliton state, i.e., it oscillates in the trap with an oscillation frequency
$\omega_{\rm osc}$ equal to the eigenfrequency $\omega_1$ of the {\it first} anomalous mode
(determined by a BdG analysis of the pertinent 1D mean-field models of Sec.~\ref{lowerd}),
and interacts with the neighboring solitons via the effective repulsive potential of
Eq.~(\ref{dx01}).

\begin{figure}
\center
\includegraphics[width=8cm]{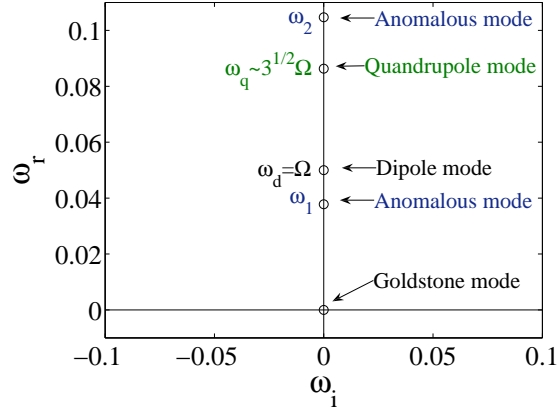}
\caption{(Color online) The lowest characteristic eigenfrequencies of the Bogoliubov
excitation spectrum for a condensate, in the dimensionality crossover regime from 3D to 1D,
carrying two dark solitons. The model used is the NPSE~(\ref{1dNPSE}), with parameters
$\Omega=0.05$, and $\mu=1.86$; the dimensionality parameter is
$d = N\Omega\alpha/\alpha_{\perp}=0.82$. The lowest characteristic eigenfrequencies
of the Bogoliubov excitation spectrum: shown are the one located at the origin,
the one at $\Omega = 0.05$, the one at $\sqrt{3}\Omega=0.087$ (corresponding to the
Goldstone mode, the Kohn mode and the quadrupole mode, respectively), as well as
the two anomalous modes, one with $\omega_{1}=0.756\Omega= 0.0378$ and one with
$\omega_{2}= 2.094 \Omega =0.1047$.}
\label{npse2ds}
\end{figure}

In order to further elaborate on the above, let us consider --- as an example --- the
two-dark soliton state of the NPSE model of Eq.~(\ref{1dNPSE}). This state (which, in the
linear limit, corresponds to the second excited state of the quantum harmonic oscillator)
can be obtained as a nonlinear stationary state of the system by means of, e.g., a
Newton-Raphson method. The pertinent configuration, has the form of two overlapping dark
solitons, placed at $z_i = \pm 2.185$ ($i=1,2$) with a fixed relative distance
$\delta z_0 =4.37$. The corresponding Bogoliubov excitation spectrum, namely the spectral plane
$(\omega_{r}, \omega_{i})$ of the eigenfrequencies $\omega \equiv \omega_{r}+i \omega_{i}$,
is shown in Fig.~\ref{npse2ds}: as it can clearly be observed, among the lowest
eigenfrequencies (such as the ones at $\omega=0$, $\omega_d=\Omega$ and
$\omega_q \approx \sqrt{3}\Omega$, corresponding to the Goldstone, dipole and quadrupole mode,
respectively), there exist two anomalous modes with eigenfrequencies
$\omega_{1}=0.756\Omega= 0.0378$ and $\omega_{2}= 2.094 \Omega =0.1047$.

According to the analysis of Sec.~\ref{BdGanalysis}, {\it small} displacements of the
dark solitons from their equilibrium points lead to the in-phase and out-of-phase
oscillatory motion of the dark soliton pair (see Fig.~\ref{inout}), with the respective
oscillation frequencies being equal to the eigenfrequencies $\omega_{1}$ and $\omega_2$
of the two anomalous modes. Importantly, the value of the eigenfrequency of the first
anomalous mode, $\omega_1=0.0378$, is quite close to the oscillation frequency of a
{\it single} dark soliton, $\omega_{\rm osc}=0.0375$, in the same setup (i.e., with the
same parameter values), with the percentage difference being $\approx 1.6\%$. This
generic example suggests that, generally, the dynamics of the two dark soliton state
can be described by an effective Lagrangian for the two solitons, namely
$L_{\rm eff} = T-V$; here, $T$ and $V$ are the kinetic and potential energies,
respectively, depending on the soliton centers, $z_i$ ($i=1,2$), and soliton
velocities, $\dot{z}_i \equiv dz_i/dt$, as follows,
\begin{equation}
T \equiv \sum_{i=1}^2 \frac{1}{2} (\dot{z}_i)^2, \,\,\,\,\,\,
V \equiv \sum_{i=1}^2 \frac{1}{2}\omega_{\rm osc}^2 z_i^2 + V_{\rm int}(z_2-z_1),
\label{Lagr2ds}
\end{equation}
where $V_{\rm int}(z_2-z_1)$ (with $z_2-z_1 \equiv 2z_0$) is the repulsive potential of
Eq.~(\ref{dx01}). Then, the evolution of the soliton centers can readily be determined by
the Euler-Lagrange equations $d(\partial_{\dot{z}_i}L_{\rm eff})/dt-\partial_{z_i}L_{\rm eff}=0$.
The latter, may be simplified upon using the approximate form of the repulsive potential
[cf. Eq.~(\ref{dx01ap})], thus leading to the following equations of motion:
\begin{eqnarray}
\ddot{z}_1 & = &-\omega^2_{\rm osc}z_1 - 8n_{0}^{3/2}\exp[-2\sqrt{n_{0}} (z_2-z_1)],
\label{sys1} \\
\ddot{z}_2 & = &-\omega^2_{\rm osc}z_2 +8n_{0}^{3/2}\exp[-2\sqrt{n_{0}} (z_2-z_1)],
\label{sys2}
\end{eqnarray}
where we have assumed well-separated, almost black solitons [i.e., in Eq.~(\ref{dx01ap}) we
have set $B \approx 1$]. Apparently, the above analysis can readily be generalized for multiple
solitons \cite{draft6}, with each one interacting with its neighbors. Importantly, if
$\omega_{\rm osc}$ in the system of Eqs.~(\ref{sys1})--(\ref{sys2}) was considered to be
unknown, then it would be possible to be directly obtained in the form of the characteristic
frequencies of the {\it normal modes} of this system (see details in Ref.~\cite{draft6});
these characteristic frequencies coincide to the ones determined via the BdG analysis.
This results justifies {\it a posteriori} the considered decomposition of the principal
physical mechanisms (oscillations and interactions of solitons) characterizing the system.

We should also note that apart from the case of small-amplitude oscillations of two well
separated, almost black solitons, the more general case of the dynamics of $n$-interacting
dark solitons (which may also perform large amplitude oscillations --- see Sec.~\ref{large}
below) is possible using the full set of the above mentioned Euler-Lagrange equations; the
latter, lead to the following $n$-coupled equations of motion \cite{draft6}:
\begin{equation}
\ddot{z}_i - \sum_{k=1}^{n} \left( \frac{\partial^2 V}{\partial z_k \partial \dot{z}_i } \dot{z_k}+
\frac{\partial^2 V}{\partial \dot{z}_k \partial \dot{z}_i} \ddot{z}_k \right) +
\frac{\partial V}{\partial z_i} =0,
\label{eqm}
\end{equation}
where $V\equiv \sum_{i=1}^{n} V_i$ is the potential energy,
$V_i= \sum_{i\ne j}^{n} n_0 B_{ij}^2/\{2 \sinh^2[\sqrt{n_0} B_{ij}(z_i-z_j)]\}$
is the interaction potential felt by the $i$-th soliton due to the presence of the
other solitons, while $z_{ij} = (1/2)(z_i - z_j)$ and $B_{ij}=(1/2)(B_i + B_j)$ denote,
respectively, the relative coordinate and the average depth for solitons $i$ and $j$.

\subsubsection{Large amplitude oscillations and experimental observations.}
\label{large}

\begin{figure}
\centering
\includegraphics[width=8cm]{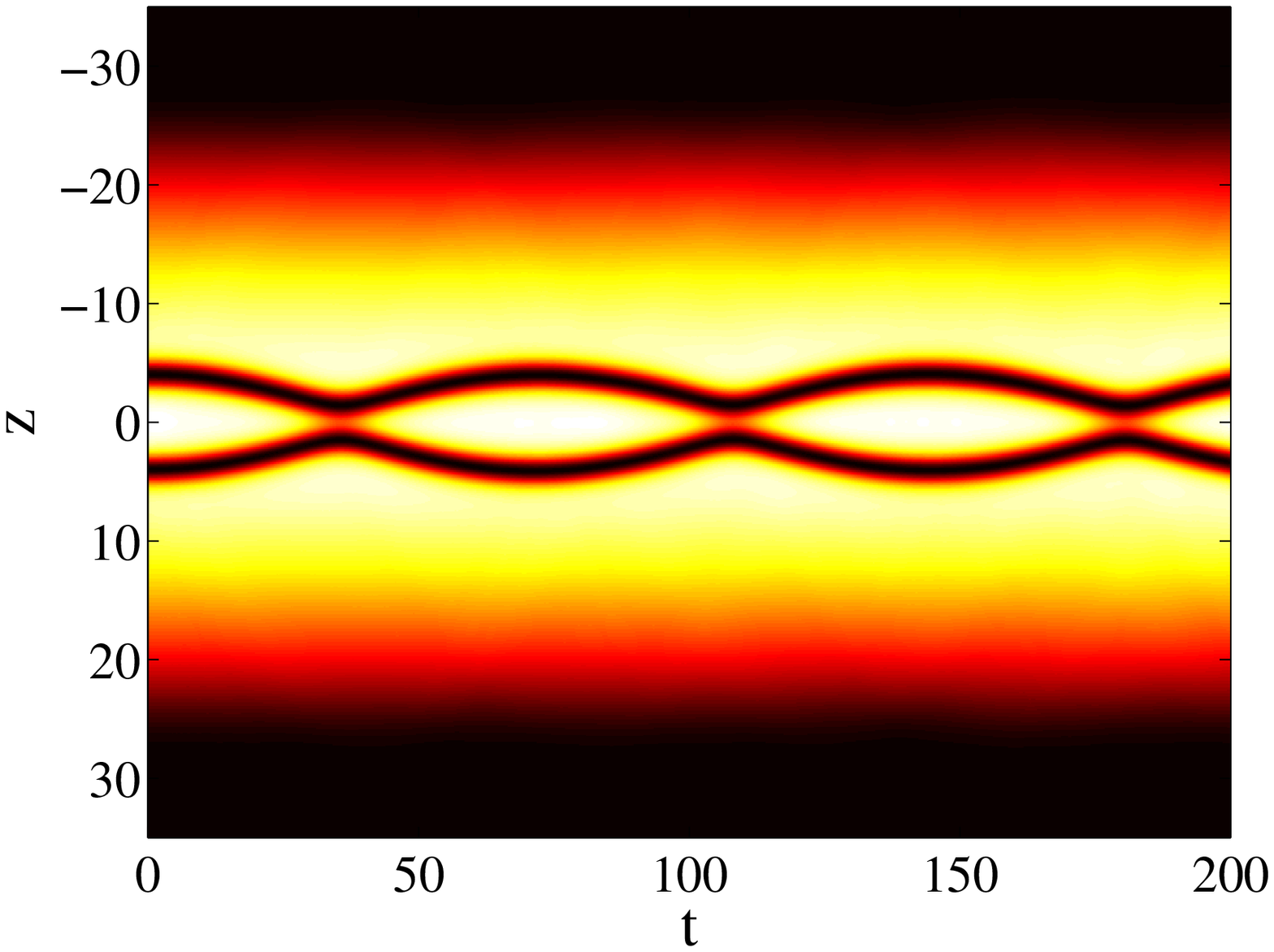}
\includegraphics[width=8cm]{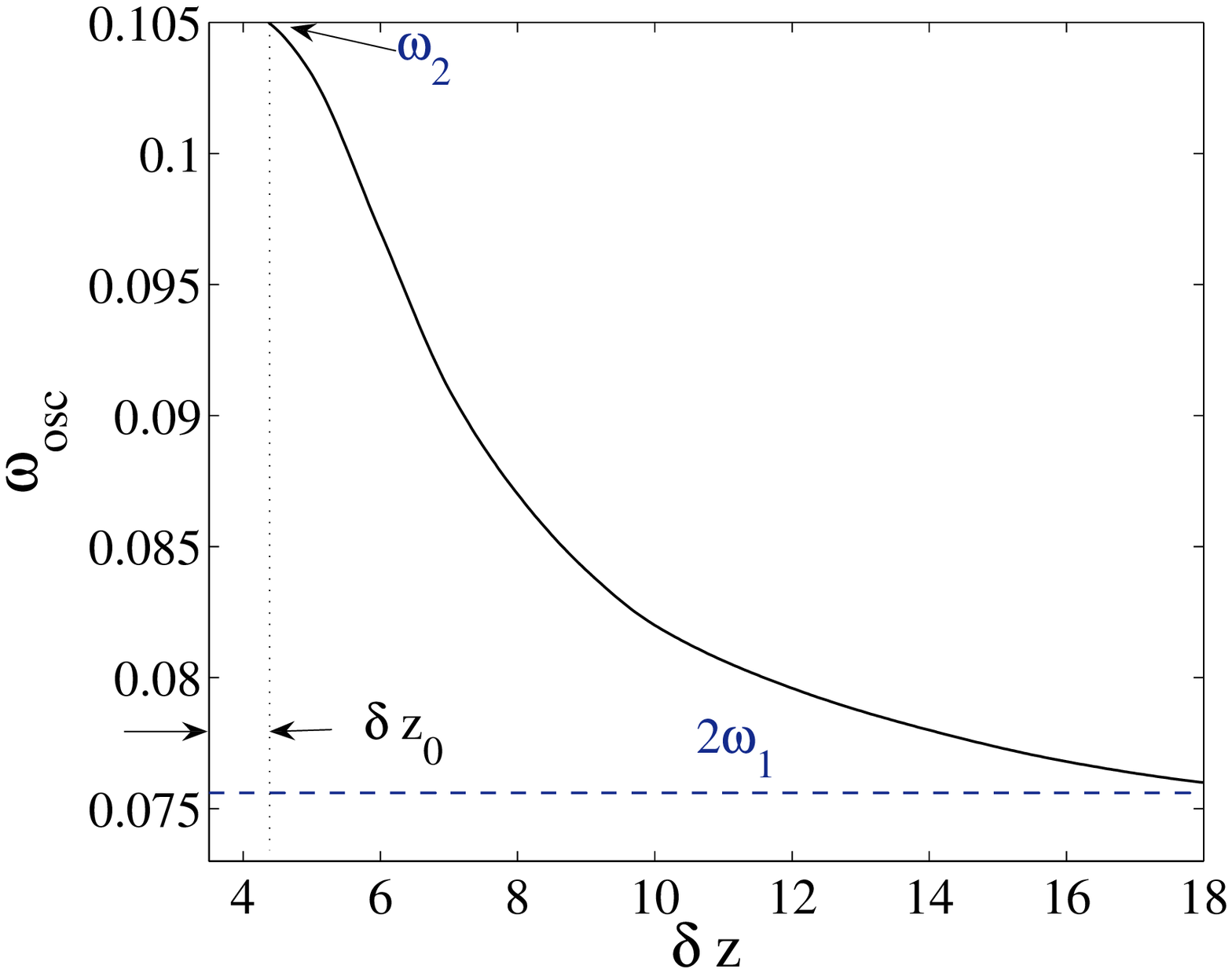}
\caption{(Color online) Top panel: Spatio-temporal contour plot of the density of
a cigar-shaped BEC confined in a trap of strength $\Omega=0.05$. The condensate is in the
dimensionality crossover regime from 3D to 1D, and the model used is the NPSE Eq.~(\ref{1dNPSE}).
The dark solitons, initially placed at $z=\pm 4$ (i.e., $\delta z =8 > \delta z_0 = 4.37$),
oscillate out-of-phase with a frequency
$\omega_{\rm osc}=1.74 \Omega < \omega_{2}= 2.094 \Omega$.
Bottom panel: The oscillation frequency of the dark solitons, $\omega_{\rm osc}$, as a
function of their initial relative distance $\delta z$. For $\delta z \rightarrow \delta z_{0}$
(corresponding to the stationary state), we obtain $\omega_{\rm osc} \rightarrow \omega_{2}$,
while for $\delta z \gg \delta z_{0}$, we obtain $\omega_{\rm osc}\rightarrow 2\omega_{1}$.
}
\label{inout}
\end{figure}

Let us now return to the above example of the two-dark soliton state of Eq.~(\ref{1dNPSE}) and
consider again the out-of-phase oscillation (we will use the parameter values of
Fig.~\ref{npse2ds}). In this case, if the initial soliton separation is significantly larger
than $\delta z_0 =4.37$ --- or, in other words, if the displacements of solitons around their
equilibrium positions are {\it not} small --- then $\omega_{\rm osc}$ differs from (in fact,
it is quite smaller than) $\omega_2$: considering, e.g., that $\delta z = 8$ (corresponding
to initial locations of the soliton centers $z=\pm 4$ --- see Fig.~\ref{inout}), the
out-of-phase oscillation of the two solitons is characterized by a frequency
$\omega_{\rm osc}=1.74\Omega < \omega_{2}= 2.094 \Omega$.
A qualitative explanation for this difference is the following: As discussed above,
the evolution of two initially overlapping dark solitons can be effectively described
by the equations of motion (\ref{sys1})-(\ref{sys2}), in the presence of the repulsive
potential which depends exponentially on their relative distance. If the relative distance
between the two solitons is not significantly different than the one pertaining to the
corresponding stationary state, i.e., $\delta z \approx \delta z_{0}$, the effective
repulsive force is strong and, as a result, their motion is strongly affected by their
coupling. On the other hand, if their initial separation becomes larger (as, e.g., in
the case of the example under consideration, with $\delta z=8$), the repulsive force
becomes exponentially small and, as a result, the motion of each individual soliton is
not significantly affected by the presence of the other.

In the bottom panel of Fig.~\ref{inout} we show the oscillation frequency of the
dark soliton --- obtained by the direct integration of the NPSE model of Eq.~(\ref{1dNPSE}) ---
as a function of the initial relative distance between the two solitons. It is clear that
the oscillation frequency, which takes values in the interval $\delta z \ge \delta z_{0}$
(the value corresponding to the stationary state), exhibits two different asymptotic regimes:
when the initial soliton separation is small enough, $\delta z \rightarrow \delta z_{0}$
(i.e., for strong coupling between the two solitons), the oscillation frequency tends
to the eigenfrequency $\omega_{2}$ of the largest anomalous mode; on the other hand,
when the initial soliton separation is large enough, $\delta z \gg \delta z_{0}$ (i.e.,
when the solitons are actually decoupled), the oscillation frequency tends to
$2 \omega_{1}$; the latter value can be explained by the fact that the period of
oscillation of each individual soliton is the half of the one that would correspond
to a single soliton oscillation.

As concerns relevant experiments, an oscillating and interacting dark soliton pair in
a $^{87}$Rb BEC, in the dimensionality crossover regime between 1D and 3D, was experimentally
observed in the Heidelberg experiment of Ref.~\cite{kip}. In this experiment, large amplitude
oscillations were induced by the method of matter-wave interference (see Sec.~\ref{interf} below).
The dependence of the oscillation frequency on the distance between the two solitons (see the
left panel of Fig.~\ref{inout}) was found and compared with experimental data: the agreement
between the theoretical predictions (based on a study of the NPSE model) and the experimentally
observed oscillation frequencies was excellent. In accordance to the analysis of this Section,
considerable upshifts --- up to $16\%$ --- of the soliton oscillation frequency from the value
of $\Omega/\sqrt{2}$ were observed in the study of Ref.~\cite{kip}, and they were quantitatively
attributed to the dimensionality of the system and the soliton interactions.


\section{Matter-wave dark solitons in various settings and parameter regimes}
\label{exp}

\subsection{Matter-wave dark solitons in multi-component condensates.}
\label{multi}

Multi-component ultracold atomic gases and BECs may be composed by two or more atomic gases, which may
have the form of mixtures of:
(a) two different spin states of the same atom species
(so-called {\it pseudo-spinor} condensates) \cite{Myatt1997a,proximity,chap01:stamp};
(b) different Zeeman sub-levels of the same hyperfine level (so-called {\it spinor} condensates)
\cite{Stenger1998a};
(c) different atom species (so called {\it heteronuclear} mixtures) \cite{Modugno2001a};
(d) degenerate boson-fermion clouds \cite{40Kfermion}; (e) purely degenerate fermion
clouds \cite{Li-fermion1} (see also \cite{book1,book2,BECBOOK} for reviews and references therein).
Such multi-component systems support various types of matter-wave soliton complexes, with the type of
soliton in one species being the same or different to the type of soliton in the other species.
Here, of particular interest are the so-called {\it vector} solitons with the one component being
a dark soliton, which have mainly been studied in the context of two-component and spinor condensates.

\subsubsection{Dark solitons in two-component condensates.}

Generally speaking, a mixture of $\mathcal N$ purely bosonic components can be described in the framework
of mean-field theory by a system of $\mathcal N$ coupled GP equations, which can be expressed in the
following dimensionless form:
\begin{equation}
i \frac{\partial \psi_n}{\partial t}=-\frac{1}{2} \nabla^2 \psi_n
+ V_n({\bf r}) \psi_n + \sum_{k=1}^{\mathcal N}\left[ g_{n,k}|\psi_k|^2 \psi_n
-\kappa_{n,k} \psi_k + \Delta_{n,k} \psi_n  \right].
\label{2C}
\end{equation}
Here, $\psi_n$ is the wave function of the $n$-th component ($n=1,\dots,{\mathcal N}$),
$V_n({\bf r})$ is the trapping potential confining the $n$-th component,
$\Delta_{n,k}$ is the chemical potential difference between components $n$ and $k$,
the nonlinearity coefficients $g_{n,k}=g_{k,n}$ characterize inter-atomic collisions, while
the linear coupling coefficients $\kappa_{n,k}=\kappa_{k,n}$ account for
spin state inter-conversion, usually induced by a spin-flipping resonant electromagnetic wave
(see, e.g., Ref.~\cite{NZ}). In some works (see, e.g., Refs.~\cite{fermionGPE1,fermionGPE2,fermionGPE3})
fermionic mixtures are also described in the framework of the mean-field theory, with the self-interacting
nonlinear terms being replaced by $g_{n,n}|\psi_n|^{4/3}\psi_n$. Notice that in the GP Eqs.~(\ref{2C})
both the energy $E$ and the total number of atoms,
$N\equiv \sum_{k=1}^{\mathcal N} N_{k}=\sum_{k=1}^{\mathcal N}\int |\psi_k|^2 d\mathbf{r}$,
are conserved; furthermore, in the absence of linear inter-conversions ($\kappa_{n,k}=0$),
the number of atoms of each component $N_{k}$ is separately conserved.

Let us consider the case of two bosonic species (${\mathcal N}=2$), and assume that the system is
homogeneous ($V_n=0$). If, additionally, there is no spin-state inter-conversion ($\kappa_{n,k}=0$)
and chemical potential difference ($(\Delta\mu_{n,k})=0$), then the binary mixture is {\it immiscible}
provided that the following {\it immiscibility condition} holds \cite{miscibility},
\begin{equation}
\Delta \equiv (g_{12}g_{21}-g_{11}g_{22})/g_{11}^{2} > 0,
\label{Delta}
\end{equation}
where the, so-called, miscibility parameter $\Delta$ takes in practice the values
$\Delta \approx 9 \times 10^{-4}$ or $\Delta \approx 0.036$ for a mixture of two spin
states of a $^{87}$Rb BEC \cite{Myatt1997a,proximity} (see also Ref.~\cite{CompSep})
or a $^{23}$Na BEC \cite{chap01:stamp}, respectively. The condition (\ref{Delta}) indicates that
if the mutual repulsion between species is stronger than the repulsion between atoms of the same
species then the two species do not mix. In such a case, the two species tend to separate by
filling two different spatial regions, thus forming a ``ball'' and ``shell'' configuration
(see, e.g., Ref.~\cite{proximity} for relevant experimental results). This way, the ground state
of the system --- i.e., the state minimizing the energy --- may take the form of {\it domain-wall}
solutions of the GP Eqs.~(\ref{2C}) \cite{pu1,shenoy,esry,Timmermans,Marek,healt}.
In accordance to the experimental observations, these solutions represent configurations of
the following form: in the Thomas-Fermi limit (where kinetic energy is negligible), one species
occupies the region around the trap center, and it is separated by two domain-walls from side domains
occupied by the other species; on the other hand, kinetic energy favors a configuration where
a single domain-wall at the trap center separates two domains occupied by the different species \cite{Marek}.
The dynamics of phase-separation of two-component BECs has been studied in various works both theoretically
(see, e.g., Refs.~\cite{merhasin,jap,navarro} and also Refs.~\cite{decon,rabiswitch} for proposed
applications) and experimentally \cite{CompSep,ingu,papp}. Importantly, magnetic-field Feshbach resonances
can be used to controllably change the the inter-species \cite{ingu} or the intra-species \cite{papp}
scattering length, and thus controllably change the (im)miscibility between the two species \cite{papp}.

Apart from domain-walls, a trapped two-component quasi-1D BEC supports vector solitons, with the one
component being a dark soliton; in such a case, typically, the other component may be a dark soliton
\cite{vectordark,aguero,hadi,ourship,jap,obsantos1,obsantos2,epjd,kinez1,kinez2} or a bright soliton
\cite{BA01,jap,kinez2,schap,csf,laksh}. Additionally, apart from such {\it dark--dark} and {\it dark--bright}
solitons, {\it dark--antidark}
\footnotemark[1]
\footnotetext[1]{An anti-dark soliton is actually a dark soliton with reverse-sign amplitude, i.e., it has
the form of a hump (instead of a dip) on top of the background density (see, e.g., Refs.~\cite{opt1,opt2}).}
solitons have also been predicted to exist in BECs, either in a stationary form \cite{epjd} or as a
dynamical entity resulting from instabilities \cite{hadi,jap}. Below we will focus on the most generic
vector matter-wave solitons, namely the dark--dark and dark--bright ones (note that the latter have also
been observed experimentally \cite{hamburg}), presenting results corresponding to the simplest possible setup,
which solely includes the traditional time-independent harmonic trap. Notice that in the absence of the trap,
vector solitons of the above mentioned types have been extensively studied in the context of nonlinear
optics: there, multi-component solitons occur when fields of one frequency, or one polarization, become
coupled to fields of other frequencies, or other polarizations (see, e.g., the review \cite{kivpr} and
references therein). Mathematically speaking, the existence (and stability) of multi-component optical
solitons (and also matter-wave solitons in the miscible case) can be understood by the fact that the
relevant coupled NLS equations rely on the so-called {\it Manakov system} \cite{manakov}:
the latter, has the form of a vector NLS equation, namely,
\begin{equation}
i \partial_t {\bf u} = -\frac{1}{2}\partial_{z}^2 {\bf u} \pm |{\bf u}|^2 {\bf u},
\label{man}
\end{equation}
where ${\bf u}(z,t)=(u_1(z,t), u_2(z,t), \cdots, u_n(z,t))$ is a $n$-component vector.
This system is known to be completely integrable \cite{intman1,intman2,intman3} (in fact, it can be
integrated by extending the IST method that has been used to integrate the scalar NLS equation \cite{zsb,zsd})
and admits such vector $N$-soliton solutions \cite{ralak,shepkiv,park}.

As shown in Refs.~\cite{obsantos1,obsantos2}, the dynamics and interaction of {\it dark--dark} solitons
in a two-component quasi-1D BEC can be studied by means of a variational approach; in the case of
equal chemical potentials, the latter is based on the use of the following ansatz for the single-component
soliton wave-functions,
\begin{eqnarray}
\psi_1(z,t)&=& B \tanh \left[B\left( z-z_0(t) \right) \right] +iA,
\label{2dsans1} \\
\psi_2(z,t)&=&B\tanh \left[B\left( z+z_0(t) \right) \right] \mp iA,
\label{2dsans2}
\end{eqnarray}
where $2z_0(t)$ denotes the relative distance between the two solitons, and the $\mp$ signs correspond,
respectively, to a kink--anti-kink state (where the solitons' phase fronts are facing each other) and
a kink--kink state (where the solitons' phase fronts are in the same direction). Both the miscible and
immiscible cases where studied in Refs.~\cite{obsantos1,obsantos2} and the main results of the analysis
can be summarized as follows. In the miscible case ($g_{11}=g_{22}=g_{12}$), and for the kink--anti-kink
state, the trajectories in the $(z_0, \dot{z}_0)$ phase-plane  are either periodic surrounding the center
$(0,0)$ [indicating the formation of a bound state (``soliton molecule'')], or free [indicating acceleration
(deceleration) of the approaching (outgoing) solitons]; contrary, in the case of the kink-kink state, where
solitons move in the same direction, the solitons form a bound state which can never be broken. On the other
hand, in the immiscible case (i.e., when domain-walls are present), it was shown that if a dark soliton
exceeds a critical velocity then it can be transferred from one component to the other at the domain-wall;
on the other hand, for lower velocities, multiple reflections within the domain were observed. In such a case,
the soliton is accelerated after each reflection and eventually escapes from the domain.

As mentioned above, {\it dark--bright} matter-wave solitons in a quasi-1D binary BEC are also possible.
Particularly, in the miscible case (with all nonlinearity coefficients $g_{n,k}$ being normalized to unity),
the wave functions $\psi_{\rm d}(z,t)$ and $\psi_{\rm b}(z,t)$ of the dark and bright soliton components
may be expressed in the following form \cite{BA01},
\begin{eqnarray}
\psi_{\rm d}(z,t) &=& \sqrt{\mu}\cos\phi
\tanh \left \{ \kappa \left[ z-z_0(t) \right] \right\} + i \sqrt{\mu}\sin\phi
\label{dbans1}, \\
\psi_{\rm b}(z,t) &=& \sqrt{\frac{1}{2}N_{\rm b}\kappa}\,
{\rm sech}\left \{ \kappa \left[ z-z_0(t) \right] \right\}\exp(i\theta_{\rm b}).
\label{dbans2}
\end{eqnarray}
Here, $\mu_{\rm d}=\mu$ and $\mu_{\rm b}=\mu+\Delta $ are the chemical potentials of the dark and bright
components, $\phi$ is the dark soliton's phase angle, $z_0$ denotes the solitons' center,
$N_{\rm b} = \int_{-\infty}^{+\infty} |\psi_{\rm b}|^2 dz$ is the normalized number of atoms of the
bright soliton, $\kappa = \sqrt{\mu \cos^2\phi+(N_{\rm b}/4)^2} - N_{\rm b}/4$ is the inverse width of
the bright soliton, and $\theta_{\rm b}=(\kappa \tan\phi) x  + [\kappa^2(1-\tan^2\phi)/2-\Delta]t$
is the bright soliton's phase. According to the analysis of Ref.~\cite{BA01}, if the external trapping
potentials $V_{\rm d}$ and $V_{\rm b}$ for the dark and bright solitons are slowly varying on the soliton
scale $\kappa^{-1}$, then the dynamics of the dark-bright soliton can be described by the effective
particle approach of Sec.~\ref{adiabatic}. Particularly, assuming that the solitons are sufficiently slow,
a multiple-time-scale boundary-layer theory --- similar to the one used in Ref.~\cite{fr2} --- leads to the
following equation of motion for the soliton center,
\begin{equation}
\frac{d^2 z_{0}}{dt^2}=-\frac{1}{2} V_{\rm d}'(z_0)
-\frac{N_{\rm b} [V_{\rm d}'(z_0)-2V_{\rm b}'(z_0)]}{8\sqrt{\mu+(N_{\rm b}/4)^2 - V_{\rm d}(z_0)}},
\label{eqmdb}
\end{equation}
where $V_{\rm d,b}'(z_0) \equiv \partial V_{\rm d,b}/\partial z_{0}$. In the limit $N_{\rm b} \rightarrow 0$,
Eq.~(\ref{eqmdb}) is reduced to Eq.~(\ref{eqmd}) (recall that the latter predicts dark soliton oscillations
with a frequency $\Omega/\sqrt{2}$), while the motion of the vector soliton becomes more sensitive to the
presence of the bright component as $N_{\rm b}$ is increased. For example, in the case of equal harmonic
traps of strength $\Omega$, such that $V_{\rm b}=V_{\rm d}\ll \mu$ (i.e., for soliton motion near the trap
center), the oscillation frequency of the dark--bright soliton resulting from Eq.~(\ref{eqmdb}) reads,
\begin{equation}
\Omega_{\rm osc} = \frac{\Omega}{\sqrt{2}}
\left( 1- \frac{N_{\rm b}}{4\sqrt{\mu+(N_{\rm b}/4)^{2}}} \right)^{1/2}.
\label{BAp}
\end{equation}
It is clear that Eq.~(\ref{BAp}) shows that the oscillation frequency is down-shifted as compared to the
characteristic value of $\Omega/\sqrt{2}$, i.e., the dark-bright pair executes slower oscillations, as
the bright component is enhanced.

The predictions of Ref.~\cite{BA01} can directly be compared to the findings of a Hamburg experiment
\cite{hamburg}, where long-lived dark--bright matter-wave solitons were observed in a two-component
quasi-1D $^{87}$Rb BEC. In particular, using the phase-imprinting method, a dark soliton was created in
one spin state of the BEC and the density dip was filled by atoms, forming the bright soliton, in another
spin state of the BEC (note that the number of atoms $N_{\rm b}$ of the bright soliton was $\approx 10\%$
of the total number of atoms). The created dark--bright soliton was then observed to perform slow oscillations
with a frequency $0.24 \Omega$, which is quite smaller than the frequency of the corresponding single
dark soliton in the same setting. Moreover, due to the initial state preparation, an extra dark soliton
was generated, which was allowed to interact with the co-existing dark--bright soliton; it was observed
that this individual dark soliton was reflected off the slower dark--bright one, with the process
resembling a hard-wall reflection.

\subsubsection{Dark solitons in spinor condensates.}

The spin degree of freedom of spinor BECs gives rise to important new phenomena (including, among others,
the formation of spin domains \cite{Stenger1998a}, spin textures \cite{spintext} and vortices \cite{textvort},
as well as spin oscillations \cite{spinosc}), which are not present in other types of BECs. Generally,
a spinor condensate formed by atoms with spin $F$ can be described in the framework of mean-field theory
by a $(2F+1)$-component macroscopic wave function; accordingly, a spinor $F=1$ condensate is characterized
by a vector order parameter, with the three components corresponding to the values of the vertical spin
projection, $m_{F}=-1,0,+1$. In a quasi-1D setting, the pertinent system of coupled GP equations for the
wave functions $\psi_{\pm 1,0}(z,t)$ can be expressed in the following dimensionless form (see, e.g.,
Refs.~\cite{bdspinor,ofyspin,spindw}),
\begin{eqnarray}
i\partial _{t}\psi _{\pm 1} &=& \mathcal{H} \psi _{\pm 1}+\delta (|\psi
_{\pm 1}|^{2}+|\psi _{0}|^{2}-|\psi _{\mp 1}|^{2})\psi _{\pm 1}
+\delta \psi _{0}^{2}\psi _{\mp 1}^{\ast },
\label{dvgp1} \\
i\partial _{t}\psi _{0} &=& \mathcal{H} \psi _{0}+\delta (|\psi
_{-1}|^{2}+|\psi _{+1}|^{2})\psi _{0}
+2\delta \psi _{-1}\psi _{0}^{\ast  }\psi_{+1}.
\label{dvgp2}
\end{eqnarray}
Here, $\mathcal{H} \equiv -(1/2)\partial _{x}^{2}+(1/2)\Omega^{2}z^{2}+n$, with
$\Omega$ being the normalized trap strength and
$n=|\psi _{-1}|^{2}+|\psi_{0}|^{2}+|\psi _{+1}|^{2}$ the total density, while
$\delta \equiv (a_{2}-a_{0})/(a_{0}+2a_{2})$ where $a_{0}$ and $a_{2}$ are the $s$-wave scattering lengths
in the symmetric channels with total spin of the colliding atoms $F=0$ and $F=2$,
respectively. Actually, the parameter $\delta$ represents the ratio of the strengths of the spin-dependent
and spin-independent interatomic interactions, and may take negative or positive
values for {\it ferromagnetic} or {\it anti-ferromagnetic} (alias {\it polar}) spinor BECs, respectively.
Typically, in the relevant cases of $^{87}$Rb and $^{23}$Na atoms with $F=1$,
$\delta =-4.66\times 10^{-3}$ \cite{kempsp} and $\delta =+3.14\times 10^{-2}$ \cite{greenesp};
nevertheless, the above values may in principle be modified by employing the so-called
confinement-induced Feshbach resonance \cite{cfrm}.

In the limiting case of $\delta=0$ (and in the absence of the potential), the system of
Eqs.~(\ref{dvgp1})--(\ref{dvgp2}) is reduced to the completely integrable Manakov system.
On the other hand, as shown in Ref.~\cite{wadspin}, another completely integrable version
of Eqs.~(\ref{dvgp1})--(\ref{dvgp2}) corresponds to the case $\delta=1$ (i.e., for interatomic
and anti-ferromagnetic interactions of equal magnitude): in this case, the resulting matrix NLS
equation with non-vanishing boundary conditions is completely integrable by means of the IST method
\cite{spinist} and admits exact analytical vector $N$-dark soliton solutions (i.e., single- and
multiple-vector dark solitons of the dark--dark--dark type in terms of the $m_F =-1,0,+1$ spinor
components) \cite{wad2} (see also Ref.~\cite{ieda}). The one-dark soliton state of this system can
be classified as (a) ferromagnetic (i.e., with nonzero total spin), which has domain-wall shaped wave
functions, and (b) polar (i.e., with zero total spin), characterized by the familiar hole soliton
profile. Note that the collisions of two solitons give rise to interesting spin-dependent phenomena,
such as spin-mixing or spin-transfer \cite{wad2}.

In the physically relevant case of small $\delta$, mixed dark--bright solitons of the dark--dark--bright or
bright--bright--dark type (again in terms of the $m_F =-1,0,+1$ spinor components) were also predicted to
occur in anti-feromagnetic spinor $F=1$ BECs \cite{bdspinor}. In the small-ampliude limit (and in the
absence of the trap), these solitons were found to obey the completely integrable
{\it Yajima-Oikawa system} \cite{yo}, by means of which it was found that the functional form of the
dark and bright components is similar to the one in Eqs.~(\ref{dbans1})--(\ref{dbans2}). Numerical simulations
in Ref.~\cite{bdspinor} demonstrated that, for small-amplitudes, such dark--bright solitons feature genuine
soliton behavior (i.e., they propagate undistorted and undergo quasi-elastic collisions), while for moderate
and large amplitudes (and also for large values of $\delta$) they can exist as long-lived objects as well.
Furthermore, for sufficiently small number of atoms of the bright soliton, the bright component(s)
are guided by the dark one(s), and the vector soliton performs harmonic oscillations; the oscillation
frequency is different for small- and moderate-amplitude solitons, and it is respectively given by:
\begin{equation}
\omega_{\rm osc} = \frac{\Omega}{\sqrt{2}}(1-\alpha_0 \sqrt{\delta}) - \epsilon_0, \qquad
\omega_{\rm osc} = \Omega_{\rm osc} (1-\alpha_1 \delta^2) + \epsilon_1.
\label{bdsp}
\end{equation}
In the above expression, $\Omega_{\rm osc}$ is given by Eq.~(\ref{BAp}), while the constants
$\alpha_{0,1}$ and $\epsilon_{0,1}$ (with $\epsilon_{0,1} \ll \alpha_{0,1}$) depend on the normalized
number of atoms of the bright component. It is clear that the characteristic oscillation frequency of
the dark soliton ($\Omega/\sqrt{2}$), as well as the result of Ref.~\cite{BA01}, is modified by the
spin-dependent interactions that are present in the case of a spinor $F=1$ BEC.


\subsection{Matter-wave interference and dark solitons.}
\label{interf}

Matter-wave interference experiments (see, e.g., the seminal work of Ref.~\cite{science})
are known to demonstrate, apart from self-interference, the interference between two BECs
confined in a trap, divided into separate parts by means of a barrier potential induced by
a laser beam. In particular, the BECs are left to expand and overlap forming interference fringes,
similar to the ones known in optics. Much interest has been drawn to a better understanding of this
fundamental phenomenon, especially as concerns the coherence properties of the interfering BECs.
In that regard, it is worth mentioning that the (incorrect) assumption that ``when the interfering
BECs have fixed atom numbers, there can be no phase'', was resolved --- shortly after the
experimental realization of BECs \cite{bec1,bec2,bec3} --- in Ref.~\cite{castindal}. On the other hand,
since most of the relevant experimental findings can be quantitatively reproduced in the framework
of the GP mean-field theory \cite{rhorl}, we will proceed by adopting this approach in order
to discuss the connection between dark solitons and matter-wave interference.

An interesting variation of the interference process, which is naturally attributed to the
inherent nonlinearity of BECs due to interatomic interactions, is that --- under certain conditions --- the
collision of two initially separated condensates can lead to the creation of dark solitons. This
``nonlinear interference'' effect was first observed in simulations \cite{motion1}, and was subsequently
analyzed theoretically \cite{scott}. Other studies, basically relying on the  self-interference of BECs,
have also been proposed as well \cite{genrds1,genrds2,genrds3,nate,interfe}. Importantly, relevant recent
experiments employing this, so-called, {\it matter-wave interference method} have already been reported,
demonstrating the generation of vortices \cite{brian} (see also theoretical work in
Refs.~\cite{nate,Nate2,Nate3}) and dark solitons \cite{kip,technion,draft6,enghoef} (see also the
experiment of Ref.~\cite{jo}).

To get a deeper insight into the physics of the matter-wave interference process, let us follow the
arguments of Ref.~\cite{scott} and consider the interference between two separated quasi-1D BECs colliding
in the presence of a harmonic trap. There exist two different regimes characterizing this process, namely
a linear and a nonlinear one, depending on the competition between the kinetic and the interaction energies.
In the linear regime, the kinetic energy of the condensates exceeds the nonlinear interaction energy of
the atoms. In this case, and at any time $t$, the total wave function of the system can be well approximated
by a linear superposition of the wave functions that each individual condensate would have at $t$.
The two initially {\it well-separated} BECs interfere at the trap center, produce a linear interference
pattern, and then separate again regaining their initial shape. The fringe spacing $l$ of the interference
pattern is determined by the $k$ vector that each individual condensate would have if performing a dipole
oscillation alone in the trap and, at the time of maximum overlap, $l=\pi/D(\hbar/2m\omega_z)$
(here $D$ is the initial distance between the condensates). It is clear that the higher the kinetic
energy is, the higher the number of fringes and the smaller the fringe spacing is. Approximating the
individual wave functions in the TF limit, it can be found that the kinetic energy (estimated from the
curvature of a $\cos^2$ interference pattern) exceeds the peak nonlinear energy (at the center of
the fringes) when the initial distance between the two BECs exceeds  critical distance, namely $D>D_c$,
where $D_c$ is given by \cite{scott}:
\begin{equation}
D_c = \pi \left(12\pi \frac{N \hbar a}{m \omega_z}\right)^{1/3}.
\label{dc}
\end{equation}
Here, $N$ is the number of atoms, $a$ the $s$-wave scattering length, $\omega_z$ the longitudinal trap
frequency, and $m$ the atomic mass. If the above condition is not fulfilled, namely $D<D_c$, then the
system enters in the nonlinear regime. In the latter, the interference pattern consists of stable fringes
with a phase jump of order of $\pi$ across them, which can naturally be identified as genuine dark solitons.
Notice that in the nonlinear regime, the initially individual condensates instead of reforming as separate
objects, they form a combined condensate undergoing a quadrupole oscillation.

The results of Heidelberg experiments \cite{kip,draft6} can be compared directly to the above theoretical
predictions. In these experiments, dark solitons were created by releasing a $^{87}$Rb BEC from a double-well
trap into a harmonic trap in the dimensionality crossover regime from 1D to 3D. For the parameters used, the
initial distance of the individual condensates was approximately five times smaller than the critical distance
and the trap frequencies were ramped, with ramping times chosen so as to minimize the excitation of the
quadrupole mode. It was shown that, in accordance to the observations of Ref.~\cite{jo}, the number of
created solitons is even for a zero phase-difference between the two initially separated condensates,
while it is odd for a phase-difference close to $\pi$. If the phase-difference is exactly equal to $\pi$,
a standing (black) dark soliton in the middle of the trap is always created. Notice that the total number
of the created solitons depends on the momentum of the merging condensates, which may be controlled by
varying, e.g., the distance between the condensate fragments, the number of atoms, or the aspect ratio of
the trap \cite{enghoef,kip,draft6}.


\subsection{BEC superfluidity and dark solitons}
\label{drag}

A flow past an obstacle is known to be one of the most fundamental contexts for studying superfluidity.
Particularly, according to the Landau criterion for superfluidity \cite{landausf}, a superfluid flow past
an obstacle, is stable (unstable) for group velocities smaller (larger) than the speed of sound. Actually,
breakdown of superfluidity is caused by the opening of channels for emission of excitations in the fluid,
whose formation manifests itself as an effective dissipation. In the BEC context, early experiments from
the MIT group \cite{onofrio2,onofrio1} demonstrated the onset of dissipation induced by the motion of an
obstacle (in the form of a strongly repulsive dipole beam). From a theoretical standpoint, the problem can
be studied by using a NLS (or a GP) equation that includes a localized external potential of the form
$V({\bf r}-vt)$ (with $V({\bf r}) \rightarrow 0$ as $|{\bf r}| \rightarrow \infty$) accounting for the
presence of the obstacle moving with velocity $v$; this potential may be naturally superimposed to the
usual trapping potential confining the condensate. In relevant earlier studies \cite{frisch}, where the
NLS equation as a model of superflow was used, vortex formation induced by the superfluid flow around
an obstacle was predicted.

The lower-dimensional setting, namely the 1D flow of a repulsive NLS fluid in the presence of an obstacle,
was also studied \cite{hakim,pavloff2002} (see also Ref.~\cite{loeb}). Specifically, in Ref.~\cite{hakim}
it was shown that below an obstacle-dependent critical velocity, there exists a steady dissipationless
flow solution, which disappears at the critical velocity by merging with an unstable solution in
a saddle-node bifurcation. This unstable solution represents the transition state for emission of dark
solitons, which are repeatedly generated above the critical velocity. In fact, the onset of dissipation
corresponds to nonstationary flow with a wake asymptotically extending upstream to infinity, and downstream
periodic emission of dark solitons \cite{pavloff2002}. Note that in both Refs.~\cite{hakim}
and \cite{pavloff2002} the critical velocity was found to be smaller than the speed of sound,
a result that may be explained by the fact that, in the region of the potential, the local fluid velocity
can reach values higher than the local sound velocity (critical velocity values smaller than the speed
of sound were also observed in Refs.~\cite{onofrio2,onofrio1}).

The above results paved the way for a better understanding of the BEC flow past an obstacle and inspired
further investigations \cite{radou1,dsimp,mcs}. Importantly, in an recent experiment \cite{engels}, the
BEC flow induced by a broad, penetrable barrier (in the form of a laser beam) swept through an elongated
$^{87}$Rb condensate was systematically studied: it was demonstrated that at slow barrier speeds the flow
is stable, at intermediate speeds becomes unstable and dark soliton generation is observed, while at
faster speeds, remarkably, soliton formation completely ceases. Both repulsive and attractive barriers
were used in the experiment of Ref.~\cite{engels}, and were found to lead to dark soliton formation;
additionally, it was also found that the critical velocity for the breakdown of the BEC superfluidity
and soliton generation was smaller than the speed of sound. Note that in a recent work \cite{lesz},
velocity regimes similar to the ones found in Ref.~\cite{engels} were analytically predicted by using
a hydrodynamic approach.

As shown theoretically \cite{hakim,pavloff2002,radou1,dsimp,mcs} and demonstrated experimentally \cite{engels},
dark solitons (and vortices) are formed if the size of the ``hypersonic'' obstacle is of the order of,
or greater than, the characteristic healing length of the condensate. On the other hand, if the size
of the obstacle is much smaller than the healing length, the main loss channel, which opens at supersonic
velocities of the obstacle, corresponds to the Cerenkov emission of Bogoliubov's excitations \cite{ap04}.
Notice that in the case of large hypersonic obstacles, two dispersive shock waves, which start propagating
from the front and the rear parts of the obstacle, are formed. Far from the obstacle, the shock front
gradually transforms into a linear ``ship wave" located outside the Mach cone
\cite{carusotto,gegk07,gk07,gsk08}, whereas the rear zone of the shock is converted into
a ``fan'' of oblique dark solitons located inside the Mach cone \cite{ek06,egk06,egk07b}
(see also relevant experimental results in Refs.~\cite{carusotto,cornell05}). An important result
reported in Ref.~\cite{kp08} is that although such dark solitons are unstable in higher-dimensional
settings with respect to transverse perturbations (see Sec.~\ref{snaking}), the instability
becomes {\it convective} --- rather than being absolute --- for sufficiently large flow velocities
and, thus, dark solitons are effectively stable in the region around the obstacle.

The flow of a {\it multi-component} BEC past an obstacle was also studied, and the cases of a
two-component \cite{hadi,krav,ourship} and a spinor $F=1$ condensate \cite{spindrag} were analyzed.
It is interesting to note that, as shown in Ref.~\cite{hadi} in the case of a two-component BEC,
the existence of two different speeds of sound provides the possibility for three dynamical regimes:
when both components are subcritical, nucleation of coherent structures does not occur; when both
components are supercritical they both form dark solitons in 1D and vortices or rotating vortex dipoles
in 2D; in the intermediate regime, the nucleation of a dark--anti-dark soliton in 1D or a vortex-lump
configuration in 2D is observed. Furthermore, as shown in Ref.~\cite{ourship}, dark solitons can be
convectively stabilized in the 2D setting at sufficiently high values of the obstacle velocity,
similarly to the case of one-component BECs \cite{kp08}.


\subsection{Matter-wave dark solitons in optical lattices.}
\label{OLs}

Bose-Einstein condensates loaded into periodic optical potentials, so-called {\it optical lattices} (OLs),
have attracted much attention as they demonstrate rich physical properties and nonlinear dynamics
(see, e.g., Refs.~\cite{BECBOOK,pandim,brkon,blochol,chap01:markus,mlas} for reviews).
Optical lattices are generated by a pair of laser beams forming a standing wave which induces
a periodic potential; thus, for a BEC confined in an optical lattice, the trapping potential
in the GP model can be regarded as a superposition of a harmonic (magnetic or optical) trap
and a periodic potential. Particularly, in a quasi-1D setting (generalization to higher-dimensional cases
is straightforward) --- cf. Eq.~(\ref{gpe1d_u}) --- the trap takes the following
dimensionless form (see, e.g., Ref.~\cite{revnonlin}):
\begin{equation}
V(z)=\frac{1}{2}\Omega^2 z^2 + V_0 \cos^2 (kz),
\label{mtol}
\end{equation}
Here, $\Omega$ and $V_0$ denote, respectively, the harmonic trap and OL strengths,
$L \equiv \pi/k = (\lambda/2)\sin(\phi/2)$ is the periodicity of the lattice, with $\lambda$ being the
common wavelength of the two interfering laser beams, and $\phi$ the angle between them. In some cases
(as, e.g., in the experiments of Refs.~\cite{chap01:gap,mko2}), the harmonic potential is very weak as
compared to the optical lattice and, thus, it can be ignored. Then, the stationary states of
the pertinent GP equation --- including solely the OL potential --- can be found in the form of
infinitely extended waves, with the periodicity of the OL, known as {\it nonlinear Bloch waves}
(see, e.g., Ch.~6 in Ref.~\cite{BECBOOK} and references therein). In the same case (i.e., in the absence
of the harmonic potential), if the OL is very deep (compared to the chemical potential), the strongly
spatially localized wave functions at the lattice sites are approximated by Wannier functions
(see, e.g., Ref.~\cite{panwan}) and the {\it tight-binding approximation} can be applied; then,
the continuous GP equation is reduced to the {\it discrete NLS (DNLS) equation} (see, e.g.,
Refs.~\cite{pandim,revnonlin,brkon} and Ch.~13 in Ref.~\cite{BECBOOK}, as well as
Refs.~\cite{pandnls1,pandnls2} for reviews for the DNLS model). Dark solitons, which may
naturally exist in all of the above settings and combinations thereof, have been studied both in
combined harmonic and OL potentials \cite{parker3,gt1,weol,gt2} and in optical lattices (in the absence
of the harmonic trap). In the latter case, various studies have been performed in the frameworks of
the continuous GP equation, as well as its tight-binding approximation counterpart
\cite{bbbb1,yulin,bbbb2,bbbb3,dsolyuri}. Notice that matter-wave dark solitons have also been studied
in double-periodic {\it optical superlattices}
\footnotemark[1]
\footnotetext[1]{Such a potential has the form $V(z)=V_1 \cos(k_1 z)+ V_2 \cos(k_2 z)$, where $k_1$
and $k_2>k_1$ are the primary and secondary lattice wavenumbers, and $V_1$ and $V_2$ are the associated
sublattice amplitudes \cite{chap01:peil}.}
\cite{dsolyuri,super}, while there exists a vast amount of work concerning dark solitons in periodic
media arising in various contexts, such as nonlinear optics \cite{dls1,kivchub,johkiv,dls2,susjoh,manos},
solid-state physics \cite{kontak} and the theory of  nonlinear waves \cite{dmitrypanos}.

\subsubsection{Dark solitons in combined harmonic and OL potentials.}

The stability of matter-wave dark solitons in the combined harmonic and OL potential, was first studied
in Ref.~\cite{weol} by a means of a BdG analysis that was performed in the framework of both the continuous
quasi-1D GP equation and its DNLS counterpart. It was found that in the discrete model stationary dark
solitons located at the minimum of the harmonic trap are, generally, subject to a weak oscillatory instability,
which manifests itself as a shift of the soliton from its initial location,
accompanied by quasi-periodic oscillations. On the other hand, in the continuous GP model, dark solitons may
be stable, with the (in)stability determined by the period and amplitude of the OL. In any case, the dark
solitons are robust and if the oscillatory instability is present, it sets in at large times.

The dynamics of dark solitons in the combined harmonic and OL potential can be studied upon distinguishing
physically relevant cases, depending on the competition of the characteristic spatial scales of the
problem \cite{gt1}. Particularly, assuming that the harmonic trap varies slowly on the soliton scale,
i.e., $w = 1/\cos \phi \sim \xi \ll \Omega^{-1}$ (where $w$ is the soliton width for chemical potential
$\mu=1$, $\phi$ is the soliton phase angle, and $\xi$ the healing length), the following three cases
can readily be identified:
(a) the case of a long-period OL, with $L \gg \xi$,
(b) the case of a short-period OL, with $L \ll \xi$, and
(c) the intermediate case, with $L \sim \xi$.
Then, if the OL strength is sufficiently small, the soliton dynamics in cases (a) and (b) can be treated
in the framework of the adiabatic approximation (see Sec~\ref{adiabatic}). Particularly, as shown in
Ref.~\cite{gt1}, case (a) can be studied by means of the Hamiltonian approach of the perturbation theory,
and case (b) by means of a multi-scale expansion method (treating $k^{-1}$ as a small parameter);
this way, it can be shown that in both cases the dark soliton behaves as an effective classical particle,
performing harmonic oscillations in the presence of the trap of Eq.~(\ref{mtol}). The oscillation frequency,
which is different from its characteristic value $\Omega/\sqrt{2}$, is  modified by the presence of the
lattice according to the equations:
\begin{equation}
\omega_{\rm osc} = \sqrt{\frac{1}{2}\Omega^2-V_0 k^2}, \qquad
\omega_{\rm osc} = \frac{\Omega}{\sqrt{2}}\left(1- \frac{7}{256} \frac{V_0^2}{k^4}\right),
\label{ofmtol}
\end{equation}
for cases (a) and (b), respectively. As concerns the more interesting case (c)
(see Refs.~\cite{parker3,gt1}), it can be shown that if the dark soliton is initially placed quite
close to the bottom of a well of the OL potential, it remains there for a rather long time; eventually,
however, it escapes due to the radiation-loss mechanism, and then performs large-amplitude oscillations
in the condensate. Furthermore, if the harmonic trap is weak enough, the soliton eventually decays.
In fact, as discussed in Ref.~\cite{parker3}, the OL causes a dynamical instability (because the dark
soliton has to ``traverse'' the potential humps caused by the lattice) resulting in a faster decay of
the soliton than if it was evolving in the presence of the harmonic trap only: the presence of the
lattice dephases the sound waves emitted by the soliton, hence reducing the effectiveness of the
soliton to get stabilized by reabsorbing the sound waves (see discussion in Sec.~\ref{radiation}).
Nevertheless, according to the observations of Ref.~\cite{gt1} that the soliton may remain stationary
for a relatively long time, in Ref.~\cite{gt2} (see also Ref.~\cite{gt1}) it was proposed that
a time-dependent OLs may either (i) capture a moving dark soliton or (ii) capture and drag
a stationary soliton, bringing it to a pre-selected final destination. Notice that the transfer
mechanism is robust as long as adiabaticity of the process is ensured
(i.e., for sufficiently small speeds of the moving OL).

\subsubsection{Dark solitons in optical lattices and superlattices.}

As mentioned above, dark solitons in OLs and superlattices have also been studied, in the absence of the
harmonic trap, in the frameworks of the continuous and discrete NLS equations. Particularly, in
Ref.~\cite{bbbb1}, a DNLS model was derived in the tight-binding approximation --- i.e., for a
single-isolated band in the Floquet-Bloch spectrum --- which was used to study matter-wave dark solitons.
Later, in Ref.~\cite{yulin}, a continuous coupled-mode model was used to study the existence and stability
of, so-called, {\it dark lattice solitons}, while a more general analysis was presented in Ref.~\cite{bbbb2};
in that work, a continuous GP model with periodic potential was shown to support stable stationary dark
solitons for both attractive and repulsive interatomic interactions, which were found numerically
\cite{bbbb2} (see also relevant results in \cite{bbbb3}).

In a more recent work \cite{dsolyuri}, where both regular optical lattices and superlattices were considered,
it was shown that each type of nonlinear Bloch wave can serve as a stable background supporting dark solitons.
This way, different families of dark solitons, originating within the bands of the Floquet-Bloch spectrum,
were found and their dynamical properties were analyzed. In particular, considering the continuous analogue
of the Peierls-Nabarro potential (see, e.g., Ref.~\cite{frenkelyuri}) in discrete lattices, it was shown that
the mobility and interaction properties of the dark solitons can be effectively controlled by changing the
structure of the optical superlattice; moreover, following the ideas of Ref.~\cite{gt2}, time-dependent
superlattices were also shown to control the static and dynamical properties of matter-wave solitons
\cite{super}.

Here we should point out that all the above mentioned studies on the dynamics of matter-wave dark solitons
in optical lattices were carried out in the framework of the GP mean-field theory. Nevertheless, it is worth
emphasizing that the GP equation is inadequate for dealing with several important aspects of ultracold
bosons in optical lattices, such as the superfluid-to-Mott insulator phase transition (see, e.g.,
Refs.~\cite{SMIT,SMIT2}) or, more generally, strong correlation effects (see, e.g., the
review \cite{mlas}). Thus, more recently, studies on the quantum dynamics of dark solitons have started
to appear. In that regard, it is relevant to mention that matter-wave dark solitons were studied in the
context of the Bose-Hubbard model \cite{carrentds1,carrentds2}, and it was found that dark soliton
collisions become inelastic, in strong contrast to the predictions of mean-field theory. A conclusion
of the above works \cite{carrentds1,carrentds2} is that the lifetime and collision properties of matter-wave
dark solitons in optical lattices may provide a clear signature of quantum effects. Additionally, in another
recent work \cite{lewDS}, dark solitons were studied in the vicinity of the superfluid-to-Mott insulator
transition; particularly, in this work \cite{lewDS}, antisymmetric eigenstates corresponding to standing
solitons, as well as propagating solitons created by phase-imprinting, were presented and the soliton
characteristics were found to depend on quantum fluctuations.

From the viewpoint of experiments, trains of stationary dark solitons were observed in a $^{87}$Rb
condensate confined in a $3$D harmonic trap and a 1D OL \cite{scottOL}. The underlying mechanism for the
formation of such structures were multiple Bragg reflections caused by displacing the harmonic trap and,
thus, setting the BEC into motion. Due to the dimensionality of the system, the solitons were found to be
subject to the snaking instability, giving rise to the subsequent formation of vortex rings
(see Sec.~\ref{snaking}), similarly to the observations of the pertinent JILA experiment (but without the
OL) \cite{bpa}.

\subsection{Matter-wave dark solitons at finite temperatures.}
\label{finiteT}

So far, we have considered the stability and dynamics of matter-wave dark solitons at
zero temperature, $T=0$. Nevertheless, as experiments are obviously performed at finite
temperatures, it is relevant to consider the dissipative instability of dark solitons induced
by the thermal excitations that naturally occur. This problem was first addressed in Ref.~\cite{fms},
where a kinetic-equation approach, together with a study of the Bogoliubov-de Gennes (BdG) equations,
was used. In this work, it was found that the dark soliton center obeys an equation of motion which
includes an {\it anti-damping} term [similarly to Eq.~(\ref{linearized})], which is nonzero (zero)
for finite (zero)-temperature. The behavior of the solutions of this equation of motion incorporating
the anti-damping term can be used to explain --- at least qualitatively --- the soliton dynamics
observed in experiments: solitons either decay fast (for high temperatures) \cite{han1,nist,han2}
or perform oscillations (for low temperatures) \cite{hamburg,hambcol,kip,draft6} of growing amplitude
and eventually decay, so that the system finally relaxes to its ground state
(see also discussion below for the role of the anti-damping term).

Dark soliton dynamics in BECs at finite temperatures was also studied in other works by means
of different approaches. In particular, in Ref.~\cite{npppd}, the problem was treated in the framework
of a mean-field model, namely the so-called {\it dissipative GP equation} (see below); this equation
incorporates a damping term, first introduced phenomenologically \cite{lp} and later justified from
a microscopic perspective (see, e.g., the review \cite{npprev}). On the other hand, in
Refs.~\cite{npp1,npp2} the same problem was studied numerically, using coupled Gross-Pitaevskii and
quantum Boltzmann equations, which include the mean field coupling and particle exchange between the
condensate and the thermal cloud. Furthermore, in Refs.~\cite{zurek3,us}, finite-temperature dynamics
of dark solitons was studied by means of the so-called {\it stochastic GP equation} (see, e.g.,
Ref.~\cite{npprev}), while in Ref.~\cite{ruoste} quantum effects on dark solitons were additionally
studied in the framework of the truncated Wigner approximation (see, e.g.,
Refs.~\cite{steelolsen,sinatra} for this approach); we also note that in the recent work \cite{gangardt},
the dissipative dynamics of a dark soliton at temperatures $T$, lower than the chemical potential $\mu$
of the background Bose liquid, was studied. In the work \cite{us}, it was shown that for sufficiently
low temperatures and certain parameter regimes, averaged dark soliton trajectories obtained by the
stochastic GP equation, are in a very good agreement with results obtained by the dissipative GP model.
Thus, the results of Ref.~\cite{us} indicate that the use of the dissipative GP equation in studies
of dark solitons in finite-temperature BECs
(a) can reasonably be justified from a microscopic perspective, and
(b) allows for an analytical description of the problem, by employing techniques exposed
in Sec.~\ref{adiabatic}, provided that the dynamics of the thermal cloud does not play the dominant role.

To be more specific, we follow Ref.~\cite{us}, and express, at first, the dissipative GP model in the
following dimensionless form,
\begin{eqnarray}
(i-\gamma) \partial_{t}\psi = \left[\frac{1}{2} \partial_{z}^{2}
+ V(z) + |\psi|^2 - \mu \right]\psi,
\label{dgpe}
\end{eqnarray}
where units are the same to the ones used for Eq.~(\ref{dim1dgpe}), and the dimensionless dissipation
parameter $\gamma$ can be connected with temperature by means of the relation
$\gamma \propto (ma^2 k_{B}T)/(\pi \hbar^2)$, where $k_B$ is Boltzmann's constant.
Dark matter-wave soliton dynamics can be studied analytically in the framework of Eq.~(\ref{dgpe}),
by employing the Hamiltonian approach of the perturbation theory for dark solitons (see Sec.~\ref{hamap}).
In particular, we assume that the condensate dynamics involves a fast scale of relaxation of the background
to the ground state, while the dark soliton subsequently evolves on top of the relaxed ground state;
then, it is possible to derive the perturbed NLS Eq.~(\ref{pnlsd}) for the dark soliton wave function
$\psi_{s}$, with a perturbation $Q(\psi_{s})$ [cf. Eq.~(\ref{pertQ})] incorporating the additional term
$2\gamma \mu \partial_t \psi_{s}$. Then, following the procedure of Sec.~\ref{hamap}, we end up with
the following equation of motion for the dark soliton center $z_0$ \cite{us}:
\begin{equation}
\frac{d^{2}z_{0}}{dt^{2}} = \left[\frac{2}{3}\gamma  \frac{dz_{0}}{dt}
- \left( \frac{\Omega}{\sqrt{2}}\right)^2 z_0 \right].
\left[1 - \left(\frac{dz_{0}}{dt}\right)^2 \right]
\label{nl_em}
\end{equation}
In the case of nearly-black solitons with $dz_0/dt$ sufficiently small, Eq.~(\ref{nl_em}) can be reduced
to the following linearized form (similar to the equation of motion of Ref.~\cite{fms}):
\begin{equation}
\frac{d^{2}z_{0}}{dt^{2}} - \frac{2}{3}\gamma \mu \frac{dz_{0}}{dt}
+ \left( \frac{\Omega}{\sqrt{2}}\right)^2 z_0 = 0.
\label{linem}
\end{equation}
In the limiting case of zero temperature, $\gamma =0$, Eq.~(\ref{linem}) is reduced to Eq.~(\ref{eqmd})
[for the harmonic trap $V(z)=(1/2)\Omega^{2} z^{2}$]. On the other hand, at finite temperatures,
$\gamma \ne 0$, Eq.~(\ref{linem}) incorporates the anti-damping term ($\propto -dz_0/dt$).
Although it may sound counter-intuitive, this term
describes the dissipation of the dark soliton due to the interaction with the thermal cloud:
in fact, this term results in the acceleration of the soliton towards the velocity of sound, i.e.,
the soliton becomes continuously grayer and, eventually, the soliton state transforms to the ground state
of the condensate.

Explicit solutions of Eq.~(\ref{linem}) can readily be obtained in the form of $z_0 \propto \exp(s_{1,2} t)$,
where $s_{1,2}$ are the roots of the auxiliary equation $s^2 -(2/3)\gamma\mu s + (\Omega/\sqrt{2})^2 = 0$
and are given by:
\begin{equation}
s_{1, 2} = \frac{1}{3}\gamma\mu \pm \left(\frac{\Omega}{\sqrt{2}}\right) \sqrt{\Delta},
\,\,\,\,\,
\Delta =  \left( \frac{\gamma}{\gamma_{cr}} \right)^2 -1,
\,\,\,\,\,
\gamma_{cr}=\frac{3}{\mu}\left(\frac{\Omega}{\sqrt{2}}\right).
\label{s12}
\end{equation}
In Ref.~\cite{us} (see also relevant work in Ref.~\cite{dcds}), the temperature dependence of these
eigenvalues associated with the dark soliton dynamics was compared to the temperature dependence
of the eigenvalues of the pertinent anomalous mode of the system \cite{us}, and the agreement between the
two was found to be excellent. Both the motion eigenvalues [cf. Eq.~(\ref{s12})] and the anomalous mode
eigenvalues (derived by a BdG analysis) undergo Hopf bifurcations as the dissipation (temperature)
is increased/decreased --- leading to an exponential/oscillatory instability of the dark soliton --- with
the respective bifurcation diagrams being almost identical. A similar situation occurs in the case of
multiple-dark solitons as well: as shown in Ref.~\cite{dcds}, eigenvalues derived by coupled equations
of motions [similar to the ones in Eqs.~(\ref{sys1})--(\ref{sys2})] for two- or three-dark solitons,
were again found to be almost identical to the ones of the anomalous modes of the system.


\section{Conclusions and perspectives}
\label{conclusions}

We have presented the recent progress on the study of dark solitons in atomic BECs, including
analytical, numerical and experimental results. In fact, although the main body of this work was
basically devoted to the theoretical aspects of this topic, we have tried to connect the theoretical
results to pertinent experimental observations. In that regard, we have particularly tried to highlight
the close connection between theory and experiments and the reasonable agreement between the two.

Matter-wave dark solitons were predicted to occur in BECs as early as 1971 \cite{tsuzuki}, but were
observed in experiments only 28 years later, in 1999 \cite{han1}. Although, till then, dark solitons
had already a relatively long history in the context of nonlinear optics (where they were first observed
in experiments on 1987 \cite{fiber1} and studied extensively in theory during the next years \cite{kivpr}),
one can readily realize an emerging interest in them: during the last decade, there have been more than
ten experiments on dark solitons
\cite{han1,nist,dutton,bpa,han2,scottOL,ginsberg2005,engels,hamburg,hambcol,kip,technion,draft6},
and half of them have been conducted very recently \cite{engels,hamburg,hambcol,kip,technion,draft6},
with an unprecedented control over both the condensate and the solitons. As the experimental developments
continuously inspire --- and, at the same time, are guided by --- a huge number of relevant theoretical
works, one may expect that the interest in matter-wave dark solitons will still be growing in the near future.

Although there has been a tremendous progress on our understanding of matter-wave dark solitons in atomic
BECs over the last years, many important issues remain to be addressed or studied in a more systematic way.
A relevant list is appended below.

\begin{itemize}

\item
{\it Beyond Mean-Field.}
Matter-wave dark solitons, being fundamental nonlinear macroscopic excitations of BECs, play an important
role at probing the properties of the condensates at the mesoscale (see, e.g., discussion in
Ref.~\cite{anglin}). In that regard, a quite interesting research direction is the study of these
nonlinear structures, both in theory and in experiments, in various settings and regimes where
{\it thermal} and {\it quantum} effects are important. In fact, mean-field theory can only account
for {\it averaged} results (e.g., soliton decay times \cite{us}), whereas recent experiments
\cite{hamburg,kip,draft6} indicate shot-to-shot variations that could be accounted for by stochastic
approaches.
There exist various experimentally relevant settings --- such as the ones where the number of atoms is small,
particularly those in optical lattices or at very low temperatures --- which enhance the importance of
quantum fluctuations; therefore, the latter should be appropriately included. One interesting question
concerns, for example, the issue of the filling of the dark soliton due to averaging based on thermal
or quantum fluctuations \cite{dz,dz22,dz23} and its relation to the measurement process \cite{Dziarmaga2002}.
Extending this argument, one could use dark soliton experiments to test the regimes of validity of
conventional mean-field theories, a very interesting and fundamental topic in its own right.
From a theoretical standpoint, the above directions seem to be a natural next step in the study of BECs and
their excitations; in fact, relevant work --- based on various approaches {\it beyond the mean-field
approximation} --- has already started
(see, e.g., Refs.~\cite{carrentds1,carrentds2,lewDS,us,ruoste,gangardt})
and is expected to continue even more intensively in the near future.

\item
{\it Mathematical analysis.} Even in the framework of the mean-field approximation, there exist
several theoretical problems which remain unsolved or should be investigated in more detail. A pertinent
example is the study of the persistence and stability of dark solitons in the presence of confining or
periodic potentials: as mentioned in Sec.~\ref{pers}, rigorous results have only been obtained for small,
bounded and decaying potentials \cite{pelpan}, while an analysis of other cases is still missing.
Furthermore, there is still work to be done as concerns the development of perturbation theories
for multiple dark solitons, for dark solitons in multi-component systems, dissipative systems, and others.

\item
{\it Further experiments.} From the viewpoint of experiments, the recent observations of
{\it long-lived} matter-wave dark solitons \cite{hamburg,hambcol,kip,draft6} suggest many other possible
experimental investigations. In fact, there are many interesting problems related to dark solitons, which
require experimental studies. These include (a) the influence of thermal and quantum fluctuations on dark
solitons (as indicated above), (b) investigation of states composed by a large number of dark solitons
(including, so-called, ``soliton gases'' --- see, e.g., Refs.~\cite{kamdsolgas,trillodsolgas}),
(c) observation of vector solitons, such as dark--dark and dark--anti-dark solitons in two-component BECs
or vector solitons with at least one component being a dark soliton in spinor BECs
\cite{bdspinor,obsantos1,epjd,wad2,ieda}, (d) interactions of dark solitons with potential
barriers and studies of the reflectivity/transmittivity of dark solitons \cite{analogies,fr1,BilPav1,parker1},
(e) manipulation of dark solitons in collisionally inhomogeneous environments \cite{constgofx,gofx},
or by means of time-dependent optical lattices \cite{gt1,gt2,super}, and others.

\item
{\it Applications.} Apart from basic theory and relevant experiments, an important question concerns
possible {\it applications} of matter-wave dark solitons. Although there exist some works indicating
the importance of dark solitons in atomic matter-wave interferometers in the nonlinear regime
\cite{appl1,appl2,jo,appl3,interfe} --- a direction which is expected to further be explored --- other
potential applications (similar to the ones related to optical dark solitons \cite{kivpr,book}) remain
to be investigated. As an example we note that matter-wave bright--dark {\it vector solitons} (which have
already been observed \cite{hamburg}) in pseudo-spinor or spinor BECs may provide the possibility
of {\it all-matter-wave waveguiding}: in such a situation, the dark soliton component could build an
effective conduit for the bright component, similar to the all-optical waveguiding proposed in nonlinear
optics \cite{kivpr}. Waveguides of this kind would be useful for applications, such as quantum
switches and splitters emulating their optical counterparts \cite{bld}.

\item
{\it Ultracold Fermi gases.} We finally note that, so far, matter-wave dark solitons have mainly been
studied in the context of ultra-cold Bose gases. Nevertheless, recent progress in the area of
{\it ultra-cold Fermi gases} (see, e.g., Refs.~\cite{fermiRMP,fermiket} for recent reviews), suggest
that (similarly to vortices) dark solitons may be relevant in this context as well. In fact, pertinent
theoretical studies have already started to appear \cite{fermiadh,fermipit,fermihuang}, but there is
still much work to be done towards this direction, both in theory and in experiments.


\end{itemize}



\section*{Acknowledgments}

The author is grateful to Peter Schmelcher who initiated the idea of this review, as well as to
Panos Kevrekidis, Ricardo Carretero-Gonz\'{a}lez, Giorgos Theocharis, Hector Nistazakis and
Vassos Achilleos for discussions, suggestions and help towards the completion of the paper.
Furthermore, it is a great pleasure for the author to acknowledge the invaluable contribution
of all his collaborators in the topic of this work; especially, apart from the above mentioned
colleagues and students, the author names especially Fotis Diakonos, Yuri Kivshar, Volodya Konotop,
Boris Malomed, Markus Oberthaler, Dmitry Pelinovsky, and Nick Proukakis. This work was partially
supported by the Special Account for Research Grants of the University of Athens.


\end{document}